\documentclass[aps,prb,reprint,floatfix,longbibliography,superscriptaddress]{revtex4-2}

\usepackage{graphicx}
\usepackage{dcolumn}
\usepackage{bm}
\usepackage{amsmath,amssymb,bm}
\usepackage[colorlinks=true]{hyperref}
\usepackage{xcolor}

\begin{document}

\title{Parity- and Chirality-Selected Dirac Masses in Hybrid Moir\'e--1D Superlattices}

\author{Hanzhou Tan}
\affiliation{Arts and Sciences, NYU Shanghai, Shanghai, 200124, China}
\affiliation{NYU-ECNU Institute of Physics at NYU Shanghai, Shanghai, 200062, China}

\author{Pilkyung Moon}
\email{pilkyung.moon@nyu.edu}
\affiliation{Arts and Sciences, NYU Shanghai, Shanghai, 200124, China}
\affiliation{NYU-ECNU Institute of Physics at NYU Shanghai, Shanghai, 200062, China}
\affiliation{Department of Physics, New York University, New York, New York 10003, USA}
\affiliation{Department of Physics, Hanyang University, Seoul, 04763, Korea}

\date{\today}

\begin{abstract}
We show that a hybrid moir\'e--1D superlattice gives electrical control over whether a Dirac--Dirac momentum resonance opens a charge-neutrality gap, and through which mass channel. 
In twisted bilayer graphene subject to a layer-dependent unidirectional scalar potential, the charge-neutrality gap is controlled by the Fourier parity of the resonance-selected dressed interlayer harmonic: odd selected harmonics gap same-chirality Dirac-cone pairs, whereas even selected harmonics gap opposite-chirality pairs. A dressed-harmonic two-cone theory shows that this rule follows from two ingredients: the selected harmonic fixes the Pauli sector of the intercone matrix, and the dressed transverse velocities fix the relative chirality of the interacting cones. Because the 1D modulation can reverse the transverse velocity of one layer, it can electrically switch the relative chirality and hence the active mass channel, e.g., turning an even resonance from gapless to gapped. Full-wave continuum miniband calculations confirm both the exact-resonance rule and a finite near-resonant gapped window with quantitative fabrication tolerances. Far away from resonance, the same dressing yields strongly anisotropic but gapless Dirac bands. Hybrid moir\'e--1D superlattices therefore provide a route to programmable mass selection in coupled Dirac systems.
\end{abstract}

\maketitle

\section{Introduction}

Twisted moir\'e Dirac systems provide a powerful setting for miniband engineering, including flat bands and correlated phenomena near the magic angle~\cite{LopesdosSantos2007_PRL,BistritzerMacDonald2011_PNAS,cao2018correlated,cao2018unconventional}, but their low-energy reconstruction is largely fixed once the twist angle is chosen. Unidirectional electrostatic superlattices, by contrast, offer a gate-defined handle on reshaping Dirac cones and generating strong anisotropy~\cite{Park2008_NatPhys_1D,BreyFertig2009_PRL,Li2021_NatNano_1DSL}. Yet robust single-particle gaps at charge neutrality remain difficult in either setting~\cite{po2018origin,Zou2018_PRB,Park2008_NatPhys_1D,BreyFertig2009_PRL}.

A distinct line of work has studied one-dimensional bilayer moir\'e patterns generated by shear, strain, or twist, where the superlattice is encoded in a spatial modulation of the interlayer hopping or other matrix-valued interlayer couplings, revealing non-Abelian-gauge, confinement, narrow-band, and mass-inversion physics~\cite{SanJose2012_PRL,Gonzalez2016_PRB,TimmelMele2020_PRL}. However, bringing two Dirac cones into momentum resonance does not by itself guarantee a gap: the resonant intercone coupling must contain the Pauli component that anticommutes with the effective Dirac kinetic term. This obstruction is explicit in strongly coupled double-walled nanotubes and related coupled-Dirac systems, where exact momentum matching can still leave the spectrum gapless~\cite{PhysRevB.91.035405}.

Distinct from those earlier frameworks, here we show that a hybrid moir\'e--1D superlattice---a layer-dependent scalar superlattice applied on top of a preexisting moir\'e superlattice---provides electrical control over whether a Dirac--Dirac momentum resonance opens a charge-neutrality-point (CNP) gap. 
Using twisted bilayer graphene (TBG) subject to a layer-dependent unidirectional scalar modulation as a representative platform, we show that the scalar potential dresses the layer wave functions, converting each bare first-star TBG tunneling matrix into a family of dressed interlayer tunneling components carrying 1D Fourier momenta.
At a given resonance, momentum conservation selects a single such component; its Fourier parity fixes the leading Pauli sector, while the dressed transverse velocities fix the relative chirality of the two cones. The result is a parity--chirality selection rule: odd selected harmonics gap same-chirality Dirac-cone pairs, whereas even selected harmonics gap opposite-chirality pairs. Because the modulation can reverse the dressed transverse velocity of one layer, the relative chirality and hence the active mass channel become electrically switchable. Full-wave continuum minibands further show a finite near-resonant gapped window, while away from resonance the same dressing produces strongly anisotropic gapless Dirac bands.

The even--odd label introduced here should be distinguished from other even--odd distinctions known in graphene Dirac-coupling mechanisms. In the TBG--1D mechanism studied here, this label refers to the Fourier component of a scalar-dressed interlayer tunneling matrix selected by Dirac--Dirac momentum matching, rather than to a structural commensuration class or to a directly imposed intervalley potential. Consequently, parity alone is not yet the gap-opening criterion: it becomes a mass-selection rule only after the relative chirality of the two dressed cones is specified.

The paper is organized as follows. Section~II introduces the continuum model and the resonant design space. Section~III derives the dressed-harmonic description and the renormalized exact-resonance two-cone model. Section~IV presents the parity--chirality selection rule together with electrical chirality switching, and Sec.~V analyzes the finite near-resonant gapped window. Section~VI discusses Chern triviality, the off-resonant anisotropic regime, and the relation to known even--odd Dirac-coupling mechanisms, and Sec.~VII concludes. The appendices collect the numerical implementation, detailed derivations, and feasibility estimates.

\section{Model and resonant design space}

This section fixes the laboratory-frame geometry, defines the hybrid TBG--1D continuum Hamiltonian, and then identifies the local resonance conditions in the $\mathbf G_{\rm 1D}$ design space. All vectors below are expressed in the fixed laboratory frame. We label graphene valleys by $\xi=\pm1$ and write $\omega=e^{i2\pi/3}$ and $R(\phi)=\bigl(
\begin{smallmatrix}
\cos\phi & -\sin\phi\\
\sin\phi & \cos\phi
\end{smallmatrix}\bigr)$.

\subsection{Geometry and momentum conventions}
\label{sec:reference_graphene_lattice}

We start from a reference monolayer graphene lattice with lattice constant $a\approx 0.246~{\rm nm}$, primitive vectors
\begin{equation}
\mathbf a_1^{(0)}=a(1,0),\qquad
\mathbf a_2^{(0)}=a(1/2,\sqrt3/2),
\end{equation}
and reciprocal vectors $\mathbf b_i^{(0)}$ satisfying $\mathbf a_i^{(0)}\cdot\mathbf b_j^{(0)}=2\pi\delta_{ij}$. The two layers are generated by symmetric rotations,
\begin{equation}
\theta_1=-\theta/2,\qquad \theta_2=+\theta/2,
\end{equation}
so that
\begin{equation}
\mathbf a_i^{(l)}=R(\theta_l)\mathbf a_i^{(0)},
\qquad
\mathbf b_i^{(l)}=R(\theta_l)\mathbf b_i^{(0)}.
\end{equation}
The sublattice coordinates in the reference layer are fixed as
\begin{equation}
\boldsymbol\tau_A=(0,0),\qquad
\boldsymbol\tau_B=\frac{\mathbf a_1^{(0)}-2\mathbf a_2^{(0)}}{3},
\end{equation}
with $\boldsymbol\tau_X^{(l)}=R(\theta_l)\boldsymbol\tau_X$. This choice fixes the Bloch-basis gauge used below for the tunneling phases.

The reference Dirac point is taken as
\begin{equation}
\mathbf K_\xi^{(0)}
=
-\xi\,\frac{2\mathbf b_1^{(0)}+\mathbf b_2^{(0)}}{3},
\end{equation}
and the layer Dirac points are
\begin{equation}
\mathbf K_\xi^{(l)}=R(\theta_l)\mathbf K_\xi^{(0)}.
\end{equation}
The Dirac-point mismatch is therefore
\begin{equation}
\Delta\mathbf K_\xi(\theta)
\equiv
\mathbf K_\xi^{(1)}-\mathbf K_\xi^{(2)}.
\label{eq:S1:DeltaK_def}
\end{equation}
For the symmetric layer rotation used throughout,
\begin{equation}
\Delta\mathbf K_\xi
=
k_\theta(0,\xi),
\qquad
k_\theta
=
\frac{8\pi}{3a}\sin\frac{\theta}{2}.
\end{equation}
This expression is kept in its exact trigonometric form in Sec.~II; no small-$\theta$ expansion is needed here.

The moir\'e reciprocal vectors are
\begin{equation}
\mathbf G_i^{\rm M}
=
\mathbf b_i^{(1)}-\mathbf b_i^{(2)},
\end{equation}
and the corresponding real-space moir\'e vectors satisfy
\begin{equation}
\mathbf L_i^{\rm M}\cdot\mathbf G_j^{\rm M}=2\pi\delta_{ij},
\qquad
|\mathbf L_i^{\rm M}|=\frac{a}{2\sin(\theta/2)}.
\end{equation}
The first-star interlayer tunneling momenta are
\begin{equation}
\mathbf q_0^\xi
=
\mathbf K_\xi^{(1)}-\mathbf K_\xi^{(2)},
\quad
\mathbf q_j^\xi
=
(C_3)^j\mathbf q_0^\xi,
\quad
j=0,1,2,
\label{eq:S:q_j_xi}
\end{equation}
where $C_3=R(2\pi/3)$ acts on momentum vectors. Thus $\mathbf q_0^\xi=\Delta\mathbf K_\xi$ by definition.

\subsection{Hybrid TBG--1D continuum model}
\label{sec:hybrid_continuum_model}
\label{sec:energy_scales_and_dimensionless_parameters}

We use the layer--sublattice basis
\begin{equation}
\Psi
=
(\psi_\mathrm{1A},\psi_\mathrm{1B},\psi_\mathrm{2A},\psi_\mathrm{2B})^\mathrm{T}.
\end{equation}
The single-valley continuum Hamiltonian is
\begin{equation}
H_\xi(\mathbf r)
=
\begin{pmatrix}
h_\xi^{(1)}(-i\nabla)+V_1(\mathbf r)\sigma_0
&
T_\xi^\dagger(\mathbf r)
\\
T_\xi(\mathbf r)
&
h_\xi^{(2)}(-i\nabla)+V_2(\mathbf r)\sigma_0
\end{pmatrix}.
\label{eq:prb_model}
\end{equation}
Here the intralayer Dirac block is
\begin{equation}
h_\xi^{(l)}(\mathbf q)
=
-\hbar v_{\rm F}
\begin{pmatrix}
0 &
(\xi q_x-iq_y)e^{-i\theta_l}
\\
(\xi q_x+iq_y)e^{i\theta_l}
&
0
\end{pmatrix},
\label{eq:prb_h_intralayer}
\end{equation}
where $\mathbf q$ is measured from $\mathbf K_\xi^{(l)}$ in the laboratory frame. Equivalently,
\begin{equation}
h_\xi^{(l)}(\mathbf q)
=
-\hbar v_{\rm F}
\left(\xi q_x\sigma_x^{(l)}+q_y\sigma_y^{(l)}\right),
\end{equation}
with
\begin{equation}
\begin{aligned}
\sigma_x^{(l)}&=(\cos\theta_l)\sigma_x+(\sin\theta_l)\sigma_y,\\
\sigma_y^{(l)}&=-(\sin\theta_l)\sigma_x+(\cos\theta_l)\sigma_y.
\end{aligned}
\label{eq:prb_sigma_layer}
\end{equation}

The Bloch basis underlying Eq.~\eqref{eq:prb_h_intralayer} and the tunneling phases below is
\begin{equation}
|\mathbf k,X,l\rangle
=
\frac{1}{\sqrt N}
\sum_{\mathbf R^{(l)}}
e^{i\mathbf k\cdot(\mathbf R^{(l)}+\boldsymbol\tau_X^{(l)})}
|\phi_{p_z}(\mathbf R^{(l)}+\boldsymbol\tau_X^{(l)})\rangle,
\end{equation}
with $X=A,B$ and $\mathbf k=\mathbf K_\xi^{(l)}+\mathbf q$ in the valley-$\xi$ sector. In this gauge, the first-star interlayer tunneling is
\begin{equation}
T_\xi(\mathbf r)
=
\sum_{j=0}^2
T_j^\xi e^{i\mathbf q_j^\xi\cdot\mathbf r}.
\label{eq:prb_Txi}
\end{equation}
The relaxed two-parameter convention used in the full-wave numerical calculations is
\begin{equation}
T_j^\xi
=
\begin{pmatrix}
u_{\rm AA}
&
u_{\rm AB}\omega^{-\xi j}
\\
u_{\rm AB}\omega^{\xi j}
&
u_{\rm AA}
\end{pmatrix},
\qquad
j=0,1,2.
\label{eq:prb_Tj_relaxed}
\end{equation}
The analytical derivation uses the rigid limit $u_{\rm AA}=u_{\rm AB}\equiv t_0$ unless stated otherwise, whereas the full-wave calculations use the relaxed model~\cite{NamKoshino2017_PRB}. Relaxation changes the relative AA and AB amplitudes, but it does not alter the leading first-star harmonic organization used below to identify the selected Pauli sector. For real $u_{\rm AA}$ and $u_{\rm AB}$, each $T_j^\xi$ is Hermitian; the full Hamiltonian is Hermitian because the off-diagonal blocks in Eq.~\eqref{eq:prb_model} are $T_\xi$ and $T_\xi^\dagger$.

The added one-dimensional scalar potential is layer dependent but sublattice diagonal:
\begin{equation}
\begin{aligned}
V_l(\mathbf r)
&=
\frac{|v_l|}{2}
\cos(\mathbf G_{\rm 1D}\cdot\mathbf r+\phi_l)
\\
&=
\frac{v_l}{4}e^{i\mathbf G_{\rm 1D}\cdot\mathbf r}
+
\frac{v_l^*}{4}e^{-i\mathbf G_{\rm 1D}\cdot\mathbf r},
\qquad
v_l=|v_l|e^{i\phi_l}.
\end{aligned}
\label{eq:prb_Vl}
\end{equation}
The corresponding real-space period vector can be chosen as
\begin{equation}
\mathbf L_{\rm 1D}
=
\frac{2\pi\mathbf G_{\rm 1D}}{G_{\rm 1D}^2},
\qquad
G_{\rm 1D}=|\mathbf G_{\rm 1D}|.
\end{equation}
Unlike in strain- or shear-induced one-dimensional bilayer moir\'e structures, where the superlattice is encoded in the spatial modulation of interlayer hopping or other matrix-valued interlayer couplings~\cite{SanJose2012_PRL,Gonzalez2016_PRB,TimmelMele2020_PRL}, here $T_\xi(\mathbf r)$ remains the standard first-star TBG tunneling and the additional 1D superlattice enters through the block-diagonal scalar potentials $V_l(\mathbf r)$.

We define the natural kinetic energy scales
\begin{equation}
\mathcal E_\theta
=
\hbar v_{\rm F}k_\theta,
\qquad
\mathcal E_{\rm 1D}
=
\hbar v_{\rm F}G_{\rm 1D},
\end{equation}
and the dimensionless couplings
\begin{equation}
u_0
\equiv
\frac{t_0}{\mathcal E_\theta},
\qquad
u_l
\equiv
\frac{v_l}{\mathcal E_{\rm 1D}}
=
|u_l|e^{i\phi_l}.
\label{eq:prb_dimless}
\end{equation}
Here $t_0$, $u_{\rm AA}$, and $u_{\rm AB}$ have dimensions of energy; $u_0$ and $u_l$ are dimensionless.

Finally, for later use we define the local frame associated with the 1D modulation. Writing
$\hat{\mathbf e}_\parallel=\mathbf G_{\rm 1D}/G_{\rm 1D}\equiv(n_x,n_y)$ and $\hat{\mathbf e}_\perp=\hat{\mathbf z}\times\hat{\mathbf e}_\parallel=(-n_y,n_x)$,
\begin{equation}
\begin{aligned}
q_\parallel&=\mathbf q\cdot\hat{\mathbf e}_\parallel=n_xq_x+n_yq_y,\\
q_\perp&=\mathbf q\cdot\hat{\mathbf e}_\perp=-n_yq_x+n_xq_y,\\
\sigma_\parallel^{(l)}&=n_x\sigma_x^{(l)}+n_y\sigma_y^{(l)},\\
\sigma_\perp^{(l)}&=-n_y\sigma_x^{(l)}+n_x\sigma_y^{(l)}.
\end{aligned}
\label{eq:prb_local_axes}
\end{equation}
The sign convention is fixed by Eq.~\eqref{eq:prb_local_axes}. In the analytic frame $\mathbf G_{\rm 1D}=G_{\rm 1D}\hat{\mathbf y}$, one has $q_\parallel=q_y$, $q_\perp=-q_x$, $\sigma_\parallel^{(l)}=\sigma_y^{(l)}$, and $\sigma_\perp^{(l)}=-\sigma_x^{(l)}$, so $q_\perp\sigma_\perp^{(l)}=q_x\sigma_x^{(l)}$. Unless stated otherwise, we focus on layer-symmetric in-phase modulation, $u_1=u_2$, which defines the default same-chirality parameter slice used in the exact-resonance spectra below. Layer-asymmetric modulation will be used in Sec.~\ref{sec:selection_rule} to switch the relative chirality.
\subsection{Resonant design space}
\label{sec:configuration_space}

For a fixed twist angle $\theta$, the Dirac-point mismatch $\Delta\mathbf K_\xi$ is fixed. The experimentally tunable geometric variable in the hybrid structure is therefore the magnitude and direction of the 1D reciprocal vector $\mathbf G_{\rm 1D}$. The leading design branch analyzed quantitatively in the main text is the nearest-cone Dirac--Dirac resonance, in which the shortest valley-conserving first-star separation is supplied by an integer multiple of $\mathbf G_{\rm 1D}$:
\begin{equation}
\Delta\mathbf K_\xi
=
\xi n_{\rm 1D}\mathbf G_{\rm 1D},
\qquad
n_{\rm 1D}\in\mathbb Z.
\label{eq:prb_resonance}
\end{equation}
This condition fixes both the magnitude and direction of $\mathbf G_{\rm 1D}$ for a chosen integer $n_{\rm 1D}$, so exact nearest-cone resonances are isolated points in the two-dimensional $\mathbf G_{\rm 1D}$ design space.
In real space, Eq.~\eqref{eq:prb_resonance} is equivalent to
\begin{equation}
\mathbf L_{\rm 1D}
=
-\frac{3}{2}n_{\rm 1D}\mathbf L_1^{\rm M}
=
\frac{3a}{4\sin(\theta/2)}
\begin{pmatrix}
0\\
n_{\rm 1D}
\end{pmatrix}.
\label{eq:prb_L1D_resonance}
\end{equation}
Therefore the resonant 1D period $\mathbf{L}_\mathrm{1D}$ is collinear with a moir\'e real-space lattice vector $\mathbf{L}_1^\mathrm{M}$ in the nearest-cone family [Figs.~\ref{fig:geometry_design_space}(a) and \ref{fig:geometry_design_space}(b)]. See Appendix~\ref{sec:S_real_space_stacking_intuition} for a registry-based interpretation and Fig.~\ref{fig:reciprocal_periodicity} for the reciprocal-space periodicity of the representative resonant configurations with $n_{\rm 1D}=1,2,3$.

It is useful to introduce, at this purely kinematic stage, the integer label for the 1D reciprocal-vector transfer required by momentum conservation. Since $\mathbf q_0^\xi=\Delta\mathbf K_\xi$, the nearest-cone resonance condition can be rewritten as
\begin{equation}
\mathbf q_0^\xi+p_0\mathbf G_{\rm 1D}=\mathbf 0,
\qquad
p_0=-\xi n_{\rm 1D}.
\label{eq:prb_p0_def}
\end{equation}
Here $p_0$ is only a momentum-conservation index: it counts how many 1D reciprocal vectors are carried by the resonant process. Its Hamiltonian meaning appears in Sec.~\ref{sec:main_dressed_harmonic}, where the scalar-potential dressing turns the interlayer tunneling into Fourier components $\tilde T_{j,p}^\xi$ and the component with this same index, $\tilde T_{0,p_0}^\xi$, becomes the direct resonant matrix element. Thus, throughout the nearest-cone branch, the phrase ``$n_{\rm 1D}$ parity'' is shorthand for the parity of the selected integer $p_0$.

\begin{figure*}[t]
  \centering
  \includegraphics[width=\textwidth]{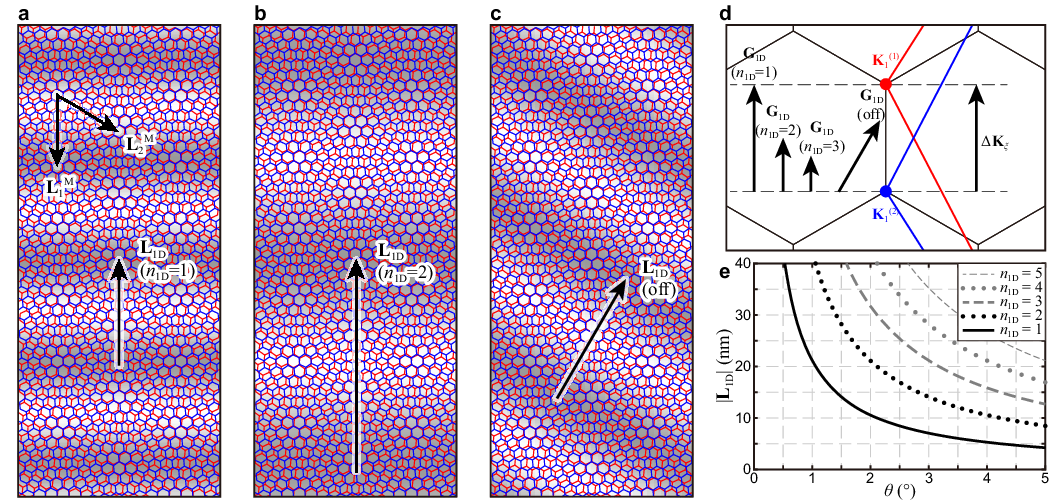}
  \caption{\textbf{a}--\textbf{c}. Lattice structures and unidirectional potential profiles of TBG--1D ($\theta=10^\circ$). Red and blue lines denote the lattices of layers 1 and 2, respectively, and the unidirectional shading represents the scalar potential in Eq.~\eqref{eq:prb_Vl}. Panels \textbf{a} and \textbf{b} show nearest-cone resonant configurations with $n_{\rm 1D}=1$ and $2$, while panel \textbf{c} shows an off-resonant configuration. \textbf{d}. Reduced Brillouin zone of TBG near $\mathbf K_1^{(l)}$ and representative $\mathbf G_{\rm 1D}$ vectors for $n_{\rm 1D}=1,2,3$ and an off-resonant choice. \textbf{e}. Resonant real-space period $|\mathbf L_{\rm 1D}|$ for $n_{\rm 1D}=1$ to $5$ as a function of twist angle. 
  The shaded potential profiles illustrate one choice of the registry phases $\phi_l$; Appendix~\ref{sec:S_real_space_stacking_intuition} discusses the corresponding real-space registry visualization of the nearest-cone even--odd label.  
  }
  \label{fig:geometry_design_space}
\end{figure*}

The nearest-cone condition in Eq.~\eqref{eq:prb_resonance} is not the most general local momentum resonance. A 1D momentum transfer can also compensate a moir\'e-reciprocal translated copy of any first-star separation:
\begin{equation}
\begin{gathered}
\mathbf q_j^\xi+\mathbf G_{\rm M}+p\,\mathbf G_{\rm 1D}=0,
\qquad p\in\mathbb Z,\\
\mathbf G_{\rm M}=m_1\mathbf G_1^{\rm M}+m_2\mathbf G_2^{\rm M},
\qquad m_{1,2}\in\mathbb Z,
\end{gathered}
\label{eq:prb_umklapp_resonance}
\end{equation}
with $j=0,1,2$. The integer $p$ in Eq.~\eqref{eq:prb_umklapp_resonance} has the same kinematic meaning as $p_0$ in Eq.~\eqref{eq:prb_p0_def}; it is the number of 1D reciprocal vectors carried by the resonant process. Branches that might look like fractional-$n_{\rm 1D}$ resonances if described only by Eq.~\eqref{eq:prb_resonance} are instead included in Eq.~\eqref{eq:prb_umklapp_resonance} with an integer $p$ and a nonzero moir\'e reciprocal vector $\mathbf G_{\rm M}$. For these generalized branches, the robust parity label is $p$, not an apparent fractional nearest-cone index (Appendix~\ref{sec:S_moire_umklapp_resonances}).

The main numerical results focus on the nearest-cone branch $j=0$, $\mathbf G_{\rm M}=0$, because it is the leading local resonance in both momentum scale and harmonic order (Appendix~\ref{sec:S_moire_umklapp_resonances}). In the first-star TBG geometry, the shortest vectors in the set $\mathbf q_j^\xi+\mathbf G_{\rm M}$ have magnitude $k_\theta$ and are precisely the nearest-cone separations. Translated branches with $|\mathbf q_j^\xi+\mathbf G_{\rm M}|>k_\theta$ either require a larger $|\mathbf G_{\rm 1D}|$ for the same $|p|$ or, at a fixed long-period design, a larger $|p|$. In the weak-to-moderate modulation regime such higher-$|p|$ harmonic components are smaller, so Eq.~\eqref{eq:prb_resonance} gives the dominant branch to analyze quantitatively. Once a resonant cone pair is selected and all other channels are remote, the local two-cone mass criterion is the same in structure: the selected dressed interlayer component supplies a Pauli sector, and the relative chirality determines whether that sector acts as a Dirac mass.

Deviations from a nearest-cone resonant point are measured by the momentum mismatch
\begin{equation}
\delta\mathbf q_\xi^{(n_{\rm 1D})}
\equiv
\Delta\mathbf K_\xi
-
\xi n_{\rm 1D}\mathbf G_{\rm 1D}.
\label{eq:S1:mm_def}
\end{equation}
The near-resonant regime, where $\delta\mathbf q_\xi^{(n_{\rm 1D})}$ is small but nonzero, is analyzed in Appendix~\ref{sec:near_resonance}.

Finally, momentum resonance and global periodicity are distinct notions. The combined moir\'e--1D structure is globally periodic only if the reciprocal vectors generate a rank-2 lattice, i.e.,
\begin{multline}
\exists\,(m_1,m_2,m_3)\in\mathbb Z^3\setminus\{(0,0,0)\}:\\
m_1\mathbf G_1^{\rm M}
+
m_2\mathbf G_2^{\rm M}
+
m_3\mathbf G_{\rm 1D}
=
\mathbf 0.
\label{eq:global_periodicity}
\end{multline}
When Eq.~\eqref{eq:global_periodicity} holds, a reduced Brillouin zone exists for the combined reciprocal lattice. Otherwise the geometry is quasiperiodic, even if a local Dirac--Dirac resonance of Eq.~\eqref{eq:prb_umklapp_resonance} is present.

\section{Dressed-harmonic theory at exact resonance}
\label{sec:main_dressed_harmonic}

At the nearest-cone resonance of Eq.~\eqref{eq:prb_resonance}, we introduced the kinematic momentum-transfer label, $p_0$ [Eq.~\eqref{eq:prb_p0_def}].
Here we reveal the meaning of $p_0$ in the Hamiltonian by using a dressed representation. 
The intralayer step follows the standard unitary treatment of monolayer graphene under a unidirectional scalar potential~\cite{Park2008_NatPhys_1D,BreyFertig2009_PRL}. The essential extension in the present bilayer problem is that the transformation is layer resolved: it acts on the two sides of each interlayer tunneling matrix and converts the bare first-star matrices into scalar-dressed 1D Fourier components, one of which is selected at a Dirac--Dirac resonance.

The reduction below uses the local momentum and Pauli-axis convention of Eq.~\eqref{eq:prb_local_axes}. We keep the 1D dressing nonperturbatively in the amplitudes $u_l$ through Bessel coefficients, expand the low-energy Hamiltonian to linear order in the reduced momentum $\mathbf q$ about the resonant cones, and integrate out nonresonant moir\'e channels to leading order in $u_0^2$. The Pauli-sector classification is then taken at leading order in small twist angle $\theta$; the finite-$\theta$ terms omitted from the displayed selection-rule formula are $\mathcal O(t_0\theta)$. Eliminated 1D strip sidebands produce only $\mathcal O(\hbar v_{\rm F}q_\perp^2/G_{\rm 1D})$ corrections, so the main-text theory is controlled within the linear Dirac regime.

\subsection{Layer-dependent unitary rotation and harmonic dressing}

For the analytic derivation we choose the local frame in which $\mathbf G_{\rm 1D}=G_{\rm 1D}\hat{\mathbf y}$. Then $q_\parallel=q_y$ and $q_\perp\sigma_\perp^{(l)}=q_x\sigma_x^{(l)}$, even though the global geometry remains defined in the fixed laboratory frame of Sec.~\ref{sec:reference_graphene_lattice}. We apply the layer-dependent sublattice rotation
\begin{equation}
U_l(y)
\equiv
\exp\!\left[\frac{i}{2}\alpha_l(y)\sigma_y^{(l)}\right],
\quad
\alpha_l'(y)
\equiv
\frac{2}{\hbar v_{\rm F}}V_l(y),
\label{eq:prb_Ul}
\end{equation}
which, for the scalar potential of Eq.~\eqref{eq:prb_Vl}, gives
\begin{equation}
\alpha_l(y)=|u_l|\sin(G_{\rm 1D}y+\phi_l).
\label{eq:prb_alpha}
\end{equation}
The block-diagonal transformation $U(y)=\mathrm{diag}[U_1(y),U_2(y)]$ maps the original Hamiltonian to the isospectral form
\begin{equation}
\begin{aligned}
\tilde H_\xi(\mathbf r)
&=U^\dagger(y)H_\xi(\mathbf r)U(y) \\
&=
\begin{pmatrix}
\tilde h_\xi^{(1)}(-i\nabla) & \tilde T_\xi^\dagger(\mathbf r)\\
\tilde T_\xi(\mathbf r) & \tilde h_\xi^{(2)}(-i\nabla)
\end{pmatrix}.
\end{aligned}
\label{eq:prb_dressed_full}
\end{equation}
Because $U(y)$ is unitary, Eq.~\eqref{eq:prb_dressed_full} is Hermitian and has the same spectrum as $H_\xi$.

The scalar potential is absorbed into the dressed intralayer kinetic terms,
\begin{equation}
\begin{aligned}
\tilde h_\xi^{(l)}(-i\nabla)=
-\hbar v_{\rm F}\bigl[
&\xi\bigl(\cos\alpha_l(y)\sigma_x^{(l)}-\sin\alpha_l(y)\sigma_z\bigr)(-i\partial_x) \\
&+\sigma_y^{(l)}(-i\partial_y)
\bigr].
\end{aligned}
\label{eq:prb_dressed_intra}
\end{equation}
The periodic functions in Eq.~\eqref{eq:prb_dressed_intra} have the Fourier expansions
\begin{equation}
\begin{aligned}
\cos\alpha_l(y)&=\sum_{p\in\mathbb Z}C_{l,p}e^{ipG_{\rm 1D}y},\\
\sin\alpha_l(y)&=\sum_{p\in\mathbb Z}S_{l,p}e^{ipG_{\rm 1D}y},
\end{aligned}
\label{eq:prb_CS_def}
\end{equation}
with
\begin{equation}
\begin{aligned}
C_{l,p}&=
\begin{cases}
J_p(|u_l|)e^{ip\phi_l}, & p\ \text{even},\\
0, & p\ \text{odd},
\end{cases} \\
S_{l,p}&=
\begin{cases}
0, & p\ \text{even},\\
-iJ_p(|u_l|)e^{ip\phi_l}, & p\ \text{odd}.
\end{cases}
\end{aligned}
\label{eq:prb_CS}
\end{equation}
Thus $\cos\alpha_l$ contains only even harmonics and $\sin\alpha_l$ only odd harmonics. All intralayer harmonic mixing in the dressed representation enters through the transverse kinetic term proportional to $q_\perp$.

The bilayer-specific step is that the same transformation also makes the interlayer tunneling periodic in the 1D coordinate:
\begin{equation}
\begin{aligned}
\tilde T_\xi(\mathbf r)
&=\sum_{j=0}^2 \tilde T_j^\xi(y)e^{i\mathbf q_j^\xi\cdot\mathbf r}, \\
\tilde T_j^\xi(y)
&\equiv U_2^\dagger(y)T_j^\xi U_1(y)
=\sum_{p\in\mathbb Z}\tilde T_{j,p}^\xi e^{ipG_{\rm 1D}y}.
\end{aligned}
\label{eq:prb_Tdress}
\end{equation}
After the expansion in Eq.~\eqref{eq:prb_Tdress}, the kinematic matching condition in Eq.~\eqref{eq:prb_p0_def} selects a definite Fourier component of the dressed tunneling. For the nearest-cone branch, the only $j=0$ component with zero residual momentum between the two folded central cones is
\begin{equation}
\mathcal M^{(n_{\rm 1D})}\equiv \tilde T_{0,p_0}^\xi,
\qquad
p_0=-\xi n_{\rm 1D}.
\label{eq:prb_M_selected}
\end{equation}
Components with $p\neq p_0$ carry a residual momentum $(p-p_0)\mathbf G_{\rm 1D}$ and therefore do not directly couple the two central cone sectors.

The nonselected Fourier components, together with the $j=1,2$ first-star channels, are remote in this branch: At the leading order retained below, their smooth low-energy effect is included through the remote-channel self-energy. General moir\'e-Umklapp resonances satisfying Eq.~\eqref{eq:prb_umklapp_resonance} are described by the same logic after replacing the nearest-cone tuple $(0,\mathbf 0,p_0)$ by the selected tuple $(j,\mathbf G_{\rm M},p)$ in the appropriate folded basis.

\subsection{Central-harmonic projection, resonant harmonic, and quasiparticle weight}

In the dressed representation the Hamiltonian is periodic along the 1D direction. 
For a fixed reduced momentum $\mathbf q=(q_\perp,q_\parallel)$, momenta whose longitudinal components differ by integer multiples of $G_{\rm 1D}$ are folded into the same 1D reduced zone; we label these sideband sectors by $m\in\mathbb Z$. 
The two low-energy Dirac cones involved in the nearest-cone resonance are the central sectors, $m=0$, of the two layers. 
Projecting the dressed intralayer Hamiltonian onto these central sectors gives, to linear order in $\mathbf q$,
\begin{equation}
\begin{aligned}
\tilde h_{\xi,{\rm CH}}^{(l)}(\mathbf q)
&=
-\hbar v_{\rm F}\bigl[
\xi \eta_l q_\perp\,\sigma_\perp^{(l)}+q_\parallel\,\sigma_\parallel^{(l)}
\bigr],\\
\eta_l&\equiv J_0(|u_l|).
\end{aligned}
\label{eq:prb_CH}
\end{equation}
This central-harmonic intralayer block keeps the 1D velocity dressing exactly in $|u_l|$ through $J_0$. 
The eliminated $m\neq0$ sidebands are separated by momenta of order $G_{\rm 1D}$ and couple back to the central sector only through the transverse kinetic term, so their leading correction is $\mathcal O(\hbar v_{\rm F}q_\perp^2/G_{\rm 1D})$ and does not affect the linear Dirac theory.

Together with the selected central-cone matrix element $\mathcal M^{(n_{\rm 1D})}=\tilde T_{0,p_0}^\xi$ defined in Eq.~\eqref{eq:prb_M_selected}, Eq.~\eqref{eq:prb_CH} gives the minimal two-cone problem at exact resonance. 
For a generalized resonance of Eq.~\eqref{eq:prb_umklapp_resonance}, the same local two-cone analysis applies after replacing $\tilde T_{0,p_0}^\xi$ by the selected component $\tilde T_{j,p}^\xi$ in the appropriate folded basis.

The full finite-$\theta$ expression for $\tilde T_{0,p}^\xi$ is given in Eq.~\eqref{eq:S:T0_evenodd_p} of Appendix~\ref{sec:exact_resonance}. In the main text we need only its leading small-angle Pauli support. With
\begin{equation}
u_\pm=u_1\pm u_2=|u_\pm|e^{i\phi_\pm},
\quad
\mathcal J_{\pm,p}=J_p(|u_\pm|/2)e^{ip\phi_\pm},
\label{eq:prb_Jpm_def}
\end{equation}
one obtains
\begin{equation}
\tilde T_{0,p}^\xi =
\begin{cases}
t_0\bigl(\mathcal J_{-,p}\sigma_0+\mathcal J_{+,p}\sigma_x\bigr)
+\mathcal O(t_0\theta),
& p\ \text{even},\\[1mm]
t_0\bigl(\mathcal J_{-,p}\sigma_y+i\mathcal J_{+,p}\sigma_z\bigr)
+\mathcal O(t_0\theta),
& p\ \text{odd}.
\end{cases}
\label{eq:prb_parity}
\end{equation}
Thus the selected harmonic lies, to leading order, in $\mathrm{span}\{\sigma_0,\sigma_x\}$ for even $p$ and in $\mathrm{span}\{\sigma_y,\sigma_z\}$ for odd $p$. For the nearest-cone branch, the parity of $p_0=-\xi n_{\rm 1D}$ is the parity of $n_{\rm 1D}$, so $n_{\rm 1D}$ fixes the leading Pauli sector of the resonant coupling. For a generalized moir\'e-Umklapp branch, the robust parity label is instead the selected integer harmonic $p$ in Eq.~\eqref{eq:prb_umklapp_resonance}. The omitted finite-angle terms are $\mathcal O(t_0\theta)$ admixtures to the displayed Pauli sectors and are kept in the appendix derivation.

Then, we integrate out the nonresonant moir\'e channels. Linearizing the remote-channel self-energy to first order in $(E,\mathbf q)$ and retaining the leading $\mathcal O(u_0^2)$ contribution gives the quasiparticle weight
\begin{equation}
Z=\frac{1}{1+6u_0^2},
\label{eq:prb_Z}
\end{equation}
and the dressed velocities
\begin{equation}
\begin{aligned}
v_\perp^{(l)}&=Z v_{\rm F}\bigl[J_0(|u_l|)-3u_0^2\bigr],\\
v_\parallel^{(l)}&=Z v_{\rm F}(1-3u_0^2).
\end{aligned}
\label{eq:prb_vperp}
\end{equation}
Applying the same quasiparticle factor once to the summed self-energy and resonant coupling gives the renormalized two-cone Hamiltonian
\begin{widetext}
\begin{equation}
H_\xi^{(\mathrm{2c})}(\mathbf q)=
\begin{pmatrix}
-\hbar\bigl(\xi v_\perp^{(1)}q_\perp\,\sigma_\perp^{(1)}+v_\parallel^{(1)}q_\parallel\,\sigma_\parallel^{(1)}\bigr) &
\bigl(Z\mathcal M^{(n_{\rm 1D})}\bigr)^\dagger\\
Z\mathcal M^{(n_{\rm 1D})} &
-\hbar\bigl(\xi v_\perp^{(2)}q_\perp\,\sigma_\perp^{(2)}+v_\parallel^{(2)}q_\parallel\,\sigma_\parallel^{(2)}\bigr)
\end{pmatrix}.
\label{eq:prb_two_cone}
\end{equation}
\end{widetext}
Appendices~\ref{sec:dressed_rep} and \ref{sec:exact_resonance} provide the reduced-zone derivation, the remote-channel self-energy, and the full validity discussion.

Equation~\eqref{eq:prb_two_cone} is Hermitian by construction and organizes the exact-resonance problem into three logically distinct ingredients. For the nearest-cone branch, momentum matching fixes the interacting cone pair and the integer 1D momentum-transfer label $p_0$; the parity of this selected harmonic fixes the leading Pauli sector of the resonant interlayer matrix $\mathcal M^{(n_{\rm 1D})}$; and the dressed transverse velocities $v_\perp^{(l)}$ fix the relative chirality of the two cones. The following section applies this two-cone Hamiltonian to determine when the selected Pauli sector acts as a Dirac mass rather than as a node-shifting channel.

\section{Selection rule and electrical chirality switching}
\label{sec:selection_rule}

In a coupled Dirac Hamiltonian, Eq.~\eqref{eq:prb_two_cone}, a charge-neutrality gap opens only if $\mathcal M^{(n_{\rm 1D})}$ contains a Pauli channel that anticommutes with the dressed Dirac kinetics. In the present problem this mass-selection question is controlled by two ingredients: the Fourier parity of the selected harmonic, which fixes the leading Pauli sector of $\mathcal M^{(n_{\rm 1D})}$, and the relative chirality of the two dressed cones, which fixes which Pauli sector is a mass.

\begin{figure*}[t]
  \centering
    \includegraphics[width=\textwidth]{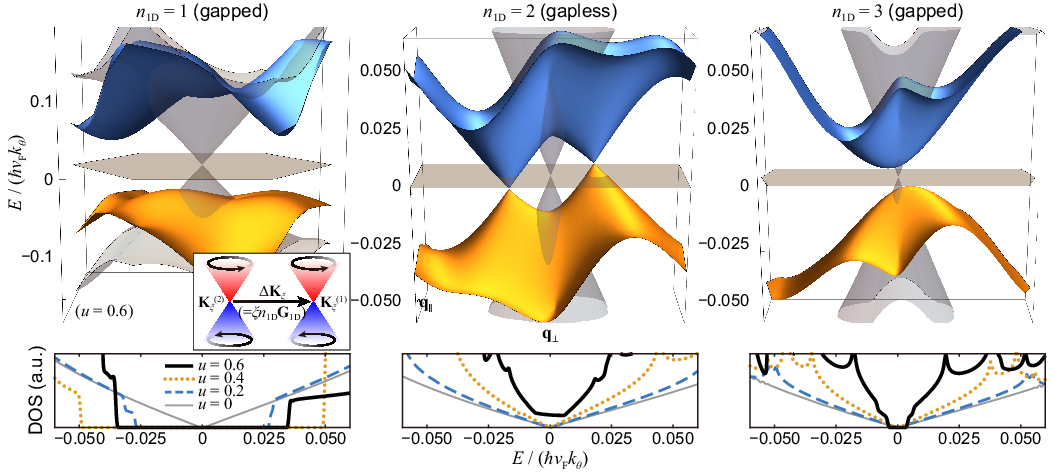}
  \caption{Low-energy spectra at exact Dirac--Dirac resonance in the default same-chirality class, $\chi_{\rm rel}=+1$. Columns correspond to $n_{\rm 1D}=1,2,3$. Top row: low-energy minibands for $u_1=u_2=u=0.6$. Gray cones indicate the unhybridized (i.e., pristine TBG) cones, and the gray horizontal planes represent the rBZ. Inset: schematic of $\chi_\mathrm{rel}=+1$ Dirac cones, separated by $\Delta \mathbf{K}_\xi$, and brought into resonance by the unidirectional modulation $\xi n_\mathrm{1D}\mathbf{G}_\mathrm{1D}$. Bottom row: density of states for $u=0,0.2,0.4,0.6$. Odd resonances ($n_{\rm 1D}=1,3$) are gapped at charge neutrality, whereas the even resonance ($n_{\rm 1D}=2$) has no leading mass channel and remains gapless in this slice.}
  \label{fig:prb_selectivity}
\end{figure*}

The full-wave minibands near the CNP in Fig.~\ref{fig:prb_selectivity}, computed at fixed twist angle with layer-symmetric modulation, show a selective resonant response as $\mathbf G_{\rm 1D}$ (i.e., $n_\mathrm{1D}$) varies. 
The full-wave Hamiltonian, momentum-space truncation, and gap/DOS diagnostics used for these spectra are summarized in Appendix~\ref{sec:full_wave_model}.
The odd nearest-cone resonances $n_{\rm 1D}=1$ and $3$ open robust CNP gaps, whereas the even resonance $n_{\rm 1D}=2$ has no leading mass channel and remains gapless in this layer-symmetric slice. 
Thus, after momentum conservation has selected the central intercone matrix element $\mathcal M^{(n_{\rm 1D})}$, one must still ask whether its Pauli content is a mass for the two Dirac cones that it couples.

The Pauli labels of the intercone matrix must be interpreted in a fixed sublattice convention. We use the same layer-sublattice convention in which the dressed tunneling components $\tilde T_{j,p}^{\xi}$, and hence the selected matrix $\mathcal M^{(n_{\rm 1D})}$, were defined. In this convention we write
\begin{equation}
Z\mathcal M^{(n_{\rm 1D})}
=
c_0\,\sigma_0+c_x\,\sigma_x+c_y\,\sigma_y+c_z\,\sigma_z .
\label{eq:prb_pauli_decomp}
\end{equation}
The kinetic blocks of Eq.~\eqref{eq:prb_two_cone} then define the cone chiralities through the sign of the transverse Dirac velocity,
\begin{equation}
\begin{aligned}
\chi_l&=\xi_l\,\mathrm{sgn}\,v_\perp^{(l)},\\
\chi_{\rm rel}
&\equiv \chi_1\chi_2
=
\xi_1\xi_2\,\mathrm{sgn}\!\left[v_\perp^{(1)}v_\perp^{(2)}\right],
\end{aligned}
\label{eq:prb_chi_rel}
\end{equation}
where $v_\parallel^{(l)}>0$ in the parameter regime considered here. For the valley-conserving resonances emphasized in this work, $\xi_1=\xi_2=\xi$, and therefore $\chi_{\rm rel}=\mathrm{sgn}[v_\perp^{(1)}v_\perp^{(2)}]$.

With this convention fixed, the mass-channel assignment is direct. For a same-chirality pair, $\chi_{\rm rel}=+1$, a $\sigma_z$ component of $Z\mathcal M^{(n_{\rm 1D})}$ anticommutes with the in-plane kinetic Pauli matrices and supplies the local Dirac mass channel, while the remaining leading components primarily shift or split Dirac nodes. For an opposite-chirality pair, $\chi_{\rm rel}=-1$, one may align the chirality of one cone by a constant sublattice rotation. With the lower-left intercone block convention of Eq.~\eqref{eq:prb_two_cone}, choosing the rotation $W=\sigma_y$ on the layer-1 spinor transforms the intercone matrix as
\begin{equation}
\begin{aligned}
Z\mathcal M^{(n_{\rm 1D})} &\rightarrow
Z\mathcal M^{(n_{\rm 1D})}W \\
&=
c_0\,\sigma_y
+i c_x\,\sigma_z
+c_y\,\sigma_0
-i c_z\,\sigma_x .
\end{aligned}
\label{eq:prb_Mshuffle}
\end{equation}
Thus the original-basis $\sigma_x$ component becomes the $\sigma_z$-like mass channel in the chirality-aligned opposite-chirality problem.

Combining the chirality-dependent mass channel above with the leading small-angle Pauli-sector rule in Eq.~\eqref{eq:prb_parity} gives the central selection rule. Provided the relevant mass-channel coefficient is not accidentally zero, the leading local CNP mass is present precisely when
\begin{equation}
\text{leading local CNP mass}\neq 0
\Longleftrightarrow
\begin{cases}
p\ \text{odd}, & \chi_{\rm rel}=+1,\\[1mm]
p\ \text{even}, & \chi_{\rm rel}=-1,
\end{cases}
\label{eq:prb_selection_p}
\end{equation}
where $p$ is the integer harmonic selected by the local resonance. In the nearest-cone branch, $p=p_0=-\xi n_{\rm 1D}$, whose parity is the parity of $n_{\rm 1D}$. Equation~\eqref{eq:prb_selection_p} therefore gives a nonzero leading local CNP mass for odd $n_{\rm 1D}$ when $\chi_{\rm rel}=+1$, and for even $n_{\rm 1D}$ when $\chi_{\rm rel}=-1$.

Equation~\eqref{eq:prb_selection_p} is a leading-order statement in the small-angle first-star continuum theory. Accidental zeros of the relevant Bessel coefficient, node-shifting channels, and finite-angle $\mathcal O(t_0\theta)$ Pauli admixtures can affect the detailed gap size, but not the organizing parity--chirality principle. The layer-symmetric spectra of Fig.~\ref{fig:prb_selectivity} lie in the $\chi_{\rm rel}=+1$ class and therefore exhibit a leading CNP mass, and hence a CNP gap, only for odd $n_{\rm 1D}$. Appendix~\ref{sec:min_two_cone_model} gives the detailed two-cone derivation.

\begin{figure*}[t]
  \centering
  \includegraphics[width=\textwidth]{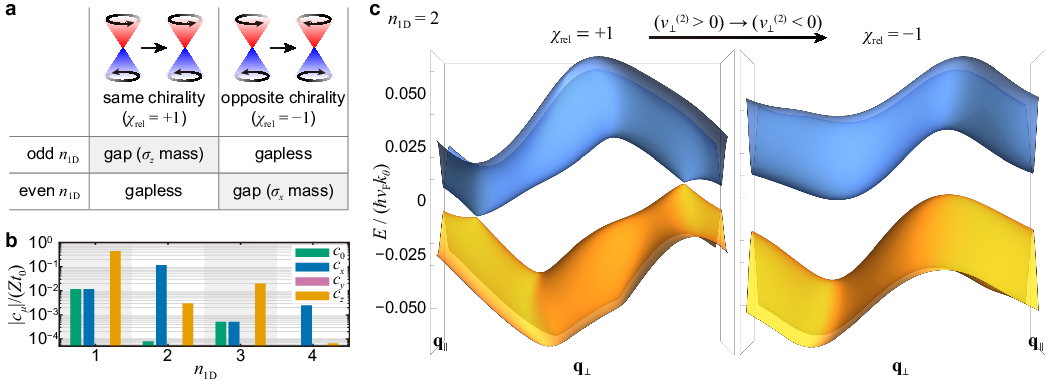}%
  \caption{Parity--chirality selection rule and electrical chirality switching. (a) Schematic summary of the leading CNP mass/gap selection rule in the resonance Dirac--Dirac interaction studied here. (b) Pauli decomposition of the selected resonant harmonic, $Z\mathcal M^{(n_\mathrm{1D})}=Z\tilde T_{0,-\xi n_\mathrm{1D}}^\xi=\sum_{\mu=0,x,y,z} c_\mu \sigma_\mu$. 
In the Pauli convention used for $\mathcal M^{(n_{\rm 1D})}$ in Sec.~\ref{sec:selection_rule}, odd harmonics supply the dominant mass channel at the same-chirality ($\sigma_z$), whereas even harmonics supply the dominant mass channel at the opposite-chirality ($\sigma_x$).
  (c) Full-wave continuum minibands for the even resonance $n_{\rm 1D}=2$, showing that the $\chi_\mathrm{rel}=+1$ branch has no leading even-harmonic mass and remains gapless (left), whereas layer-asymmetric dressing drives $v_\perp^{(1)}v_\perp^{(2)}<0$, switches $\chi_{\rm rel}$ from $+1$ to $-1$, and opens a CNP gap (right). The minibands are shown in the same low-energy momentum window as the $n_\mathrm{1D}=2$ spectra in Fig.~\ref{fig:prb_selectivity}; the gray unhybridized Dirac cones and the rBZ outline are omitted for clarity.}
  \label{fig:prb_chirality}
\end{figure*}

The selection rule becomes a control principle because the relative chirality in Eq.~\eqref{eq:prb_chi_rel} is electrically tunable. The dressed transverse velocities in Eq.~\eqref{eq:prb_vperp} depend separately on the modulation strength in each layer, so a layer-asymmetric 1D potential can reverse the sign of one $v_\perp^{(l)}$ while keeping the sign of the other. Figure~\ref{fig:prb_chirality}(c) illustrates this mechanism for the even resonance $n_{\rm 1D}=2$: when $v_\perp^{(1)}v_\perp^{(2)}>0$, the system belongs to the $\chi_{\rm rel}=+1$ class and has no leading even-harmonic mass, whereas once $v_\perp^{(1)}v_\perp^{(2)}<0$, the system switches to $\chi_{\rm rel}=-1$ and the same even harmonic becomes a gap-opening channel.

This sign change is not a gauge artifact of a single isolated cone. For an isolated Dirac cone, the sign of $v_\perp^{(l)}$ can simply be flipped by a constant sublattice rotation. In a coupled two-cone problem, however, applying such a rotation to only one cone also transforms the intercone matrix, as in Eq.~\eqref{eq:prb_Mshuffle}, and thereby reshuffles which Pauli component acts as a mass. The physically meaningful quantity is therefore the relative sign $\mathrm{sgn}[v_\perp^{(1)}v_\perp^{(2)}]$, or equivalently $\chi_{\rm rel}$. This is the sense in which the active mass channel is electrically programmable.

The same perspective clarifies why the added 1D scalar modulation is essential. In the leading small-angle first-star convention, bare TBG does not provide the same mass-channel control. In the rigid model each $T_j^\xi$ is rank one; in the relaxed model the strict rank-one constraint is lifted, but the $\theta\to0$ bare tunneling still lies in $\mathrm{span}\{\sigma_0,\sigma_x,\sigma_y\}$ and contains no leading $\sigma_z$ mass channel for the same-chirality class. This is not an exact finite-angle constraint: when the layer Dirac Hamiltonians are rotated to a common standard frame, part of the $u_{\rm AA}\sigma_0$ sector (or $t_0\sigma_0$ in the rigid notation) can mix into a $\sigma_z$ component, reflecting the approximate mirror symmetry of the small-angle model and producing an $\mathcal O(t_0\theta)$ channel admixture~\cite{ScheerGuLian2022_PRB}. This subleading admixture does not change the leading parity--chirality selection rule. By contrast, 1D dressing in TBG--1D reshuffles the Pauli content of the resonance-selected interlayer harmonic at leading order, so that the allowed parity--chirality class can activate the corresponding mass channel.
A more explicit rank-based version of this argument is given in Appendix~\ref{sec:SI_rank_tunneling}.

For an allowed nearest-cone gapped branch, the selected mass-channel coefficient is controlled by the layer-symmetric combination $u_+=u_1+u_2$. Using $p_0=-\xi n_{\rm 1D}$, its scale is
\begin{equation}
|m_{\rm gap}|
\sim
Zt_0\left|\mathcal J_{+,p_0}\right|
=
Zt_0
\left|
J_{|p_0|}\!\left(\frac{|u_+|}{2}\right)
\right|,
\label{eq:prb_massscale}
\end{equation}
which becomes $Zt_0
\left|
J_{|n_{\rm 1D}|}(|u_l|)
\right|$ in the equal in-phase limit ($u_1=u_2\equiv u_l$) used in the layer-symmetric spectra of Fig.~\ref{fig:prb_selectivity}. 
Accidental zeros of the relevant Bessel coefficient, node-shifting channels, and finite-angle $\mathcal O(t_0\theta)$ Pauli admixtures can affect the detailed gap size, but not the organizing parity--chirality principle.
See Appendix~\ref{sec:layer_symmetric} for more details.

\section{Near-resonant gapped window and design tolerances}

\begin{figure}[t]
  \centering
  \includegraphics[width=\columnwidth]{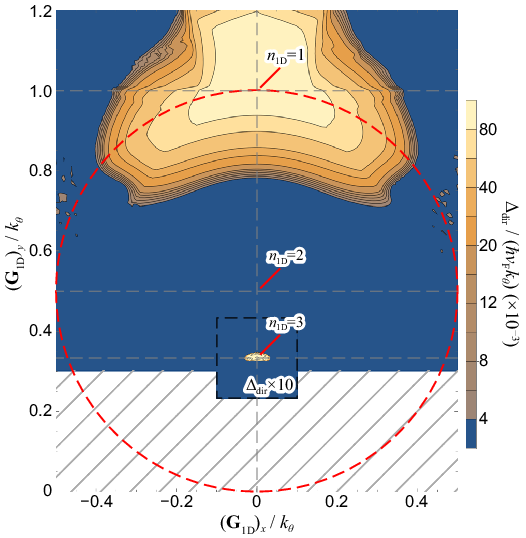}%
  \caption{Map of the direct gap at CNP, $\Delta_\mathrm{dir}$ (in units of $\hbar v_\mathrm{F} k_\theta$, log scale), versus $\mathbf G_\mathrm{1D}$ with fixed $\Delta\mathbf K_\xi$ at twist angle $\theta=2^\circ$ for a layer-symmetric (in-phase) modulation ($|u_1|=|u_2|=0.6$ and $\phi_1=\phi_2=0$), computed using a full-wave continuum model. The hatched region lies outside the numerical scan. Resonant markers indicate the $\mathbf G_\mathrm{1D}$ values for $n_\mathrm{1D}=1,2,3$, at which Dirac--Dirac matching occurs, while the red dashed circle denotes the analytic ridge $\delta q_\parallel=0$ around the $n_\mathrm{1D}=1$; the bright upper segment satisfies the full near-resonant gap condition.
 Within the black dashed box around the $n_{\rm 1D}=3$ resonance, the displayed color represents $10\Delta_{\rm dir}$; the corresponding physical gap is therefore one tenth of the colorbar value. Additional direct-gap maps for $\theta=3^\circ$ and $5^\circ$ are shown in Fig.~\ref{fig:near_resonance_gap_maps} of Appendix~\ref{sec:near_resonance}.}
  \label{fig:prb_window}
\end{figure}

Exact resonance is a measure-zero condition in the two-dimensional $\mathbf G_{\rm 1D}$ design map at fixed $\Delta\mathbf K_\xi$, but the insulating phase survives over a finite near-resonant window. Here we focus on the same-chirality gapped branch displayed in Fig.~\ref{fig:prb_window}, for which $v_\perp^{(1)}v_\perp^{(2)}>0$ and the selected odd harmonic supplies the leading mass channel. The opposite-chirality gapped branch is described by the same detuned two-cone criterion after replacing the selected mass channel according to Eq.~\eqref{eq:prb_selection_p} and using the corresponding positive velocity magnitudes. For the momentum mismatch
$\delta\mathbf q_\xi\equiv \Delta\mathbf K_\xi-\xi n_{\rm 1D}\mathbf G_{\rm 1D}$,
with components $(\delta q_\perp,\delta q_\parallel)$ in the local $(\hat{\mathbf e}_\perp,\hat{\mathbf e}_\parallel)$ frame, we define
\begin{equation}
v_\perp^{\rm eff}\equiv \sqrt{\left|v_\perp^{(1)}v_\perp^{(2)}\right|},
\qquad
v_\parallel^{\rm eff}\equiv \sqrt{\left|v_\parallel^{(1)}v_\parallel^{(2)}\right|},
\end{equation}
and the mismatch-energy scale
\begin{equation}
\varepsilon_{\rm mm}(\delta\mathbf q_\xi)
\equiv
\frac{\hbar}{2}
\sqrt{\bigl(v_\perp^{\rm eff}\delta q_\perp\bigr)^2+
      \bigl(v_\parallel^{\rm eff}\delta q_\parallel\bigr)^2 }.
\label{eq:prb_nearres}
\end{equation}
Within the detuned two-cone theory, the gap remains when
\begin{equation}
\varepsilon_{\rm mm}(\delta\mathbf q_\xi)<|m_{\rm gap}|,
\label{eq:prb_gapcriterion}
\end{equation}
which gives a filled ellipse in the mismatch coordinates $(\delta q_\perp,\delta q_\parallel)$.

The experimentally relevant map in Fig.~\ref{fig:prb_window} scans $\mathbf G_{\rm 1D}$ at fixed $\Delta\mathbf K_\xi$, for which the mapping from $\mathbf G_{\rm 1D}$ to $(\delta q_\perp,\delta q_\parallel)$ is nonlinear. Writing $\mathbf g\equiv n_{\rm 1D}\mathbf G_{\rm 1D}/k_\theta=(g_x,g_y)$, one finds
\begin{equation}
\delta q_\perp=\xi k_\theta\frac{g_x}{g},
\quad
\delta q_\parallel=\xi k_\theta\Bigl(\frac{g_y}{g}-g\Bigr),
\quad
g\equiv |\mathbf g|.
\label{eq:prb_dqg}
\end{equation}
The ridge of the gapped region is defined by $\delta q_\parallel=0$, namely
\begin{equation}
g_x^2+\Bigl(g_y-\frac12\Bigr)^2=\Bigl(\frac12\Bigr)^2,
\label{eq:prb_ridge}
\end{equation}
which is a circle centered at $(0,1/2)$ in the $\mathbf g$ coordinates. The exact-resonance point lies on this circle but is not its center, so the filled ellipse in mismatch space is converted into the thick off-centered arc seen in Fig.~\ref{fig:prb_window}. Appendix~\ref{sec:near_resonance} and Fig.~\ref{fig:near_resonance_gap_maps} provide the full derivation and the complementary fixed-$\mathbf G_{\rm 1D}$ view in which the gapped region remains an ellipse.

Using the direct-gap convention of Appendix~\ref{sec:geometric_errors}, an illustrative estimate $\Delta_{\rm dir}\simeq2|m_{\rm gap}|=10~\mathrm{meV}$ gives, at $\theta=2^\circ$, the tolerances $|\delta\theta|\lesssim0.10^\circ$, $|\delta L_{\rm 1D}/L_{\rm 1D}|\lesssim4.9\%$, and $|\delta\varphi_{\rm 1D}|\lesssim2.8^\circ$; at $\theta=3^\circ$ they become $0.069^\circ$, $2.3\%$, and $1.3^\circ$, respectively. The same selected mass channel that opens the exact resonant gap therefore also controls the width of the usable design window. 
Target periods and mass-channel magnitudes, electrostatic attenuation and higher-harmonic leakage, and geometric tolerances are discussed in Appendices~\ref{sec:S8_2}, \ref{sec:S8_3}, and \ref{sec:geometric_errors}, respectively.

\section{Discussion}
\label{sec:discussion_main}

\subsection{Chern-trivial resonant gaps and off-resonant band renormalization}
\label{sec:discussion_chern_offres}

For the single-harmonic scalar modulation studied here, all gapped CNP states found in our scans are valley-Chern trivial. Cosine-registry configurations preserve $C_{2z}T$, and away from that point the valley Chern number remains zero unless the gap closes~\cite{po2018origin,Zou2018_PRB,Zhang2019_PRB}. In other words, the selection rule identified above determines whether a Dirac mass opens at resonance, but within the present scalar-modulation setting it does not by itself produce a topological CNP gap. Appendix~\ref{sec:chern_appendix} presents the Berry-curvature diagnostics and the Fukui discretization used in the full-wave calculations.

Far away from resonance, the same 1D dressing produces a complementary gapless regime of strong anisotropic Dirac renormalization. In that regime there is no direct resonant Dirac--Dirac hybridization, but the transverse velocity can be strongly suppressed and even driven to zero along a flattening line in the $(\theta,|u_l|)$ plane. The hybrid moir\'e--1D platform therefore supports two distinct kinds of band engineering: near resonance it enables parity- and chirality-selected mass generation, whereas off resonance it reshapes the Dirac cone into a strongly anisotropic but gapless spectrum. 
Appendix~\ref{sec:off_resonance} gives a controlled L\"owdin-downfolded description of this off-resonant regime, while Appendix~\ref{sec:S8_5} summarizes the corresponding feasibility considerations.

\subsection{\texorpdfstring{Relation to known even--odd Dirac-coupling mechanisms}{Relation to known even-odd Dirac-coupling mechanisms}}
\label{sec:relation_evenodd_mechanisms}
The term ``even--odd'' appears in several graphene Dirac-coupling problems. The comparison is most useful when one identifies the microscopic object whose parity is being classified and the Dirac-matrix channel selected by that parity. In the present TBG--1D mechanism, the relevant factor is the Fourier index $p$ of a resonance-selected dressed interlayer harmonic in Eq.~\eqref{eq:prb_umklapp_resonance}. For the nearest-cone branch used in the main analysis, this index becomes $p_0=-\xi n_{\rm 1D}$, so that the parity of $n_{\rm 1D}$ fixes the leading Pauli sector of $\tilde T_{0,p_0}^{\xi}$. 
Thus, the CNP gap is governed not by harmonic parity alone, but by the combination of the selected Pauli sector and the relative chirality of the two dressed Dirac cones.

Commensurate TBG provides a classic atomic-scale benchmark. Mele showed that sublattice-exchange even and odd commensurations have distinct low-energy spectra and gap structures~\cite{Mele2010_PRB}. The shared message is algebraic: a discrete parity can constrain the sublattice structure of an intercone coupling. While the parity is fixed by atomic commensuration and sublattice-exchange symmetry in commensurate TBG, that in TBG--1D is set by the Fourier component of the interlayer tunneling generated through the layer-dependent dressing $U_l(y)$. 

Alternating twisted trilayer graphene and its multilayer generalizations provide a complementary multi-cone setting. Their alternating twist geometry aligns several momentum shifts, and the continuum Hamiltonian can be decomposed into TBG-like sectors together with dispersive Dirac sectors~\cite{Xie2022_NPJ}. In that setting, the even/odd distinction is tied to layer, mirror, or partition-sector organization, and the residual Dirac sector follows from this multilayer sector decomposition.

The closest comparison at the level of low-energy two-cone bookkeeping is monolayer graphene subject to a nearly $\sqrt{3}\times\sqrt{3}$ superlattice potential~\cite{Crepel2023_PRB}. In that problem, the superlattice folds the original graphene valleys into a common low-energy description, and selected Fourier components of an on-site potential organize the intervalley coupling according to their valley/sublattice structure. This provides a useful reference point for the even--odd language used here, because both settings ask how a superlattice-selected harmonic enters the Pauli structure of a coupled-Dirac Hamiltonian.

The TBG--1D mechanism realizes this bookkeeping through a different microscopic object and with an additional control knob. The active matrix element is generated from the pre-existing, ordinary TBG interlayer tunneling: (i) the layer-resolved scalar dressing $U_l(y)$ converts $T_j^\xi$ into dressed interlayer components $\tilde T_{j,p}^\xi$, and (ii) the Dirac--Dirac resonance selects one component of this generated interlayer tower. In addition, (iii) the same dressing renormalizes the transverse velocities separately in the two layers, so the relative chirality can be electrically switched through the sign of $v_\perp^{(1)}v_\perp^{(2)}$. Thus the shared two-cone language is implemented here as a gate-generated interlayer-dressing mechanism in which harmonic parity and dressed relative chirality jointly select the Dirac mass channel.

Taken together, these examples show that even--odd labels in graphene Dirac problems often encode constraints on intercone Pauli structure, while the microscopic object carrying the parity can vary. The contribution of the present TBG--1D setting is to realize this idea through gate-generated layer-resolved dressing of an interlayer tunneling matrix: a long-wavelength scalar modulation creates a Bessel harmonic tower of dressed interlayer components, momentum resonance selects one component, and the electrically tunable sign of $v_\perp^{(1)}v_\perp^{(2)}$ decides whether that component acts as a Dirac mass.
\section{Conclusion}

We studied how an added unidirectional scalar superlattice reshapes Dirac--Dirac coupling in a hybrid moir\'e--1D system, using twisted bilayer graphene as a representative platform. The central question was not only whether the two Dirac cones can be brought into momentum resonance, but whether the resonance-selected dressed interlayer harmonic carries the Pauli structure required to open a mass at charge neutrality. This distinction between momentum matching and mass-channel selection is the organizing idea of the work.

Our main result is an electrically switchable parity--chirality selection rule for the charge-neutrality-point gap. In the dressed-harmonic two-cone description, the Fourier parity of the resonance-selected dressed interlayer harmonic fixes its dominant Pauli content, while the dressed transverse velocities fix the relative chirality of the interacting cones. As a consequence, odd selected harmonics gap same-chirality cone pairs, whereas even selected harmonics gap opposite-chirality pairs. Because the 1D modulation independently dresses the two layers, it can reverse the transverse velocity of one layer and thereby switch the relative chirality, converting an otherwise gapless even resonance into a gapped one. Full-wave continuum minibands further show that this mechanism survives beyond exact matching as a finite near-resonant gapped window with quantitative geometric tolerances.

Viewed as a band-design problem, the mechanism separates into three complementary ingredients: momentum resonance selects the cone pair, harmonic parity selects the available Pauli channel, and layer-resolved 1D dressing selects whether that channel acts as a Dirac mass. The period and orientation of the 1D modulation therefore determine where in reciprocal space the interaction becomes active, while the layer asymmetry of the modulation determines whether the interacting cones belong to the same- or opposite-chirality class. This separation of roles is what makes chirality switching a practical control knob rather than a purely formal classification.

Taken together, these results identify hybrid moir\'e--1D superlattices as a platform in which externally designed 1D dressing can control not only where Dirac cones meet in momentum space, but also which mass channel becomes active once they do. In the present single-harmonic scalar setting the resulting resonant gaps are Chern trivial, while away from resonance the same dressing yields strong anisotropic but gapless band renormalization. More broadly, the parity--chirality viewpoint developed here provides a compact design principle for programmable mass selection and band-shape engineering in coupled Dirac systems, and suggests a useful organizing framework for future hybrid superlattice designs in other coupled-Dirac platforms.

\section*{Acknowledgements}

This work was supported 
by the NYU-ECNU Institute of Physics at NYU Shanghai
and by the Brain Pool Program through the National Research Foundation of Korea (NRF) funded by the Ministry of Science and ICT (MSIT) (Grant No. RS-2025-25446099).
This work was conducted at NYU Shanghai and Hanyang University, and was carried out on the High Performance Computing resources at NYU Shanghai. 
P.M. used ChatGPT (OpenAI) for proofreading and language editing to improve readability. The scientific content, analysis and conclusions are solely those of the authors.

% \clearpage
\appendix
% \onecolumngrid

\section{Resonance geometry and real-space interpretation}

This appendix supplements the laboratory-frame geometry, hybrid continuum Hamiltonian, energy scales, local 1D axes, nearest-cone resonance condition, momentum mismatch, and global-periodicity criterion defined in Secs.~\ref{sec:reference_graphene_lattice}--\ref{sec:configuration_space}. We collect here additional reciprocal-space illustrations, a clarification of moir\'e-Umklapp resonances beyond the nearest-cone branch, a real-space visualization of the nearest-cone parity label, and the representation choices used in the appendices.

\begin{figure}[t]
\centering
\includegraphics[width=\linewidth]{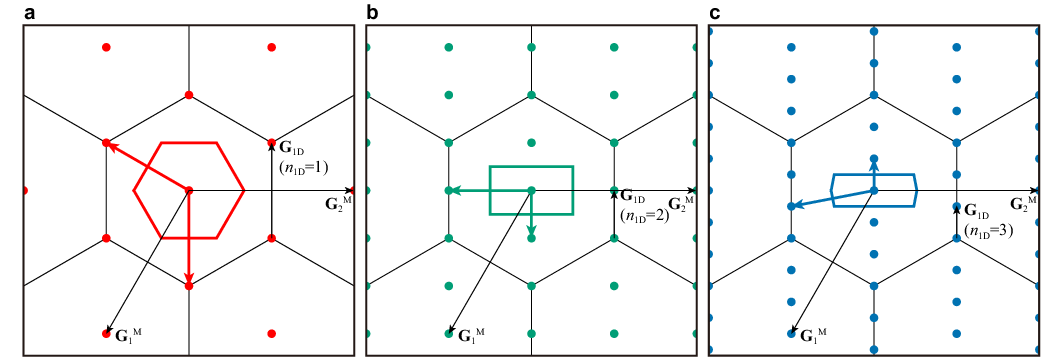}
\caption{
Periodicity in the reciprocal space of resonant configurations \textbf{a} $n_\mathrm{1D}=1$, \textbf{b} $n_\mathrm{1D}=2$, \textbf{c} $n_\mathrm{1D}=3$. $\mathbf{G}_i^\mathrm{M}$ ($i=1,2$) represent the reciprocal lattice vectors of TBG, and $\mathbf{G}_\mathrm{1D}$ show the reciprocal vector of the unidirectional potential. Hexagons in \textbf{a} and \textbf{c} and rectangle in \textbf{b} show the rBZ in each configuration.
}
\label{fig:reciprocal_periodicity}
\end{figure}

\subsection{\texorpdfstring{Moir\'e-Umklapp resonances beyond the nearest-cone family}{Moire-Umklapp resonances beyond the nearest-cone family}}
\label{sec:S_moire_umklapp_resonances}
The resonance condition used in the main numerical scans, Eq.~\eqref{eq:prb_resonance}, selects the nearest pair of valley-conserving Dirac cones. Other locally resonant pairs can be obtained by first translating one cone by a moir\'e reciprocal lattice vector and then compensating the remaining separation by an integer harmonic of the 1D potential, as summarized by Eq.~\eqref{eq:prb_umklapp_resonance}. These resonances are not described by allowing half-integer values of $n_{\rm 1D}$ in Eq.~\eqref{eq:prb_resonance}; the Fourier harmonic index is always an integer $p$.

It is also useful to record why the nearest-cone branch is the leading design branch. Choose $\xi=+1$ and $j=0$; the other first-star momenta follow by $C_3$, and the opposite valley follows by time reversal. In units of $k_\theta$,
\begin{equation}
\begin{aligned}
\mathbf q_0&=(0,1),\\
\mathbf G_1^{\rm M}&=\left(-\frac{\sqrt3}{2},-\frac32\right),\\
\mathbf G_2^{\rm M}&=(\sqrt3,0).
\end{aligned}
\label{eq:S_umklapp_shortest_vectors}
\end{equation}
Therefore
\begin{multline}
\frac{\left|\mathbf q_0+m_1\mathbf G_1^{\rm M}+m_2\mathbf G_2^{\rm M}\right|^2}{k_\theta^2}\\
=
3\left(m_2-\frac{m_1}{2}\right)^2
+
\left(1-\frac{3m_1}{2}\right)^2 .
\label{eq:S_umklapp_shortest_bound}
\end{multline}
For integer $m_1,m_2$, the minimum value is $1$, reached by the nearest first-star separations. Hence
\begin{equation}
|\mathbf q_j^\xi+\mathbf G_{\rm M}|\ge k_\theta
\label{eq:S_umklapp_shortest_result}
\end{equation}
for all moir\'e translations of a first-star separation.

For a generalized resonance, Eq.~\eqref{eq:prb_umklapp_resonance} gives
\begin{equation}
|\mathbf G_{\rm 1D}|
=
\frac{|\mathbf q_j^\xi+\mathbf G_{\rm M}|}{|p|} .
\label{eq:S_umklapp_G1D_order}
\end{equation}
Equation~\eqref{eq:S_umklapp_shortest_result} then shows that, at fixed $|p|$, translated branches cannot yield a longer 1D period than the nearest-cone branch. If one instead keeps $L_{\rm 1D}$ long by increasing $|p|$, the selected coupling involves a higher-order harmonic. For $|u_+|\ll1$,
\begin{equation}
J_{|p|}\!\left(\frac{|u_+|}{2}\right)
=
\frac{1}{|p|!}\left(\frac{|u_+|}{4}\right)^{|p|}
+\mathcal O(|u_+|^{|p|+2}),
\label{eq:S_umklapp_bessel_suppression}
\end{equation}
which makes the higher-order branch parametrically weaker in the weak-modulation limit. This is a design-scope statement, not an additional symmetry constraint: the local parity rule is still read from the selected integer $p$.

As a concrete collinear example in the present convention, Eq.~\eqref{eq:S:q_j_xi} and $\mathbf K_\xi^{(l)}=-\xi(2\mathbf b_1^{(l)}+\mathbf b_2^{(l)})/3$ imply
\begin{equation}
2\mathbf G_1^{\rm M}+\mathbf G_2^{\rm M}=-3\xi\,\mathbf q_0^\xi.
\end{equation}
Choosing the moir\'e translation
\begin{equation}
\mathbf G_{\rm M}=\xi(2\mathbf G_1^{\rm M}+\mathbf G_2^{\rm M})=-3\mathbf q_0^\xi
\end{equation}
gives a translated separation $\mathbf q_0^\xi+\mathbf G_{\rm M}=-2\mathbf q_0^\xi$. A 1D vector satisfying $\mathbf G_{\rm 1D}=2\xi\mathbf q_0^\xi$ would look like an ``$n_{\rm 1D}=1/2$'' alignment if forced into the nearest-cone formula, but it is instead an integer-harmonic Umklapp resonance of Eq.~\eqref{eq:prb_umklapp_resonance} with $p=\xi$.

Once such a cone pair is selected and the remaining channels are remote, the local two-cone algebra is the same as in Appendix~\ref{sec:exact_resonance}: the gap-opening rule is controlled by the parity of the selected integer harmonic $p$, together with the relative chirality of the two cones. The nearest-cone rule stated in terms of $n_{\rm 1D}$ is therefore the $\mathbf G_{\rm M}=0$ specialization rather than a restriction on possible moir\'e-folded resonances.

\subsection{Real-space registry and the nearest-cone parity label}
\label{sec:S_real_space_stacking_intuition}

The mass-selection rule used in the main text is derived in momentum space. This subsection records the corresponding real-space registry pattern for the nearest-cone resonance family, mainly to clarify the interpretation of Fig.~\ref{fig:geometry_design_space}. Combining the nearest-cone resonance condition with the moir\'e real-space lattice gives
\begin{equation}
\mathbf L_{\rm 1D}
=
-\frac{3}{2}n_{\rm 1D}\mathbf L_1^{\rm M}.
\label{eq:S_realspace_period_parity}
\end{equation}
Then
\begin{equation}
\mathbf L_{\rm 1D}
\equiv
\begin{cases}
\mathbf 0 \pmod{\mathbf{L}_{\rm M}}, & n_{\rm 1D}\ {\rm even},\\[2mm]
-\mathbf L_1^{\rm M}/2 \pmod{\mathbf{L}_{\rm M}}, & n_{\rm 1D}\ {\rm odd},
\end{cases}
\label{eq:S_stacking_parity_mod_moire}
\end{equation}
where $\mathbf{L}_{\rm M}$ denote the moir\'e real-space lattice generated by $\mathbf L_1^{\rm M}$ and $\mathbf L_2^{\rm M}$. 
Thus, in the nearest-cone family, advancing by one period of the imposed 1D potential returns the same potential phase at a moir\'e-equivalent registry for even $n_{\rm 1D}$, while for odd $n_{\rm 1D}$ it returns the same potential phase at a half-moir\'e-translated registry.

For a fixed real-space origin and fixed external phases $\phi_l$, this distinction can be drawn as a same-registry versus half-translated-registry pattern. For example, in the geometry illustrated in Fig.~\ref{fig:geometry_design_space}, the odd case may appear to relate inequivalent AB/BA-rich regions, whereas the even case may appear to repeat equivalent local registries. This visual alignment, however, depends on the external phases: changing $\phi_l$ translates the imposed stripes relative to the AA/AB/BA pattern without changing the momentum-resonance condition.

For this reason, the real-space registry pattern is descriptive rather than diagnostic. The invariant label used in the theoretical model in the main text is the integer momentum transfer selected by resonance. In the nearest-cone branch this integer is $p_0=-\xi n_{\rm 1D}$, so its parity can equivalently be described as the parity of $n_{\rm 1D}$. After the dressed representation is introduced, the same integer labels the selected interlayer Fourier component $\tilde T^\xi_{0,p_0}$; its Pauli sector, together with the relative chirality of the two dressed cones, determines whether a Dirac mass is generated. In generalized moir\'e-Umklapp resonances, the corresponding parity label is the integer $p$ in Eq.~\eqref{eq:prb_umklapp_resonance}, rather than an apparent fractional nearest-cone index.

\subsection{Choice of basis and representations}
In the appendices we adopt two complementary representations.
For numerical electronic-structure calculations and global spectral diagnostics (Appendix~\ref{sec:full_wave_model}), as well as for the perturbative analysis of the off-resonant regime (Appendix~\ref{sec:off_resonance}), we use the explicit-potential (non-dressed) representation in a momentum-space Bloch basis.
This representation preserves the full lattice structure and is well suited for unbiased full-wave calculations and Brillouin-zone--wide analyses.
By contrast, the unitary-rotated (dressed) representation is introduced specifically to isolate the resonant channels of the resonant regime and to enable transparent analytical derivations of the even--odd $n_{\rm 1D}$ parity rule and the associated chirality dependence (Appendices~\ref{sec:dressed_rep}--\ref{sec:near_resonance}).

The energy scales, dimensionless couplings, and local 1D momentum/Pauli-axis convention used in the appendices follow the main-text definitions in Sec.~\ref{sec:energy_scales_and_dimensionless_parameters}.

\section{Full-wave miniband calculations and diagnostics}
\label{sec:full_wave_model}

This appendix provides a compact, reproducible summary of the full-wave numerical workflow used for the resonant minibands in main-text Fig.~\ref{fig:prb_selectivity}, the chirality-switching spectra in main-text Fig.~\ref{fig:prb_chirality}(c), and the numerically evaluated direct-gap maps underlying main-text Fig.~\ref{fig:prb_window}. These calculations use the explicit-potential representation and the relaxed two-parameter moir\'e tunneling model, and therefore provide an unbiased check of the resonant mechanism derived analytically in the dressed representation of Appendices~\ref{sec:dressed_rep}--\ref{sec:near_resonance}.
We neglect spin--orbit coupling and intervalley scattering. We treat valleys $\xi=\pm1$ independently and include spin/valley degeneracies ($g_s=2$, $g_v=2$) when forming observables such as the DOS.
The neglect of intervalley scattering is justified because the momentum transfers generated by the moir\'e tunneling and by the long-wavelength 1D modulation are parametrically small compared to the $K$--$K'$ separation for $\theta\ll1$ and $|\mathbf{L}_{\rm 1D}|\gg a$.

\subsection{Continuum Hamiltonian and assumptions}
\paragraph{Continuum Hamiltonian.}
In the layer--sublattice basis $\Psi(\mathbf r)=\big(\psi_\mathrm{1A},\psi_\mathrm{1B},\psi_\mathrm{2A},\psi_\mathrm{2B}\big)^{\mathrm T}$,
the single-valley Hamiltonian $H_\xi(\mathbf r)$ takes the block form
\begin{equation}
\begin{aligned}
&H_\xi(\mathbf{r}) \\ &=
\begin{pmatrix}
h_\xi^{(1)}(-i\nabla)+V_1(\mathbf{r}) \sigma_0 & T_\xi^\dagger(\mathbf{r}) \\
T_\xi(\mathbf{r}) & h_\xi^{(2)}(-i\nabla)+V_2(\mathbf{r})\sigma_0
\end{pmatrix}.
\end{aligned}
\label{eq:S:H_total_realspace}
\end{equation}
Here $h_\xi^{(l)}(-i\nabla)=-\hbar v_\mathrm{F} \bigl(\xi(-i\nabla_x)\sigma_x^{(l)}+(-i\nabla_y)\sigma_y^{(l)}\bigr)$ is the intralayer Dirac Hamiltonian, where
$\sigma_x^{(l)} \equiv (\cos\theta_l)\sigma_x+(\sin\theta_l)\sigma_y$ and $\sigma_y^{(l)} \equiv -(\sin\theta_l)\sigma_x+(\cos\theta_l)\sigma_y$
are the layer-rotated Pauli matrices, and
\begin{equation}
T_\xi(\mathbf{r})=\sum_{j=0}^{2}T_j^\xi\,e^{i\mathbf{q}_j^\xi\cdot\mathbf{r}}
\label{eq:S:T_r}
\end{equation}
is the moir\'e first-star tunneling.
For the effective analytic models in Appendices~\ref{sec:dressed_rep}--\ref{sec:near_resonance},
we use the rigid (non-relaxed) tunneling matrices
\cite{LopesdosSantos2007_PRL, BistritzerMacDonald2011_PNAS, MoonKoshino2013_PRB, Koshino2015_NJP}
\begin{equation}
T_j^\xi=
t_0
\begin{pmatrix}
1 & \omega^{-\xi j} \\
\omega^{\xi j} & 1
\end{pmatrix},
\label{eq:Tj_mk_rigid}
\end{equation}
for transparency, where $\omega=e^{2\pi i/3}$ and $t_0 \approx 0.110\,\mathrm{eV}$,
while in the numerical calculations we incorporate lattice-relaxation effects using the standard two-parameter first-star tunneling model with distinct AA and AB/BA amplitudes
\begin{equation}
T_j^\xi=
\begin{pmatrix}
u_{\rm AA} & u_{\rm AB}\,\omega^{-\xi j} \\
u_{\rm AB}\,\omega^{\xi j} & u_{\rm AA}
\end{pmatrix},
\quad (u_{\rm AA}\le u_{\rm AB}).
\label{eq:Tj_mk_relaxed}
\end{equation}
Importantly, the even/odd harmonic structure derived from the $\omega^{\pm\xi j}$ phases is unchanged by this replacement. Therefore, the parity--chirality selection rule derived in Sec.~\ref{sec:exact_resonance} applies to both models.

\paragraph{Lattice relaxation.}
In low-angle TBG the lattice spontaneously relaxes to enlarge AB/BA domains and shrink AA regions, because the interlayer adhesion energy depends strongly on the local stacking configuration.
We use the lattice relaxation model developed by Nam and Koshino~\cite{NamKoshino2017_PRB}.
The in-plane deformation is represented by layer displacement fields $\mathbf u^{(l)}(\mathbf r)$ ($l=1,2$), which modify the local interlayer registry as
$\boldsymbol{\delta}(\mathbf r)=\boldsymbol{\delta}_0(\mathbf r)+\mathbf u^{-}(\mathbf r)$,
$\mathbf u^{-}(\mathbf r)\equiv \mathbf u^{(2)}(\mathbf r)-\mathbf u^{(1)}(\mathbf r)$,
where $\boldsymbol{\delta}_0(\mathbf r)$ is the registry field of the rigidly twisted bilayer.
The relaxed configuration minimizes the total energy
$U[\mathbf u^{(1)},\mathbf u^{(2)}]=U_E+U_B$,
given by an elastic term $U_E= \sum_{l=1}^{2}\int d^2 r\bigl[(\lambda/2)\bigl(\nabla\cdot\mathbf u^{(l)}\bigr)^2+\mu \sum_{i,j}u_{ij}^{(l)}u_{ij}^{(l)}\bigr]$, where $u_{ij}^{(l)}=(\partial_i u^{(l)}_j+\partial_j u^{(l)}_i)/2$, $\lambda,\mu$ are Lam\'e coefficients, and an adhesion term $U_B= \int d^2 r\,V\big[\boldsymbol{\delta}(\mathbf r)\big]$, where $V[\boldsymbol{\delta}]=\sum_{j=1}^{3}2V_0\cos\big(\mathbf b_j\cdot\boldsymbol{\delta}\big)$ and $\mathbf b_3=-\mathbf b_1-\mathbf b_2$.
Using the identity $\mathbf b_j\cdot\boldsymbol{\delta}_0(\mathbf r)=\mathbf G_j^{\rm M}\cdot\mathbf r$ (with $\mathbf G_3^{\rm M}=-\mathbf G_1^{\rm M}-\mathbf G_2^{\rm M}$),
the adhesion energy density becomes explicitly moir\'e periodic,
$V[\boldsymbol{\delta}(\mathbf r)]=\sum_{j=1}^{3}2V_0\cos\big(\mathbf G_j^{\rm M}\cdot\mathbf r+\mathbf b_j\cdot\mathbf u^{-}(\mathbf r)\big)$.
The optimized solution yields a pronounced AB/BA domain pattern and suppresses the spatial weight of AA stacking in the moir\'e unit cell.

In the electronic continuum Hamiltonian, the dominant consequence of this relaxation is a reduction of the AA-type Fourier component of the moir\'e interlayer tunneling relative to AB/BA.
At the first-star level this is captured by the two-parameter tunneling matrices in Eq.~\eqref{eq:Tj_mk_relaxed} with $|u_{\rm AA}|<|u_{\rm AB}|$, which we use throughout the full-wave numerical calculations.
(Out-of-plane corrugation primarily renormalizes the overall interlayer coupling scale and can be absorbed into the effective parameters $u_{\rm AA},u_{\rm AB}$; see Ref.~\cite{NamKoshino2017_PRB}.)

\subsection{Plane-wave basis and truncation}
\label{sec:planewave_basis_and_truncation}
We choose a seed Bloch momentum $\mathbf k_0$ (measured from $\Gamma$ in the laboratory frame) and generate the set of coupled momenta $\mathbf k=\mathbf k_0+\mathbf G$ with $\mathbf G=m_1\mathbf G_1^{\rm M}+m_2\mathbf G_2^{\rm M}+m_3\mathbf G_{\rm 1D}$, $m_{1,2,3}\in\mathbb Z$.
We retain all basis states satisfying the cutoff $\hbar v_{\mathrm F}|\mathbf q|\le E_{\max}$ and verify convergence of all reported spectra and gap diagnostics by increasing $E_{\max}$.

\subsection{Momentum-space matrix elements and numerical diagonalization}

We work in the valley-resolved Bloch basis $|\mathbf q,X,l\rangle$ (with $X=A,B$ and $l=1,2$), where $\mathbf q$ is measured from the valley point $\mathbf K_\xi^{(l)}$ in each layer, i.e., $\mathbf q\equiv \mathbf k-\mathbf K_\xi^{(l)}$,
and define the Bloch states as\cite{MoonKoshino2013_PRB, Koshino2015_NJP}
\begin{multline}
|\mathbf{q},X,l\rangle \\=
\frac{1}{\sqrt{N}}
\sum_{\mathbf{R}^{(l)}}
e^{i (\mathbf{q}+\mathbf{K}_\xi^{(l)}) \cdot(\mathbf{R}^{(l)}+\boldsymbol{\tau}_X^{(l)})}
|\mathbf{R}^{(l)}+\boldsymbol{\tau}_X^{(l)}\rangle.
\end{multline}
Here $\mathbf R^{(l)}$ runs over the Bravais lattice of layer $l$ and $\boldsymbol{\tau}_X^{(l)}$ denotes the sublattice coordinate.
In this basis the intralayer matrix element is
\begin{equation}
\langle \mathbf{q}',l'| h_\xi^{(l)}(-i\nabla)|\mathbf{q},l\rangle=
h_\xi^{(l)}(\mathbf{q}) \,\delta_{\mathbf{q}',\mathbf{q}}\,\delta_{l',l},
\label{eq:S:H_intra_bare}
\end{equation}
where
\begin{equation}
h_\xi^{(l)}(\mathbf q)= -\hbar v_{\mathrm F}\big(\xi q_x\,\sigma_x^{(l)}+ q_y\,\sigma_y^{(l)}\big),
\label{eq:S:H_intra_bare_2}
\end{equation}
while the moir\'e tunneling couples momenta differing by $\mathbf q_j^\xi$:
\begin{equation}
\begin{aligned}
\langle \mathbf{q}',2| 
T_\xi(\mathbf{r}) |
\mathbf{q},1\rangle &=
\sum_{j=0}^2 T_j^\xi\,
\delta_{\mathbf{q}',\mathbf{q}+\mathbf{q}_j^\xi}, \\
\langle \mathbf{q},1| 
T_\xi^\dagger(\mathbf{r}) |
\mathbf{q}',2\rangle
&=
\langle \mathbf{q}',2| 
T_\xi(\mathbf{r}) |
\mathbf{q},1\rangle^\dagger.
\end{aligned}
\label{eq:S2:T_matrix_element}
\end{equation}
Finally, the 1D scalar potential in layer $l$ couples momenta shifted by $\pm\mathbf G_{\rm 1D}$ without mixing sublattices:
\begin{multline}
\langle \mathbf{q}',l'| 
V_l(\mathbf{r}) |
\mathbf{q},l\rangle \\=
\Bigl[\frac{v_l}{4}
\delta_{\mathbf{q}',\mathbf{q}+\mathbf{G}_{\rm 1D}}
+\frac{v_l^*}{4}
\delta_{\mathbf{q}',\mathbf{q}-\mathbf{G}_{\rm 1D}} \Bigr]
\sigma_0\,\delta_{l',l}.
\label{eq:S2:V_matrix_element}
\end{multline}

For each rBZ momentum $\mathbf q$, we construct the truncated Hamiltonian matrix
$H_\xi^{\rm (full)}(\mathbf q)$ for the Hilbert space defined in Appendix~\ref{sec:planewave_basis_and_truncation} using the matrix elements in Eqs.~\eqref{eq:S:H_intra_bare}--\eqref{eq:S2:V_matrix_element}, and solve
\begin{equation}
H_\xi^{\mathrm{(full)}}(\mathbf q)\,\Psi_{n\xi}(\mathbf q) = E_{n\xi}(\mathbf q)\,\Psi_{n\xi}(\mathbf q),
\label{eq:S:H_total_numerical_bare}
\end{equation}
where $n$ labels the minibands in ascending energy order.

\subsection{Periodic superlattice geometries: rBZ sampling and direct-gap diagnostics}
\paragraph{Direct gap throughout the Brillouin zone.}
Globally periodic (rank-2) configurations have a periodic rBZ (Sec.~\ref{sec:configuration_space}).
To diagnose a direct band touching anywhere in the rBZ, we evaluate the band-edge gap function
$g_{\xi}(\mathbf q)\equiv E_{c,\xi}(\mathbf q)-E_{v,\xi}(\mathbf q)$,
where $E_{c,\xi}(\mathbf q)$ and $E_{v,\xi}(\mathbf q)$ are, respectively, the lowest conduction and highest valence energies at crystal momentum $\mathbf q$ in valley $\xi$.
We then define the direct gap as
\begin{equation}
\Delta_{\rm dir}\equiv \min_{\xi=\pm1}\ \min_{\mathbf q\in{\rm rBZ}}\, g_\xi(\mathbf q),
\label{eq:S2:Delta_true_def}
\end{equation}
which provides a conservative criterion for a band touching anywhere in the rBZ ($\Delta_{\rm dir}=0$).

\subsection{Unfolded spectral function, DOS normalization, and band opening in quasiperiodic geometries}
In rank-3 (quasiperiodic) geometries (no exact rBZ), reciprocal-space coupling generates band replicas at quasiperiodically spaced momenta, in contrast to the strictly periodic copies in a globally periodic structure.
To extract an unfolded dispersion at a chosen momentum $\mathbf k_0$ we compute the momentum-resolved spectral function
\begin{equation}
A_l(\mathbf{k}_0,\varepsilon)=\sum_{\alpha}\sum_{X=A,B}
\bigl|\langle \alpha|\mathbf{k}_0,X,l\rangle \bigr|^2\,
\delta_\eta(\varepsilon-\varepsilon_\alpha),
\label{eq:S2:spectral_function_def}
\end{equation}
where $|\alpha\rangle$ and $\varepsilon_\alpha$ are eigenstates and eigenenergies of the truncated Hamiltonian at seed momentum $\mathbf k_0$,
and $\delta_\eta$ is a normalized broadened delta function of width $\eta$ (we use a smooth, normalized broadening; results are insensitive to the specific choice when $\eta$ is smaller than the spectral features of interest).
The layer-summed spectral function is $A(\mathbf k_0,\varepsilon)=\sum_l A_l(\mathbf k_0,\varepsilon)$.

In the converged plane-wave limit, shifting the seed momentum $\mathbf k_0$ by any reciprocal-module vector $m_1\mathbf G_1^{\rm M}+m_2\mathbf G_2^{\rm M}+m_3\mathbf G_{\rm 1D}$ only relabels basis states, so physical spectra become independent of such shifts.
For rank-3 geometries we therefore sample $\mathbf k_0$ uniformly over a chosen compact reference window $\mathcal W$ and normalize by its area, verifying convergence with respect to $\mathcal W$, the $\mathbf k_0$ mesh, and the plane-wave cutoff \cite{Moon2019_PRB_QC}.
The DOS per real-space area is then defined as
\begin{equation}
D(\varepsilon)=g_s g_v \int_{\mathcal W}\frac{d^2k_0}{(2\pi)^2}\,
\sum_{\alpha}\delta_{\eta}\left(\varepsilon-\varepsilon_{\alpha}(\mathbf{k}_0)\right),
\label{eq:S2:DOS_def_area}
\end{equation}
where $g_s=2$ and $g_v=2$ are the spin and valley degeneracies.
For globally periodic configurations we set $\mathcal W={\rm rBZ}$ so that Eq.~\eqref{eq:S2:DOS_def_area} gives the total DOS per unit area; for rank-3 geometries we use a fixed finite $\mathcal W$ and explicitly verify independence of $\mathcal W$ upon convergence.

To diagnose a neutrality band opening in rank-3 geometries, we use the conservative estimator
$\Delta_{\alpha\beta}\equiv \min_{\mathbf k_0\in\mathcal W}\big[E_\alpha(\mathbf k_0)-E_\beta(\mathbf k_0)\big]$.
Here $(\alpha,\beta)$ are chosen as the lowest conduction and highest valence bands around the CNP within the truncated spectrum.

\subsection{Convergence and numerical settings}

We verify the convergence and numerical robustness by the following checks: (i) increase $E_{\max}$ until the relevant band dispersions and the CNP gap diagnostics are stable within the plotted resolution;
(ii) refine the rBZ mesh (for periodic geometries) or the seed-momentum sampling mesh (for rank-3 geometries);
(iii) for rank-3 geometries, enlarge the reference window $\mathcal W$ to the next smallest polygonal reference window and verify that $D(\varepsilon)$ and $\Delta_{\alpha\beta}$ do not change;
(iv) decrease the broadening $\eta$ in $\delta_\eta$ to confirm that spectral features are not artificially smeared.

\section{Dressed representation details}
\label{sec:dressed_rep}

Here we introduce an exact layer-dependent, position-dependent unitary transformation in sublattice space.
The transformation leaves the spectrum invariant, but reorganizes the Hamiltonian such that the 1D scalar potentials
$V_l(\mathbf r)$ are absorbed into (i) a harmonic dressing of the intralayer kinetic terms and (ii) a harmonic dressing
of the interlayer tunneling.
The parity structure of this unitary-rotated (dressed) representation provides a natural starting point for the resonant analysis in Appendices~\ref{sec:exact_resonance} and \ref{sec:near_resonance}, where the Dirac--Dirac coupling is governed by a single resonant harmonic.

\subsection{Coordinate convention for the 1D modulation}
For the analytic development we choose the 1D reciprocal vector to be parallel to the local longitudinal axis,
\begin{equation}
\mathbf{G}_\mathrm{1D} = G_\mathrm{1D} \hat{\mathbf{y}}
=
\begin{pmatrix}
0 \\ G_\mathrm{1D}
\end{pmatrix},
\label{eq:S:G_1D_parallel_to_y}
\end{equation}
so that the local coordinate $y$ coincides with the longitudinal coordinate $r_\parallel\equiv \hat{\mathbf e}_\parallel\cdot \mathbf r$ used in the main text. This is only a local analytic frame choice aligned with $\mathbf G_{\rm 1D}$; the global geometry remains defined in the fixed laboratory frame of Sec.~\ref{sec:reference_graphene_lattice}.
With Eq.~\eqref{eq:S:G_1D_parallel_to_y}, the layer-dependent scalar potential becomes
\begin{equation}
\begin{aligned}
V_l(\mathbf r)&=V_l(y)=\frac{|v_l|}{2}\cos (G_{\mathrm{1D}} y + \phi_l),
\\
v_l&=|v_l|e^{i\phi_l}.
\end{aligned}
\label{eq:S:V_y}
\end{equation}

\subsection{Layer-dependent unitary rotation and dressed Hamiltonian}
We define a layer-dependent unitary rotation in sublattice space,
\begin{equation}
U_l(y)\equiv\exp \Bigl[\frac{i}{2}\alpha_l(y)\,\sigma_y^{(l)}\Bigr],
\quad l=1,2,
\label{eq:S4:Ul_def}
\end{equation}
with the phase field $\alpha_l(y)$ chosen to satisfy
\begin{equation}
\alpha_l'(y)\equiv\frac{2}{\hbar v_\mathrm{F}}V_l(y).
\label{eq:S4:alpha_prime_def}
\end{equation}
For the $V_l(y)$ defined in Eq.~\eqref{eq:S:V_y}, $\alpha_l'(y)$ integrates to
\begin{equation}
\alpha_l(y)=|u_l|\,\sin\bigl(G_{\mathrm{1D}} y + \phi_l\bigr),
\label{eq:S4:alpha_l_def}
\end{equation}
where we fix the additive constant such that $\langle \alpha_l(y)\rangle_y=0$.
Since $\alpha_l(y+2\pi/G_{\rm 1D})=\alpha_l(y)$, $U_l(y)$ is periodic in $y$ with period $2\pi/G_{\rm 1D}$.

A related sublattice-dependent unitary transformation has previously been employed
to incorporate a unidirectional scalar potential into the intralayer Dirac Hamiltonian
of monolayer graphene
\cite{Park2008_NatPhys_1D, BreyFertig2009_PRL}.
Here we extend this transformation to a bilayer moir\'e system and show that,
in addition to dressing the intralayer kinetics,
it generates a harmonic dressing of the interlayer tunneling,
which plays a central role in the Dirac--Dirac resonance discussed below.

We apply the block-diagonal unitary transformation
$U(y)=\mathrm{diag}\big(U_1(y),U_2(y)\big)$
to the full real-space Hamiltonian $H_\xi(\mathbf r)$ and define the unitary-transformed Hamiltonian
\begin{widetext}
\begin{equation}
\begin{aligned}
\tilde{H}_\xi(\mathbf{r})
&\equiv U^\dagger(y) H_\xi(\mathbf{r}) U(y)\\
&=
\begin{pmatrix}
U_1^\dagger(y)\bigl[h_\xi^{(1)}(-i\nabla)+V_1(y)\sigma_0 \bigr]U_1(y)
&
U_1^\dagger(y)T_\xi^\dagger(\mathbf r)U_2(y)\\
U_2^\dagger(y)T_\xi(\mathbf r)U_1(y)
&
U_2^\dagger(y)\bigl[h_\xi^{(2)}(-i\nabla)+V_2(y)\sigma_0\bigr]U_2(y)
\end{pmatrix}\\
&\equiv
\begin{pmatrix}
\tilde h_\xi^{(1)}(-i\nabla) & \tilde T_\xi^\dagger(\mathbf r)\\
\tilde T_\xi(\mathbf r) & \tilde h_\xi^{(2)}(-i\nabla)
\end{pmatrix},
\end{aligned}
\label{eq:S:dressed_full_hamiltonian}
\end{equation}
\end{widetext}
where we used $U_l^\dagger \sigma_y^{(l)} (-i \partial_y)U_l=\sigma_y^{(l)}(-i\partial_y)+(\alpha_l'(y)/2)\sigma_0$.
Because $U(y)$ is unitary, $\tilde H_\xi$ is isospectral to $H_\xi$ and remains Hermitian.

\subsection{Dressed intralayer Dirac Hamiltonian, harmonic structure, and intralayer parity}
A key feature of the choice Eq.~\eqref{eq:S4:alpha_prime_def} is that the scalar potential is absorbed into the dressed intralayer kinetic terms,
\begin{equation}
\begin{aligned}
\tilde{h}_\xi^{(l)}(-i\nabla)
= -\hbar v_\mathrm{F} \bigl[
&\xi\bigl(\cos\alpha_l(y)\,\sigma_x^{(l)}-\sin\alpha_l(y)\,\sigma_z \bigr)(-i\partial_x) \\
&+\sigma_y^{(l)}(-i\partial_y)
\bigr],
\qquad l=1,2.
\end{aligned}
\label{eq:S:dressed_intralayer_H_real}
\end{equation}
Since $\alpha_l(y)$ is periodic, we expand
$\cos\alpha_l(y)=\sum_{p\in\mathbb Z}C_{l,p}\,e^{ipG_\mathrm{1D}y}$ and
$\sin\alpha_l(y)=\sum_{p\in\mathbb Z}S_{l,p}\,e^{ipG_\mathrm{1D}y}$.
Using the Jacobi--Anger expansion $e^{iz\sin t}=\sum_{p\in\mathbb Z}J_p(z)e^{ipt}$ with $t=G_{\rm 1D}y+\phi_l$,
the coefficients take the closed form
\begin{equation}
\begin{aligned}
C_{l,p}&=
\begin{cases}
J_p(|u_l|)\,e^{ip\phi_l}, & p\ \text{even},\\[1mm]
0, & p\ \text{odd},
\end{cases}\\
S_{l,p}&=
\begin{cases}
0, & p\ \text{even},\\[1mm]
-i\,J_p(|u_l|)\,e^{ip\phi_l}, & p\ \text{odd},
\end{cases}
\end{aligned}
\label{eq:S:cos_and_sin_of_dressed_intralayer}
\end{equation}
with $J_p$ the Bessel function of the first kind.
Thus, $\cos\alpha_l$ contains only even harmonics and $\sin\alpha_l$ contains only odd harmonics.
Moreover, $C_{l,-p}=C_{l,p}^\ast$ and $S_{l,-p}=S_{l,p}^\ast$, ensuring that $\cos\alpha_l$ and $\sin\alpha_l$ are real as required.

Equation~\eqref{eq:S:dressed_intralayer_H_real} makes explicit that, in the unitary-rotated basis,
all $y$-harmonic mixing in the intralayer problem enters through the $-i\partial_x$ term (i.e., $q_x$),
whereas the $-i\partial_y$ term remains diagonal in the harmonic index.
This structural fact will be central to the minimal-wave model and the parity selection rule in Appendix~\ref{sec:exact_resonance}.

\subsection{Dressed interlayer tunneling and strategy for later truncations}
The dressed interlayer tunneling inherits the same $y$-periodicity:
\begin{equation}
\begin{aligned}
\tilde{T}_\xi(\mathbf{r})
&=\sum_{j=0}^{2} \tilde{T}_j^\xi(y)\, e^{i\mathbf{q}_j^\xi \cdot \mathbf{r}}, \\
\tilde{T}_j^\xi(y) &\equiv U_2^\dagger(y)\, T_j^\xi\, U_1(y), \qquad j=0,1,2.
\end{aligned}
\label{eq:S4:dressed_T_general}
\end{equation}
Because $U_{1,2}(y)$ are periodic in $y$, we expand
$\tilde{T}_j^\xi(y)=\sum_{p\in\mathbb{Z}} \tilde{T}_{j,p}^\xi e^{ipG_\mathrm{1D}y}$.
In the resonant analysis (Appendices~\ref{sec:exact_resonance} and \ref{sec:near_resonance}) we will adopt a central-harmonic (minimal-wave) truncation tailored to the low-energy Dirac--Dirac interaction.
Within that construction, the resonant coupling is controlled by a single Fourier component of the $j=0$ channel,
while $j=1,2$ enter only through remote processes that renormalize velocities and produce subleading corrections.
Accordingly, we provide below the explicit $y$-dependence of $\tilde T_0^\xi(y)$, and we parameterize its harmonics in a form suited for extracting the resonant component in Appendix~\ref{sec:exact_resonance}.

\subsection{Explicit form of \texorpdfstring{$\tilde T_0^\xi(y)$}{T0(y)} and its harmonic building blocks}

We introduce the layer-symmetric and layer-antisymmetric combinations
\begin{equation}
\begin{aligned}
u_\pm &\equiv u_1\pm u_2 \equiv |u_\pm|e^{i\phi_\pm},\\
\alpha_\pm(y)&\equiv \alpha_1(y)\pm \alpha_2(y)=|u_\pm|\sin\bigl(G_{\mathrm{1D}} y + \phi_\pm\bigr), \\
\mathcal{J}_{\pm,p} &\equiv J_p(|u_\pm|/2) e^{ip\phi_\pm}.
\end{aligned}
\label{eq:S4:upm_def}
\end{equation}
\noindent
Then, 
\begin{equation}
\begin{aligned}
\tilde T_0^\xi(y)
=
t_0\big[
&A_I(y)\,\sigma_0
+A_x(y)\,\sigma_x \\
&+A_y(y)\,\sigma_y
+A_z(y)\,\sigma_z
\big],
\end{aligned}
\label{eq:S4:T0_Pauli_decomp}
\end{equation}
in the common $(A,B)$ sublattice basis with
\begin{equation}
\begin{aligned}
A_I(y) &= c^2\cos\frac{\alpha_-(y)}{2}+s^2\cos\frac{\alpha_+(y)}{2}+i\,s\,\sin\frac{\alpha_+(y)}{2}, \\
A_x(y) &= \cos\frac{\alpha_+(y)}{2}+i\,s\,\sin\frac{\alpha_+(y)}{2}, \\
A_y(y) &= i\,c\,\sin\frac{\alpha_-(y)}{2}, \\
A_z(y) &= -c\,\sin\frac{\alpha_+(y)}{2} \\ 
& \quad -i\,s c\,\Big(\cos\frac{\alpha_-(y)}{2}-\cos\frac{\alpha_+(y)}{2}\Big),
\end{aligned}
\label{eq:S4:A_coeffs_def}
\end{equation}
where we defined $c\equiv\cos(\theta/2)$ and $s\equiv\sin(\theta/2)$.
All $y$ dependence of $\tilde T_0^\xi(y)$ is therefore controlled by $\cos(\alpha_\pm/2)$ and $\sin(\alpha_\pm/2)$, which we expand as
$\cos (\alpha_\pm(y)/2) =\sum_{p\in\mathbb Z}C_{\pm,p}\,e^{ipG_\mathrm{1D}y}$
and
$\sin (\alpha_\pm(y)/2) =\sum_{p\in\mathbb Z}S_{\pm,p}\,e^{ipG_\mathrm{1D}y}$,
with coefficients
\begin{equation}
\begin{aligned}
C_{\pm,p}&=
\begin{cases}
\mathcal{J}_{\pm,p}, & p\ \text{even},\\[1mm]
0, & p\ \text{odd},
\end{cases} \\
S_{\pm,p}&=
\begin{cases}
0, & p\ \text{even},\\[1mm]
-i \mathcal{J}_{\pm,p}, & p\ \text{odd}.
\end{cases}
\end{aligned}
\label{eq:S:cos_and_sin_of_dressed_interlayer}
\end{equation}
Equation~\eqref{eq:S:cos_and_sin_of_dressed_interlayer} is the interlayer analogue of
Eq.~\eqref{eq:S:cos_and_sin_of_dressed_intralayer} and encodes the same even/odd harmonic parity.
In Appendix~\ref{sec:exact_resonance}, we show that this parity structure enforces the parity rule responsible for the gap opening in the resonant configuration through the Pauli component of a single resonant Fourier harmonic of $\tilde T_0^\xi(y)$.

\section{Exact-resonance reduction and selection-rule derivation}
\label{sec:exact_resonance}

We develop a compact few-wave description of the low-energy electronic structure of resonant TBG--1D,
defined by the Dirac-resonance condition Eq.~\eqref{eq:prb_resonance}.
The explicit reduction below is written for the nearest-cone representative $j=0$, $\mathbf G_{\rm M}=0$, because this is the branch used in the full-wave spectra, near-resonant maps, and feasibility estimates. A generalized moir\'e-Umklapp resonance has the same local two-cone algebra after selecting a resonant tuple $(j,\mathbf G_{\rm M},p_\star)$ satisfying Eq.~\eqref{eq:prb_umklapp_resonance}, provided the remaining channels are remote on the scale of the low-energy Dirac problem. In that case the parity rule is read from the selected integer $p_\star$, while the shorthand $n_{\rm 1D}$ parity applies only to the nearest-cone specialization $p_\star=p_0=-\xi n_{\rm 1D}$.
Starting from the unitary-rotated Hamiltonian $\tilde H_\xi(\mathbf r)$, we derive a minimal two-cone model
that captures the direct intercone resonance responsible for the resonant-gap physics.
Depending on the relative chirality between the Dirac cones renormalized by the 1D potential, only one of the two harmonic parities can host a Pauli component that opens a gap at the Dirac point.

\paragraph{Approximation scheme.}
Throughout this appendix we work to linear order in the reduced momentum $\mathbf q$ about each Dirac point.
The intralayer dressing by the 1D potential is kept nonperturbatively through the central-harmonic factor
$\eta_l=J_0(|u_l|)$.
For the interlayer sector, we retain only the two central cones (the $m=0$ sectors in each layer) and the single resonant harmonic
$\mathcal M_\xi^{(n_{\rm 1D})}$, corresponding to the $-\xi n_{\rm 1D}$th harmonic of the dressed $j=0$ tunneling $\tilde{T}_0^\xi$, with its direct 1D-dressed amplitude retained nonperturbatively in $u_l$ through the Bessel coefficients. The subsequent remote-channel self-energy corrections are still evaluated in the leading $u_0^2$ approximation.

\subsection{1D periodicity and the reduced-zone (strip) basis}

For the symmetric layer rotation $\theta_1=-\theta/2$ and $\theta_2=\theta/2$ in this work, the $\mathbf{G}_\mathrm{1D}$ for a commensurate configuration is always parallel to $\hat{\mathbf{y}}$.
In the resonant analysis, after retaining the $j=0$ channel explicitly (and treating $j=1,2$ as remote self-energy contributions, Appendix~\ref{sec:self_energ_correction}), the effective low-energy Hamiltonian is periodic in $y$ with period $2\pi/G_\mathrm{1D}$; hence we introduce the one-dimensional rBZ associated with $\mathbf G_{\rm 1D}$,
\begin{equation}
-\frac{G_{\mathrm{1D}}}{2}\le q_y < \frac{G_{\mathrm{1D}}}{2}.
\label{eq:S1-qy-strip}
\end{equation}
For a given layer $l$ and valley $\xi$, any valley-centered momentum relative to the Dirac point,
$\mathbf k-\mathbf K_\xi^{(l)}$, can be uniquely decomposed as
\begin{equation}
\mathbf k-\mathbf K_\xi^{(l)} \equiv \tilde{\mathbf q}_m
=\mathbf q+m\mathbf G_{\mathrm{1D}}
=(q_x,\ q_y+mG_{\mathrm{1D}}),
\label{eq:S1-qtilde}
\end{equation}
where $\mathbf q=(q_x,q_y)$ lies in the rBZ and $m\in\mathbb Z$ labels the corresponding harmonic (sector) index.
We therefore introduce two-component sublattice spinors
\begin{equation}
|\mathbf q,m,l\rangle \equiv
\bigl(|\mathbf q,m,A,l\rangle,\ |\mathbf q,m,B,l\rangle\bigr)^{\mathrm T},
\label{eq:S:strip_basis}
\end{equation}
with 
$|\mathbf q,m,X,l\rangle \equiv |\tilde{\mathbf q}_m,X,l\rangle
=|\mathbf q+m\mathbf G_{\mathrm{1D}},X,l\rangle$
for $X=A,B$ and $l=1,2$,
as sector basis states.

\paragraph{Layer-dependent folding.}
% (Cite Figure.)
The mapping between $(\mathbf q,m)$ and the Bloch momentum $\mathbf k$ is layer dependent because
$\mathbf K_\xi^{(1)}\neq \mathbf K_\xi^{(2)}$.
In particular, the same reduced momentum $\mathbf q$ corresponds to Bloch momenta shifted by
$\Delta\mathbf K_\xi = \mathbf K_\xi^{(1)}-\mathbf K_\xi^{(2)}$ between the two layers.
In a resonant configuration,
the Dirac point of one layer is folded into a definite sector of the other with the same $\mathbf{q}$, and this folding enables a unique low-energy
Dirac--Dirac resonance via a single Fourier component of the dressed tunneling.

\subsection{Dressed intralayer blocks in reduced-zone basis and the central-harmonic projection}

A Fourier component of the $y$-periodic dressed Hamiltonian with harmonic index $p$
couples the momentum sectors $m$ and $m+p$.
Accordingly, the dressed intralayer Hamiltonian in the reduced-zone basis is
\begin{equation}
\begin{aligned}
&\langle \mathbf q',m',l' | \tilde h_\xi^{(l)}(-i\nabla) |\mathbf q,m,l \rangle \\
&=
-\hbar v_\mathrm{F}\Big[
\xi q_x \big( C_{l,m'-m}\,\sigma_x^{(l)}-S_{l,m'-m}\,\sigma_z \big) \\
&\qquad\qquad\ +(q_y+mG_\mathrm{1D})\,\sigma_y^{(l)}\,\delta_{m',m}
\Big] \delta_{\mathbf{q}',\mathbf{q}} \delta_{l',l}.
\end{aligned}
\label{eq:S3-Hl-strip-matrix}
\end{equation}
Equation~\eqref{eq:S3-Hl-strip-matrix} makes two structural facts explicit:
(i) the longitudinal $q_y$ term is diagonal in $m$; and
(ii) all $m$-mixing arises from the transverse $q_x$ term through the harmonic coefficients $C_{l,p}$ and $S_{l,p}$.

\paragraph{Central-harmonic projection.}
To obtain a compact low-energy model, we retain only the central sector $m=0$ for the intralayer kinetics,
i.e.\ we project Eq.~\eqref{eq:S3-Hl-strip-matrix} onto $m=m'=0$.
Using $S_{l,0}=0$ and $C_{l,0}=J_0(|u_l|)$, we obtain the central-harmonic intralayer Dirac Hamiltonian
\begin{equation}
\begin{aligned}
\tilde h_{\xi,{\rm CH}}^{(l)}(\mathbf q)
&=
-\hbar v_\mathrm{F}\Big[
\xi \eta_l\,q_x\,\sigma_x^{(l)}+q_y\,\sigma_y^{(l)}
\Big],\\
\eta_l &\equiv J_0(|u_l|),
\end{aligned}
\label{eq:S5:CH_intralayer}
\end{equation}
which is nonperturbative in the 1D modulation amplitude through $\eta_l$.

\paragraph{Control of the central-harmonic projection.}
The eliminated sectors $m\neq 0$ are separated in energy by $\sim \hbar v_\mathrm{F}|m|G_{\rm 1D}$.
They couple to $m=0$ through the $q_x$ term in Eq.~\eqref{eq:S3-Hl-strip-matrix}, with matrix elements
$\sim \hbar v_\mathrm{F}|q_x|$.
Eliminating $m\neq 0$ therefore produces corrections scaling as
$(\hbar v_\mathrm{F} q_x)^2/(\hbar v_\mathrm{F} G_{\rm 1D})\sim \hbar v_\mathrm{F} (q_x^2/G_{\rm 1D})$, i.e.\ quadratic in momentum.
Hence, within a Dirac theory kept to linear order in $\mathbf q$, Eq.~\eqref{eq:S5:CH_intralayer} captures the exact $\mathcal O(q)$ intralayer kinetics in the unitary-rotated basis.

\subsection{Dressed interlayer tunneling in reduced-zone basis, single resonant harmonic \texorpdfstring{$\tilde T_{0,p_0}^\xi$}{T0p}, and leading-order parity rule}

The dressed interlayer tunneling is
\begin{equation}
\begin{aligned}
\tilde T_\xi(\mathbf r)&=\sum_{j=0}^2 \tilde T_j^\xi(y)\,e^{i\mathbf q_j^\xi\cdot\mathbf r}
\ \
\left(\tilde T_j^\xi(y)=U_2^\dagger(y)\,T_j^\xi\,U_1(y)\right),\\
\tilde T_j^\xi(y)&=\sum_{p\in\mathbb Z}\tilde T_{j,p}^\xi\,e^{ipG_{\rm 1D}y}.
\end{aligned}
\label{eq:S5:Tj_Fourier_general}
\end{equation}
Throughout this appendix we specialize to the resonance family $\mathbf G_{\rm 1D}\parallel \mathbf q_0^\xi=\Delta\mathbf K_\xi$, specifically the nearest-cone branch with $\mathbf G_{\rm M}=0$, which is the family realized by the local strip-basis convention aligned with $\mathbf G_{\rm 1D}$. The nonzero-$\mathbf G_{\rm M}$ moir\'e-Umklapp generalization is summarized in Appendix~\ref{sec:S_moire_umklapp_resonances}; it amounts to replacing $p_0=-\xi n_{\rm 1D}$ by the integer solution of Eq.~\eqref{eq:prb_umklapp_resonance}.
Then in resonant configurations, the Dirac--Dirac resonance is governed by the $j=0$ channel, and can therefore be bridged by a
single 1D harmonic.
More generally, within the nearest-cone class $\mathbf G_{\rm M}=0$, since the first-star set satisfies $\mathbf q_j^\xi=(C_3)^j\mathbf q_0^\xi$, a commensurate geometry chosen with $\mathbf G_{\rm 1D}\parallel \mathbf q_{j'}^\xi$ ($j'=1$ or 2) is described by the same strip-basis construction after the relabeling $0\to j'$, with the direct resonant harmonic supplied by $\tilde T_{j',\,p=-\xi n_{\rm 1D}}^\xi$.

\paragraph{Strip-basis selection rule (resonant $j=0$ channel).}

Using $\mathbf q_0^\xi=\xi n_{\rm 1D}\mathbf G_{\rm 1D}$ in the nearest-cone resonant geometry,
the matrix element of the $j=0$ channel is
\begin{multline}
\langle  \mathbf q',m',2 \, | \, \tilde T_0^\xi(y)e^{i\mathbf q_0^\xi\cdot\mathbf r} \, | \, \mathbf q,m,1  \rangle\\
=
\tilde T_{0,p}^\xi\, \delta_{{\mathbf{q}',\mathbf{q}}} \delta_{m',\,m+\xi n_{\rm 1D}+p}.
\label{eq:S5:sector_selection_rule}
\end{multline}
Thus, the two central Dirac cones (the $m=0$ sectors of layers 1 and 2 at the same reduced momentum $\mathbf q$)
are directly coupled only by the single resonant harmonic with $p=-\xi n_\mathrm{1D}\equiv p_0$,
\begin{equation}
 \mathcal M^{(n_\mathrm{1D})} \equiv \tilde T_{0,p_0}^\xi = \tilde T_{0,p=-\xi n_{\rm 1D}}^\xi.
\label{eq:S5:resonant_harmonic_def}
\end{equation}
We suppress the valley index $\xi$ in $\mathcal{M}^{(n_{\rm 1D})}$, because $T_0^\xi$ and its unitary-transformed counterpart are independent of $\xi$.
Since the parity of $p_0$ is the same as the parity of $n_{\rm 1D}$, the resonant index $n_\mathrm{1D}$ directly controls the Pauli content of the resonant coupling. For the generalized moir\'e-Umklapp condition Eq.~\eqref{eq:prb_umklapp_resonance}, this statement should instead be phrased in terms of the selected integer harmonic $p$.

The Fourier component $\tilde T_{0,p}^\xi$ can be written in closed form as
\begin{widetext}
\begin{equation}
\tilde{T}_{0,p}^\xi=
\begin{cases}
t_0\Big[
\big(c^2\mathcal J_{-,p}+s^2\mathcal J_{+,p}\big)\sigma_0
+\mathcal J_{+,p}\sigma_x
-i s c\big(\mathcal J_{-,p}-\mathcal J_{+,p}\big)\sigma_z
\Big], & p\ \text{even},\\[1mm]
t_0\Big[
s\,\mathcal J_{+,p}(\sigma_0+\sigma_x)
+c\,\mathcal J_{-,p}\sigma_y
+i c\,\mathcal J_{+,p}\sigma_z
\Big], & p\ \text{odd}.
\end{cases}
\label{eq:S:T0_evenodd_p}
\end{equation}
\end{widetext}
In the $\theta\to 0$ limit ($c\to 1$, $s\to 0$), Eq.~\eqref{eq:S:T0_evenodd_p} reduces to the clean parity structure
\begin{equation}
\tilde{T}_{0,p}^\xi \in
\begin{cases}
\mathrm{span}\{\sigma_0,\sigma_x\}, & p\ \text{even},\\[1mm]
\mathrm{span}\{\sigma_y,\sigma_z\}, & p\ \text{odd}.
\end{cases}
\label{eq:S:T0_parity_rule_theta0}
\end{equation}
The terms neglected in Eq.~\eqref{eq:S:T0_parity_rule_theta0} are proportional to
$s=\sin(\theta/2)$, which is parametrically small,
$s=\mathcal{O}(\theta)$, for the twist angles of interest in TBG
($\theta\lesssim5^\circ$), but can become non-negligible at larger twist angles.
In the main text, we use Eq.~\eqref{eq:S:T0_parity_rule_theta0} as the leading organizing principle for the even/odd $n_\mathrm{1D}$ resonant criterion,
and treat the $s$-dependent admixtures as subleading corrections.

\paragraph{Role of $j=1,2$ channels.}
In the $\mathbf G_{\rm 1D}\parallel \mathbf q_0^\xi$ resonance family analyzed here, the $j=1,2$ moir\'e channels carry momenta rotated away from $\mathbf G_{\rm 1D}$ and therefore do not generate a direct strip-resonant coupling between the two central cones. Their leading effect is to renormalize velocities and dressing factors via remote processes, treated in Appendix~\ref{sec:self_energ_correction} and validated against full-wave numerics (Appendix~\ref{sec:full_wave_model}).

\subsection{Self-energy correction from remote channels}
\label{sec:self_energ_correction}

We neglect intralayer couplings to the $m=\pm1$ remote intermediate states because the
$m\!\to\! m\pm1$ vertices in the dressed strip Hamiltonian are proportional to $q_x$.
Integrating out these sidebands therefore generates a self-energy $\Sigma_{\rm 1D}\sim (\hbar v_\mathrm{F} q_x)^2/(\hbar v_\mathrm{F} G_{\rm 1D})
=\mathcal O(q_x^2)$,
which lies beyond the linear Dirac theory considered here.

For the moir\'e remote channels, all remaining harmonics of $\tilde{T}_0$, namely
$\tilde{T}_{0,p\neq p_0}$, as well as the $j=1,2$ tunneling $\tilde{T}_j^\xi$,
couple the central cones only to remote intermediate states with a characteristic
energy scale $\sim \hbar v_\mathrm{F} k_\theta$.
In the unitary-transformed representation, the exact first-star remote self-energy for the central cones (before any $p$ truncation) is
\begin{widetext}
\begin{equation}
\begin{aligned}
\Sigma_{1,{\rm M}}^{\xi,{\rm exact}}(E,\mathbf q)
=
\sum_{j=0}^2\ \sum_{p\in\mathbb Z}^{\ \prime}\ 
(\tilde T_{j,p}^\xi)^\dagger\,
\Big[E-\tilde h_{\xi,{\rm CH}}^{(2)}(\mathbf q+\mathbf q_j^\xi+p\mathbf G_{\rm 1D})\Big]^{-1}
\tilde T_{j,p}^\xi,\\
\Sigma_{2,{\rm M}}^{\xi,{\rm exact}}(E,\mathbf q)
=
\sum_{j=0}^2\ \sum_{p\in\mathbb Z}^{\ \prime}\ 
\tilde T_{j,p}^\xi\,
\Big[E-\tilde h_{\xi,{\rm CH}}^{(1)}(\mathbf q-\mathbf q_j^\xi-p\mathbf G_{\rm 1D})\Big]^{-1}
(\tilde T_{j,p}^\xi)^\dagger.
\end{aligned}
\label{eq:S5:SigmaM_exact}
\end{equation}
\end{widetext}
where the prime on the $p$ sum means that the direct resonant channel $p=p_0$ already kept explicitly (Eq.~\eqref{eq:S5:resonant_harmonic_def}) is excluded from $\Sigma_{\rm M}$ to prevent double counting.

We now make controlled simplifications of Eq.~\eqref{eq:S5:SigmaM_exact} appropriate for extracting the leading $\mathcal O(u_0^2)$ smooth renormalization.
In a weak-modulation expansion $U_l(y)=e^{i\alpha_l(y)\sigma_y/2}$ with $\alpha_l=\mathcal O(u_l)$ and $\langle\alpha_l\rangle_y=0$,
one has $\tilde T_{j,0}^\xi=T_j^\xi+\mathcal O(|u_\pm|^2)$, where $T_j^\xi=\mathcal{O}(u_0)$, and $\tilde T_{j,p\neq0}^\xi=\mathcal O(|u_\pm|^{|p|})$,
while the intermediate-state energy denominators for nonresonant remote channels are always of order
$\hbar v_\mathrm{F} |\mathbf q_j^\xi+p\mathbf G_{\rm 1D}|\sim p \hbar v_\mathrm{F} k_\theta$.
Therefore, the $p\neq0$ contribution to $\Sigma_\mathrm{M}$ is suppressed as $\mathcal O(u_0^2|u_\pm|^2)$.
Thus, to leading $\mathcal O(u_0^2)$ accuracy we keep only the unit-cell averaged vertex and further approximate it by the bare $T_j^\xi$.
\begin{equation}
\tilde T_j(y) \approx \tilde T_{j,0}^\xi = \langle \widetilde T_j(y)\rangle_y,
\qquad
\tilde T_{j,0}^\xi\approx T_j^\xi.
\label{eq:S5:Tj_p0_approx_chain}
\end{equation}
Similarly, we evaluate the remote propagators with the bare Dirac operator,
$\tilde h_{\xi,{\rm CH}}^{(l)}\approx h_{\xi}^{(l)}$,
since $J_0(|u_l|)=1+\mathcal O(u_l^2)$ implies that the difference to $\Sigma_\mathrm{M}$ is
$\mathcal O(u_0^2|u_l|^2)$.
With these approximations, Eq.~\eqref{eq:S5:SigmaM_exact} reduces to
\begin{equation}
\begin{aligned}
\Sigma_{1,{\rm M}}^\xi(E,\mathbf q)
&=
\sum_{j=0}^2
(T_j^\xi)^\dagger\,
\Big[E-h_{\xi}^{(2)}(\mathbf q+\mathbf q_j^\xi)\Big]^{-1}
T_j^\xi,\\
\Sigma_{2,{\rm M}}^\xi(E,\mathbf q)
&=
\sum_{j=0}^2
T_j^\xi\,
\Big[E-h_{\xi}^{(1)}(\mathbf q-\mathbf q_j^\xi)\Big]^{-1}
(T_j^\xi)^\dagger.
\end{aligned}
\label{eq:S5:SigmaM_def}
\end{equation}
Linearizing Eq.~\eqref{eq:S5:SigmaM_def} to first order in $(E,\mathbf q)$ gives
\begin{equation}
\begin{aligned}
\Sigma_{l,{\rm M}}^\xi(E,\mathbf q)&\approx
\Sigma^{(E)}E+\Sigma_{l}^{(q)}(\mathbf q),\\
\Sigma^{(E)}&=-6u_0^2\,\sigma_0,\\
\Sigma_{l}^{(\mathbf{q})}(\mathbf q)&=-3u_0^2\,h_{\xi}^{(l)}(\mathbf q).
\end{aligned}
\label{eq:S5:Sigma_linear}
\end{equation}
To incorporate the remote-channel correction in the two-cone problem, one adds the diagonal self-energy blocks to the central-cone Hamiltonian.
Keeping only terms linear in $(E,\mathbf q)$, Eq.~\eqref{eq:S5:Sigma_linear} implies the following energy-dependent eigenvalue equation:
\begin{multline}
\begin{pmatrix}
\tilde h_{\xi,{\rm CH}}^{(1)}(\mathbf q)+\Sigma_{1}^{(\mathbf{q})}(\mathbf q) & \big(\mathcal M^{(n_{\rm 1D})}\big)^\dagger\\
\mathcal M^{(n_{\rm 1D})} & \tilde h_{\xi,{\rm CH}}^{(2)}(\mathbf q)+\Sigma_{2}^{(\mathbf{q})}(\mathbf q)
\end{pmatrix}
\Psi\\
=
(1+6u_0^2)E\,\Psi
+\mathcal O(|\mathbf q|^2,u_0^2E|\mathbf q|,u_0^2E^2).
\label{eq:S5:2c_with_Sigma_lin}
\end{multline}
Thus, the $E$ dependence can be removed by multiplying Eq.~\eqref{eq:S5:2c_with_Sigma_lin} by the quasiparticle weight
\begin{equation}
Z\equiv (\sigma_0-\Sigma^{(E)})^{-1}=\frac{1}{1+6u_0^2}.    
\label{eq:S:Z_def}
\end{equation}
See Eq.~\eqref{eq:S5:two_cone_model} for the final form.

\subsection{Minimal two-cone model and the parity--chirality selection rule on gap-opening channel}
\label{sec:min_two_cone_model}

\paragraph{Key point.}
The nearest-cone commensurate $n_{\rm 1D}$ resonance selects a single harmonic of the dressed $j=0$ moir\'{e} tunneling, $p_0=-\xi n_{\rm 1D}$, that directly hybridizes the two central Dirac cones. For a generalized moir\'e-Umklapp resonance, the same local two-cone logic applies after replacing this branch-specific $p_0$ by the selected integer harmonic $p$ solving Eq.~\eqref{eq:prb_umklapp_resonance}.
This determines the Pauli structure of the resonant intercone coupling $Z\mathcal M^{(n_{\rm 1D})}$ (to leading order in $\theta$) via the even--odd rule.
Whether such a Pauli structure opens an insulating gap depends additionally on the relative chirality of the two interacting cones, which fixes the effective mass channel in the coupled two-cone problem.
The minimal theory is used to identify symmetry-allowed gap channels and scaling; the existence of a direct gap throughout the BZ is verified by the fully periodic full-wave numerics (Appendix~\ref{sec:full_wave_model}).

\paragraph{Minimal-wave (two-cone) Hamiltonian and the resonant harmonic.}
Combining the central-harmonic dressed intralayer kinetics Eq.~\eqref{eq:S5:CH_intralayer}, the resonant harmonic Eq.~\eqref{eq:S5:resonant_harmonic_def},
and the remote-channel renormalization Eq.~\eqref{eq:S5:2c_with_Sigma_lin},
we obtain the renormalized minimal two-cone model
\begin{widetext}
\begin{equation}
H_\xi^\mathrm{(2c)}(\mathbf q)=
\begin{pmatrix}
-\hbar\big(\xi v_{x}^{(1)} q_x\,\sigma_x^{(1)}+v_{y}^{(1)} q_y\,\sigma_y^{(1)}\big) & \big(Z\mathcal M^{(n_{\rm 1D})}\big)^\dagger\\
Z\mathcal M^{(n_{\rm 1D})} & -\hbar\big(\xi v_{x}^{(2)} q_x\,\sigma_x^{(2)}+v_{y}^{(2)} q_y\,\sigma_y^{(2)}\big)
\end{pmatrix},
\label{eq:S5:two_cone_model}
\end{equation}
\end{widetext}
with
\begin{equation}
v_{x}^{(l)}\equiv Z\,v_\mathrm{F}(\eta_l-3u_0^2),\quad
v_{y}^{(l)}\equiv Z\,v_\mathrm{F}(1-3u_0^2),
\label{eq:S5:ren_recipe}
\end{equation}
where $\eta_l=J_0(|u_l|)$.
This $4\times4$ Hamiltonian is Hermitian by construction and captures the symmetry-dictated Pauli structure of the resonance.

\paragraph{Pauli decomposition and the parity (even--odd $n_\mathrm{1D}$) resonant rule.}
We write the resonant coupling as
\begin{equation}
Z\mathcal M^{(n_{\rm 1D})}
= c_0\,\sigma_0 + c_x\,\sigma_x + c_y\,\sigma_y + c_z\,\sigma_z,
\label{eq:S5:M_Pauli_decomp}
\end{equation}
with the coefficients $c_\mu$ determined by Eqs.~\eqref{eq:S5:resonant_harmonic_def} and \eqref{eq:S:T0_evenodd_p}.
To leading order in $\theta$, the parity rule implies that
\begin{widetext}
\begin{equation}
\lim_{\theta\rightarrow0} Z\mathcal{M}^{(n_\mathrm{1D})} \approx
\begin{cases}
Z t_0\big[
\mathcal{J}_{-,p_0} \,\sigma_0 + \mathcal{J}_{+,p_0} \,\sigma_x
\big], & p_0 \, (=-\xi n_{\rm 1D}) \ \text{even}, \\
Z t_0\big[
\mathcal J_{-,p_0}\,\sigma_y
+i
\,\mathcal J_{+,p_0}\,\sigma_z
\big], & p_0 \, (=-\xi n_{\rm 1D}) \ \text{odd},
\end{cases}
\label{eq:S:M_parity_rule_theta0}
\end{equation}
\end{widetext}
in the leading $\theta\to 0$ limit.

\begin{figure}[t]
\centering
\includegraphics[width=\linewidth]{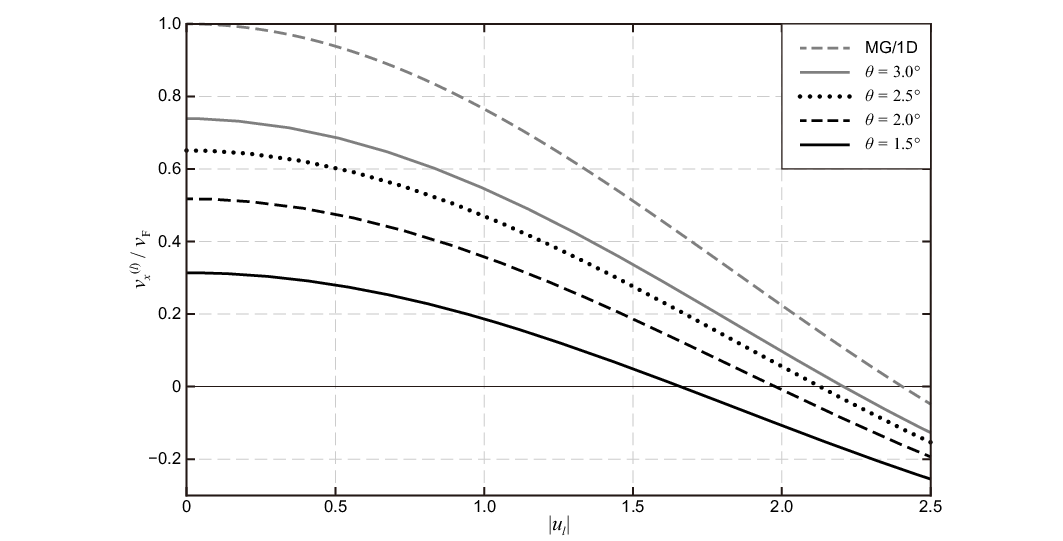}
\caption{Renormalized band velocity along the direction perpendicular to $\mathbf{G}_\mathrm{1D}$, $v_x^{(l)}$, of TBG--1D with $\theta$ from $1.5^\circ$ to $3.0^\circ$ plotted against $|u_l|$. Gray dashed line (MG/1D) shows the renormalized velocity of monolayer graphene under a unidirectional potential for reference. The unidirectional potential causes $v_x^{(l)}$ to change sign as $|u_l|$ increases. The value of $|u_l|$ at which $v_x^{(l)}=0$ decreases as $\theta$ decreases, reflecting the intrinsic velocity renormalization of TBG.
}
\label{fig:transverse_velocity_flip}
\end{figure}

\begin{figure}[t]
\centering
\includegraphics[width=\linewidth]{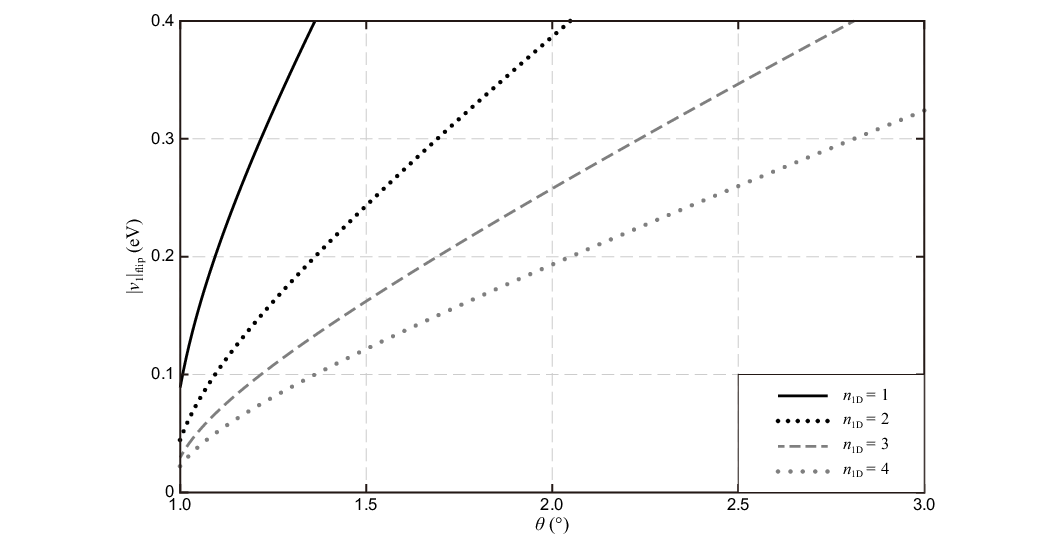}
\caption{Modulation amplitude $|v_1|$ (in eV) at which the transverse velocity $v_x^{(1)}$ first changes sign, determined by $J_0(|u_1|)-3u_0^2=0$, as a function of twist angle $\theta$ for four resonance orders $n_{\rm 1D}=1,2,3,4$.
Exceeding this threshold at fixed $|v_2|$ (layer-asymmetric modulation) reverses $\chi_{\rm rel}$ and switches the gap-opening channel [see Eq.~\eqref{eq:S5:M_shuffle_chi} and main-text Fig.~\ref{fig:prb_chirality}(c)].
}
\label{fig:chirality_flip_threshold}
\end{figure}

\paragraph{Relative chirality and the gap-opening channel.}
A gap in a Dirac system requires a mass term that anticommutes with the kinetic Dirac matrices (see, e.g., Ref.~\cite{Chiu2016_RMP}).
Gap opening in a coupled two-Dirac-cone problem depends on both the Pauli content in Eq.~\eqref{eq:S5:M_Pauli_decomp} and the relative chirality of the two cones.
As in the main text, the Pauli label of a mass channel is meaningful only after fixing a standard kinetic basis. Allowing the two cones to carry valley indices $\xi_1$ (layer 1) and $\xi_2$ (layer 2), and assuming $v_y^{(l)}>0$, we use constant sublattice rotations to write each kinetic block as
\begin{equation}
\begin{aligned}
h_l^{\rm std}(\mathbf q)
&= -\hbar\left[
\chi_l |v_x^{(l)}|q_x\sigma_x
+v_y^{(l)}q_y\sigma_y
\right],
\\
\chi_l&=\xi_l\,\mathrm{sgn}\,v_x^{(l)}.
\end{aligned}
\label{eq:S5:std_dirac_basis}
\end{equation}
In this standard two-cone Dirac basis, the relative chirality is
\begin{equation}
\chi_{\rm rel}\equiv \xi_1\xi_2\,\mathrm{sgn}\big(v_{x}^{(1)}v_{x}^{(2)}\big).
\label{eq:S5:chi_rel_def}
\end{equation}
For the valley-conserving resonant configurations treated in this work, $\xi_1=\xi_2=\xi$ and hence
$\chi_{\rm rel}=\mathrm{sgn}(v_{x}^{(1)}v_{x}^{(2)})$.
As Fig.~\ref{fig:transverse_velocity_flip} shows, 
% As Extended Data Fig.~2 shows, 
$v_x^{(l)}$ [Eq.~\eqref{eq:S5:ren_recipe}] of TBG--1D inverts its sign as $|u_l|$ increases and makes $J_0(|u_l|)-3u_0^2$ negative. 
While monolayer graphene under a unidirectional potential also exhibits similar anisotropic band renormalization (gray dashed line in 
Fig.~\ref{fig:transverse_velocity_flip})
% Extended Data Fig.~2)
\cite{Park2008_NatPhys_1D, BreyFertig2009_PRL, Li2021_NatNano_1DSL},
TBG--1D exhibits the sign inversion at much weaker $|u_l|$ because of the intrinsic band velocity renormalization by TBG (i.e., the $-3u_0^2$ term).
Accordingly, the value of $|u_l|$ at which $v_x^{(l)}$ changes sign decreases as $\theta$ decreases, that is, as $u_0$ increases.
Figure~\ref{fig:chirality_flip_threshold} shows the corresponding modulation amplitude $|v_1|$ [with $u_1=v_1/(\hbar v_\mathrm{F} G_\mathrm{1D})$] that produces the first sign change of $v_x^{(1)}$ as a function of $\theta$ for $n_\mathrm{1D}=1$--4. Exceeding this threshold at fixed $|v_2|$ (layer-asymmetric modulation) reverses $\chi_{\rm rel}$ and switches the gap-opening channel [see Eq.~\eqref{eq:S5:M_shuffle_chi} and main-text Fig.~\ref{fig:prb_chirality}(c)].
For example, if $|v_2|$ is kept weak while $|v_1|$ exceeds this threshold, the dominant gap-opening channel switches from $\sigma_z$ to $\sigma_x$.
Accordingly, the even/odd parity rule is reversed: odd $n_{\rm 1D}$ resonances become gapless, whereas even $n_{\rm 1D}$ resonances open a gap, consistent with Eq.~\eqref{eq:S:M_parity_rule_theta0}.

In the weak-to-moderate modulation regime relevant to the main text, or in the layer-symmetric modulation ($|u_1|\approx|u_2|$), $v_{x}^{(1)}v_{x}^{(2)}>0$ and therefore $\chi_{\rm rel}=+1$. 
By contrast, a strongly layer-asymmetric modulation can invert the sign of one transverse velocity (i.e., $v_x^{(1)} v_x^{(2)}<0$, main-text Fig.~\ref{fig:prb_chirality}(c)), realizing $\chi_{\rm rel}=-1$.

With the standard-basis convention of Eq.~\eqref{eq:S5:std_dirac_basis}, for $\chi_{\rm rel}=+1$ (same chirality), a $\sigma_z$ component in $Z\mathcal M^{(n_{\rm 1D})}$ provides the distinguished mass channel, whereas $\sigma_{x,y}$ primarily shift/split Dirac nodes and can generate charge-neutral pockets.
For $\chi_{\rm rel}=-1$ (opposite chirality), the mass channel is identified after the chiralities are aligned by a constant sublattice rotation on one cone, e.g.
\begin{equation}
W=\sigma_y,\quad W^\dagger\sigma_x W=-\sigma_x,\quad W^\dagger\sigma_y W=\sigma_y.
\label{eq:S5:chirality_align_and_M}
\end{equation}
However, the alignment reshuffles the Pauli components of $Z\mathcal M^{(n_\mathrm{1D})}=c_0\sigma_0+c_x\sigma_x+c_y\sigma_y+c_z\sigma_z$ as
\begin{equation}
\begin{aligned}
Z\mathcal M^{(n_\mathrm{1D})} &\to Z\mathcal M_{\rm align}^{(n_\mathrm{1D})} \\
&\equiv Z\mathcal M^{(n_\mathrm{1D})} W \\
&=
c_0\,\sigma_y
+i c_x\,\sigma_z
+c_y\,\sigma_0
-i c_z\,\sigma_x.
\end{aligned}
\label{eq:S5:M_shuffle_chi}
\end{equation}
Thus, in an opposite-chirality two-cone problem, the $\sigma_x$ term in the original standard-basis expression [Eq.~\eqref{eq:S5:two_cone_model}] serves as an effective gap-opening mass (i.e., $\sigma_z$-like) channel in the chirality-aligned expression. Therefore, the relative chirality rule implies that
\begin{multline}
\mathrm{gap\!-\!opening~(mass)~channel} \\ \quad=
\begin{cases}
\sigma_z, & \chi_\mathrm{rel}=+1,\\[1mm]
\sigma_x, & \chi_\mathrm{rel}=-1.
\end{cases}
\label{eq:S:M_chirality_rule_theta0}
\end{multline}

Note that, for any general, isolated, non-interacting Dirac cone, the sign of $v_{x}^{(l)}$ is not physical: it can be flipped by a constant sublattice rotation, Eq.~\eqref{eq:S5:chirality_align_and_M}.
Only in a coupled Dirac cone system, the sign of $v_{x}^{(l)}$ has a physical meaning: applying such a rotation to only one cone also transforms the intercone coupling,
$Z\mathcal M^{(n_\mathrm{1D})}\rightarrow Z\mathcal M^{(n_\mathrm{1D})} W$,
and therefore reshuffles Pauli components [Eq.~\eqref{eq:S5:M_shuffle_chi}].
But, if both $v_x^{(1)}$ and $v_x^{(2)}$ change their signs (which keeps $\chi_\mathrm{rel}=+1$), the intercone coupling merely changes the sign of its Pauli components,
$Z\mathcal M^{(n_\mathrm{1D})}\rightarrow WZ\mathcal M^{(n_\mathrm{1D})} W=c_0 \sigma_0 -c_x \sigma_x + c_y \sigma_y - c_z\sigma_z$, even in a coupled Dirac cone system.
Consequently, the physically meaningful quantity in a coupled system is also the relative sign, $\mathrm{sgn}(v_{x}^{(1)}v_{x}^{(2)})$, or equivalently the relative chirality $\chi_{\rm rel}$ in Eq.~\eqref{eq:S5:chi_rel_def}.

\begin{figure}[t]
\centering
\includegraphics[width=\linewidth]{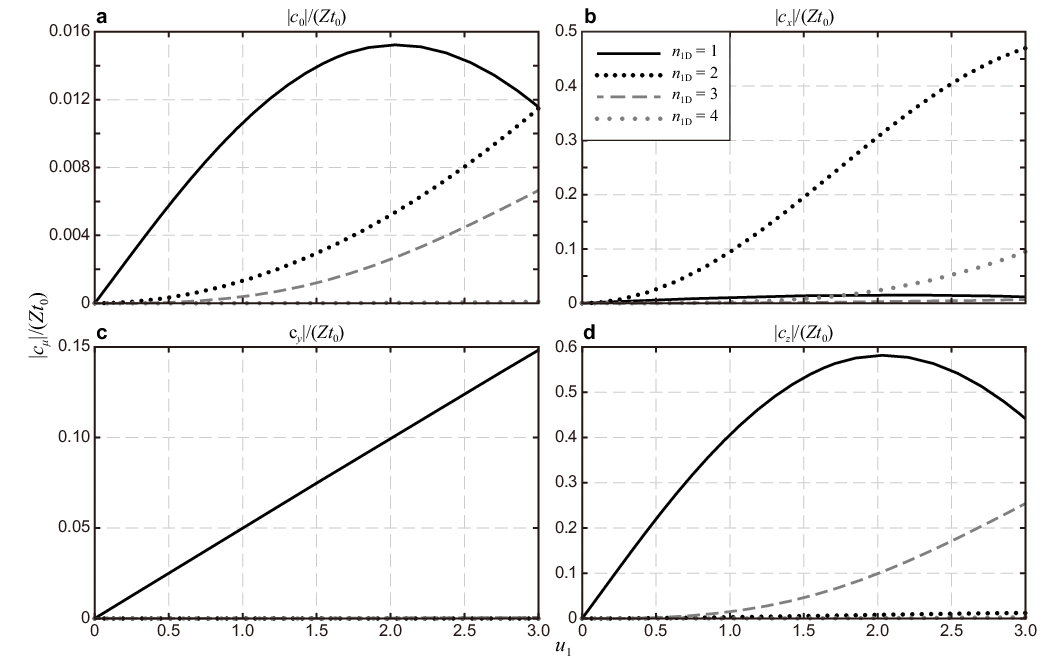}
\caption{Pauli channels \textbf{a} $|c_0|$, \textbf{b} $|c_x|$, \textbf{c} $|c_y|$, \textbf{d} $|c_z|$ (normalized by $Zt_0$) of the effective resonant inter-cone coupling $Z\mathcal M^{(n_{\rm 1D})}$ [Eq.~\eqref{eq:S5:M_Pauli_decomp}] in TBG--1D with $\theta=3^\circ$ plotted against $u_1$. Each line corresponds to $n_\mathrm{1D}=1,2,3,4$. The layer-dependent 1D modulation is taken to have a slightly asymmetric amplitude between layers, with $u_2 = 0.8u_1$, while keeping a common phase, $\phi_1 = \phi_2$.
In the standard two-cone Dirac basis of Eq.~\eqref{eq:S5:std_dirac_basis}, the gap-opening channel for $\chi_\mathrm{rel}=+1$ (same chirality), $c_z$ [Eq.~\eqref{eq:S:M_chirality_rule_theta0}], becomes substantial in TBG--1D when $n_\mathrm{1D}$ is odd, whereas the gap-opening channel for $\chi_\mathrm{rel}=-1$ (opposite chirality), $c_x$, becomes sizable when $n_\mathrm{1D}$ is even, resulting in the $n_\mathrm{1D}$-parity rule Eq.~\eqref{eq:S:M_parity_rule_theta0}. By contrast, the node-shifting channels $c_0$ and $c_y$ remain comparatively small.
}
\label{fig:pauli_channels_amplitude}
\end{figure}

We plot the magnitudes of the normalized Pauli components $|c_\mu|/(Z t_0)$ ($\mu=0,x,y,z$) of the inter-cone coupling $Z\mathcal{M}^{(n_\mathrm{1D})}$ [Eq.~\eqref{eq:S5:M_Pauli_decomp}] for TBG--1D with $\theta=3^\circ$ in 
Fig.~\ref{fig:pauli_channels_amplitude}.
% Extended Data Fig.~3.
Each line represents $n_{\rm 1D}=1$ to $4$, respectively. In this plot, we chose the layer-dependent 1D modulation with a slightly asymmetric amplitude between layers ($u_2=0.8u_1$), to show the contribution from $\mathcal{J}_{-,p}$ [$c_0$ for even $p$ and $c_y$ for odd $p$, Eq.~\eqref{eq:S:T0_evenodd_p}], while keeping a common phase, $\phi_1=\phi_2$.
In the same standard basis, the gap-opening channel for $\chi_\mathrm{rel}=+1$ (same chirality), $c_z$ [Eq.~\eqref{eq:S:M_chirality_rule_theta0}], becomes substantial in TBG--1D when $n_\mathrm{1D}$ is odd, whereas the gap-opening channel for $\chi_\mathrm{rel}=-1$ (opposite chirality), $c_x$, becomes sizable when $n_\mathrm{1D}$ is even, following the $n_\mathrm{1D}$-parity rule Eq.~\eqref{eq:S:M_parity_rule_theta0}. By contrast, the node-shifting channels $c_0$ and $c_y$ remain comparatively small.

\paragraph{(i) For $\chi_\mathrm{rel}=+1$ (same chirality):}
For the same Dirac-cone chirality, $\sigma_z$ is the gap-opening channel in the standard basis of Eq.~\eqref{eq:S5:std_dirac_basis} [Eq.~\eqref{eq:S:M_chirality_rule_theta0}].

In a TBG--1D with an odd $n_{\rm 1D}$, $Z\mathcal M^{(n_{\rm 1D})}\approx c_y\sigma_y+i c_z\sigma_z$:
as Fig.~\ref{fig:pauli_channels_amplitude}
% as Extended Data Fig.~3
shows, a gap-opening mass channel is generically present unless $c_z (\propto J_{p_0}(|u_+|/2) )=0$ accidentally.
Therefore, for $\chi_{\rm rel}=+1$ the odd-$n_{\rm 1D}$ resonant configurations
can open a local direct gap $\Delta_{\mathbf q=\mathbf 0}\approx 2|c_z|$ at $\mathbf q=\mathbf 0$
when $|c_z|\gtrsim |c_y|$ (see Appendix~\ref{sec:layer_symmetric} for the relative magnitude).

If $n_{\rm 1D}$ is even,
the coupling $Z \mathcal{M}^{(n_\mathrm{1D})}\in\mathrm{span}\{\sigma_0,\sigma_x\}$ contains no leading $\sigma_z$ mass channel and generically produces Dirac-node splitting and/or charge-neutral pockets rather than a direct gap.

\paragraph{(ii) For $\chi_\mathrm{rel}=-1$ (opposite chirality):}
For the opposite Dirac-cone chirality, the original standard-basis $\sigma_x$ component becomes the gap-opening channel after chirality alignment.
TBG--1D with an even $n_\mathrm{1D}$ has a sizable $\sigma_x$ channel 
(Fig.~\ref{fig:pauli_channels_amplitude}b),
% (Extended Data main-text Fig.~\ref{fig:prb_chirality}(c)),
thus can open a direct gap, while that with an odd $n_\mathrm{1D}$ cannot [Eq.~\eqref{eq:S:M_parity_rule_theta0}].
In other words, the intercone coupling $Z \mathcal{M}^{(n_\mathrm{1D})} =c_0\,\sigma_0 +c_x\,\sigma_x$ for even $n_\mathrm{1D}$ acquires an effective $\sigma_z$ component in the chirality-aligned basis, $c_0\,\sigma_y
+c_x\,\sigma_z$, and can therefore open a gap.

An opposite-valley pair ($\xi_1=-\xi_2$), by a short-wavelength unidirectional superlattice, also yields $\chi_{\rm rel}=-1$ even when $v_{x}^{(1)}v_{x}^{(2)}>0$, providing an alternative route to the opposite-chirality two-cone problem without tuning through a flattening point.

\begin{figure}[t]
\centering
\includegraphics[width=\linewidth]{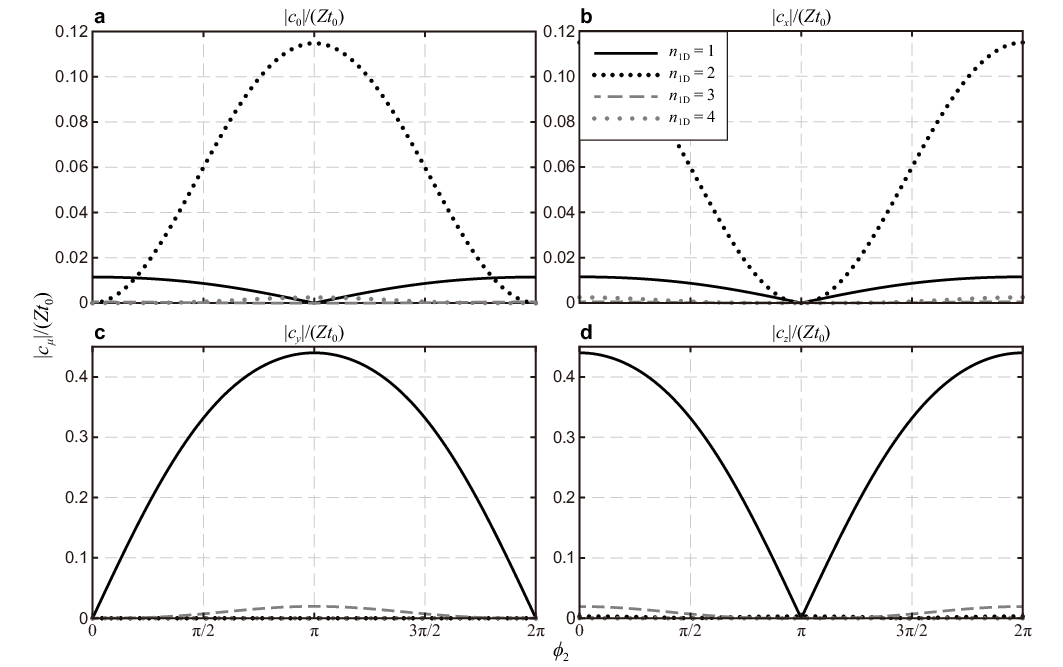}
\caption{Plot similar to Fig.~\ref{fig:pauli_channels_amplitude}, but plotted against $\phi_2$, while using $|u_1|=|u_2|=1$, $\phi_1=0$, $\theta=3^\circ$, to show the effects of a layer-asymmetric phase of the 1D modulation. The gap-opening channel for either chirality $\chi_\mathrm{rel}$ is maximized when the two modulations are in-phase ($\phi_2-\phi_1 \equiv 0$ (mod $2\pi$)) and vanishes when they are out-of-phase ($\phi_2-\phi_1 \equiv \pi$ (mod $2\pi$)).
}
\label{fig:pauli_channels_phase}
\end{figure}

\subsection{Rank structure of rigid interlayer tunneling and why Moir\'e--1D can open a Dirac gap}
\label{sec:SI_rank_tunneling}

In the rigid-lattice continuum model for pristine TBG, all interlayer tunneling matrices $T_j^\xi$ ($j=0,1,2$) are rank-one because
\begin{equation}
\det T_j^\xi = t_0^2\left(1-\omega^{-\xi j}\omega^{\xi j}\right)=0,
\end{equation}
and can be viewed as a projector in the sublattice space. Consequently, if a low-energy Dirac--Dirac resonance is governed by a single bare harmonic $T_j^\xi$---including cases where the Dirac-point mismatch is exactly compensated so that the two cones meet at the same reduced momentum, as in strongly coupled double-walled carbon nanotubes~\cite{PhysRevB.91.035405}---the $4\times4$ Hamiltonian at the Dirac point has the form
\begin{equation}
H(0)=
\begin{pmatrix}
0 & (T_j^\xi)^\dagger\\
T_j^\xi & 0
\end{pmatrix},
\end{equation}
whose spectrum contains two zero modes whenever $\mathrm{rank}(T_j^\xi)=1$. Therefore, a rank-one intercone coupling cannot open a full gap at the Dirac crossing; the spectrum remains nodal (Dirac or semi-Dirac depending on the relative chirality and detuning).

Lattice relaxation modifies the tunneling by distinguishing AA and AB/BA amplitudes. In the simplest parametrization,
\begin{equation}
\begin{aligned}
T_j^\xi
&=
\begin{pmatrix}
u_{\rm AA} & u_{\rm AB}\,\omega^{-\xi j}\\
u_{\rm AB}\,\omega^{\xi j} & u_{\rm AA}
\end{pmatrix},
\\
\det T_j^\xi &= u_{\rm AA}^2-u_{\rm AB}^2.
\end{aligned}
\end{equation}
Thus, the strict rank-one constraint is lifted when $u_{\rm AA}\neq u_{\rm AB}$. However, in the leading small-angle convention, the bare relaxed first-star tunneling at $\theta\to0$ still lies in $\mathrm{span}\{\sigma_0,\sigma_x,\sigma_y\}$ and therefore does not generate the leading same-chirality mass channel $\sigma_z$ identified in Eq.~\eqref{eq:S:M_chirality_rule_theta0}. At finite twist this should be understood as a leading-order statement rather than an exact prohibition: after untwisting the two layer Dirac Hamiltonians into a common standard basis, the $u_{\rm AA}\sigma_0$ sector can acquire a small $\sigma_z$ component of order $\mathcal O(u_{\rm AA}\theta)$, as discussed in Ref.~\cite{ScheerGuLian2022_PRB}. In pristine TBG, additional symmetries (notably $C_{2z} \mathcal T$ in a single-valley description) can still protect gapless Dirac points at charge neutrality, so relaxation alone does not generically produce a charge-neutral Dirac gap.

In the moir\'e--1D problem studied in this work, a unidirectional scalar potential 
dresses the tunneling by a layer-dependent unitary transformation $U_l(y)$, as
\begin{equation}
\begin{aligned}
\tilde T_j^\xi(y)&=U_2^\dagger(y)\,T_j^\xi\,U_1(y),
\\
\tilde T_{j,p}^\xi&=\frac{G_{\rm 1D}}{2\pi}\int_{0}^{2\pi/G_{\rm 1D}} \tilde T_j^\xi(y)
e^{-ipG_{\rm 1D}y}\, dy.
\end{aligned}
\end{equation}
While $\tilde T_j^\xi(y)$ preserves the instantaneous rank of $T_j^\xi$ at each real-space point, the Fourier component $\tilde T_{j,p}^\xi$ relevant for a Dirac--Dirac resonance is a coherent sum of inequivalent rank-one projectors whose orientation varies with $y$. As a result, $\tilde T_{j,p}^\xi$ is generically full rank. For example, for even $p$ and small twist angle one typically obtains
\begin{equation}
\begin{aligned}
\tilde T_{0,p}^\xi &\approx t_0\left(\mathcal J_{-,p}\,\sigma_0+\mathcal J_{+,p}\,\sigma_x\right),
\\
\mathcal J_{\pm,p}&=J_p(|u_\pm|/2)e^{ip\phi_\pm},
\end{aligned}
\end{equation}
so that
\begin{equation}
\det\tilde T_{0,p}^\xi \approx t_0^2\left(\mathcal J_{-,p}^2-\mathcal J_{+,p}^2\right),
\end{equation}
which is nonzero unless $|\mathcal{J}_{-,p}|=|\mathcal{J}_{+,p}|$ accidentally. Therefore, moir\'e--1D dressing provides a natural route to generate a full-rank intercone coupling, and, together with the appropriate relative chirality, enables a Dirac gap even when the bare rigid-lattice $T_j^\xi$ are rank-one.

\subsection{Layer-symmetric component of the 1D modulation and the resonant gap scale}
\label{sec:layer_symmetric}

For both chirality classes ($\chi_{\mathrm{rel}}=+1$, where the leading gapped resonances are odd $n_{\mathrm{1D}}$ with the $\sigma_z$ mass channel, and $\chi_{\mathrm{rel}}=-1$, where the leading gapped resonances are even $n_{\mathrm{1D}}$ with the $\sigma_x$ mass channel), the selected gap-opening channel is controlled by the layer-symmetric coefficient $\mathcal J_{+,p_0}$, whereas the leading node-shifting channel is controlled by the layer-antisymmetric coefficient $\mathcal J_{-,p_0}$. Using the convention fixed in Eq.~\eqref{eq:S4:upm_def}, the selected mass amplitude scales as
\begin{equation}
|c_\mu| \propto |\mathcal J_{+,p_0}|
=
\left|J_{|n_{\mathrm{1D}}|}\!\left(\frac{|u_+|}{2}\right)\right|, 
\label{eq:mass_bessel_scaling}
\end{equation}
where $p_0=-\xi n_{\rm 1D}$,
while the node-shifting term scales as
\begin{equation}
|\mathcal J_{-,p_0}|=
\left|J_{|n_{\rm 1D}|}\!\left(\frac{|u_-|}{2}\right)\right|.
\end{equation}

Accordingly, the magnitude of the resonant CNP gap is controlled by the layer-symmetric combination
\begin{equation}
\begin{aligned}
u_+ &\equiv u_1+u_2,
\\
|u_+|&=\sqrt{|u_1|^2+|u_2|^2+2|u_1||u_2|\cos(\phi_1-\phi_2)},
\end{aligned}
\label{eq:mass_vs_phase_uplus}
\end{equation}
whereas the leading node-shifting tendency is controlled by the layer-antisymmetric combination
\begin{equation}
u_- \equiv u_1-u_2.
\end{equation}
Figure~\ref{fig:pauli_channels_phase} shows that the selected mass channel is maximized and the node-shifting channel is minimized when the two layer modulations are in phase ($\phi_1=\phi_2$). Conversely, the selected mass channel is suppressed when the two modulations are out of phase ($\phi_1-\phi_2=\pi$). In the especially important case of equal in-phase layer potentials,
\begin{equation}
u_1=u_2\equiv u_l,
\qquad
\phi_1=\phi_2,
\end{equation}
one has
\begin{equation}
u_+=2u_l,
\qquad
\frac{|u_+|}{2}=|u_l|,
\end{equation}
and Eq.~\eqref{eq:mass_bessel_scaling} reduces to the compact main-text form
\begin{equation}
|m_{\rm gap}|\sim Z t_0 \left|J_{|n_{\rm 1D}|}(|u_l|)\right|.
\end{equation}
Thus the main-text expression is simply the equal in-phase limit of the general convention in Eq.~\eqref{eq:S4:upm_def}.

An equivalent viewpoint is that the resonant dressed tunneling component $\tilde T_{0,p_0}^\xi$ collects coherent contributions where the required momentum transfer $p_0\mathbf G_{\rm 1D}$ is generated before or after interlayer tunneling, or even by their convolution, through the dressing factors $U_1$ and $U_2^\dagger$. As a result, the relative gate phase $\phi_1-\phi_2$ controls constructive or destructive interference in the effective resonant coupling.

For the modulation strengths used in most of this work, $|u_+|/2$ lies on the first rising lobe of the relevant Bessel function, so increasing $|u_+|/2$ enhances the selected mass channel and hence the achievable $\Delta_{\mathrm{dir}}$ in the near-resonant gapped regime.

\subsection{Scaling of the resonant coupling with \texorpdfstring{$|n_{\rm 1D}|$}{n1D}}

For a general resonant index $n_{\rm 1D}$, the direct intercone coupling is controlled by the single harmonic
$p_0=-\xi n_{\rm 1D}$ of Eq.~\eqref{eq:S:T0_evenodd_p}.
For both even and odd $n_\mathrm{1D}$, the magnitude of $\mathcal{M}^{(n_\mathrm{1D})}$ varies with the order $|p_0|(=|n_\mathrm{1D}|)$ Bessel functions.
This implies the rapid suppression of the potential gap-opening term in $\mathcal{M}^{(n_\mathrm{1D})}$, $t_0|J_{|n_\mathrm{1D}|}(|u_+/2|)|$ in both (i) odd $n_\mathrm{1D}$ with $\chi_\mathrm{rel}=+1$ and (ii) even $n_\mathrm{1D}$ with $\chi_\mathrm{rel}=-1$, as $n_\mathrm{1D}$ increases,
$t_0\,|J_{|n_{\rm 1D}|}(|u_+|/2)|\approx t_0 (|u_+|/4 )^{|n_{\rm 1D}|} /(|n_{\rm 1D}|!)$
for weak modulation $|u_+|\ll1$ 
(Fig.~\ref{fig:pauli_channels_amplitude}).
% (Extended Data Fig.~3).

\subsection{Weak-1D limit (\texorpdfstring{$|u_l|\ll1$}{ul1}): expansions for velocities and \texorpdfstring{$\mathcal M$}{M}}

In the weak-modulation regime $|u_l|\ll1$, Bessel functions satisfy
\begin{equation}
\begin{aligned}
J_0(|u_l|)&\approx 1-\frac{|u_l|^2}{4}+\mathcal O(|u_l|^4),
\\
J_1\!\left(\frac{|u_\pm|}{2}\right) &\approx \frac{|u_\pm|}{4}+\mathcal O(|u_\pm|^3).
\end{aligned}
\label{eq:Y_Bessel_weak_1D}
\end{equation}
Using $\eta_l=J_0(|u_l|)$ and $Z=1/(1+6u_0^2)$, the renormalized velocities in Eq.~\eqref{eq:S5:ren_recipe} become
\begin{equation}
\begin{aligned}
v_{x}^{(l)} &\approx \frac{v_\mathrm{F}}{1+6u_0^2}
\left(1-3u_0^2-\frac{|u_l|^2}{4}\right),
\\
v_{y}^{(l)} &\approx \frac{v_\mathrm{F}}{1+6u_0^2}
\left(1-3u_0^2\right),
\end{aligned}
\label{eq:Y_vren_weak_1D}
\end{equation}
up to $\mathcal O(|u_l|^4)$ corrections.
For $n_{\rm 1D}=1$ (hence $p_0=-\xi$), Eq.~\eqref{eq:S:M_parity_rule_theta0} yields
\begin{multline}
Z\mathcal{M}^{(n_\mathrm{1D})}
\approx
-\xi\,\frac{Zt_0}{4}\Big[
|u_-|e^{-i\xi\phi_-}\sigma_y+i|u_+|e^{-i\xi\phi_+}\sigma_z
\Big]
\\ (|u_l|\ll1).
\label{eq:Y_M_weak_1D_leading_theta}
\end{multline}
Equations~\eqref{eq:Y_vren_weak_1D}--\eqref{eq:Y_M_weak_1D_leading_theta} provide compact checkable limits:
$v_{x}^{(l)}$ is reduced at $\mathcal O(|u_l|^2)$ through $J_0(|u_l|)$ 
(Fig.~\ref{fig:transverse_velocity_flip}), 
% (Extended Data Fig.~2), 
while the direct resonant hybridization scales linearly in $|u_\pm|$ through $J_1(|u_\pm|/2)$ 
(Figs.~\ref{fig:pauli_channels_amplitude}b and \ref{fig:pauli_channels_amplitude}d).
% (Extended Data Figs.~3b and 3d).

\subsection{Validity and limitations}

We summarize the controlled approximations used in Appendix~\ref{sec:min_two_cone_model} and their validity windows:

\noindent\textbf{Linearization in reduced momentum.} We expand near CNP to first order in $\mathbf q$ and neglect $\mathcal O(|\mathbf q|^2)$.

\noindent\textbf{Central-harmonic projection (intralayer).} Eliminating $m\neq0$ sectors in Eq.~\eqref{eq:S3-Hl-strip-matrix} produces corrections of order $\mathcal O(\hbar v_\mathrm{F} q_x^2/G_{\rm 1D})$, beyond the linear Dirac theory.
Within the $m=0$ sector, the unidirectional scalar potential is incorporated nonperturbatively through $\eta_l=J_0(|u_l|)$ in Eq.~\eqref{eq:S5:CH_intralayer}.

\noindent\textbf{Minimal-wave truncation (interlayer).} We retain only the two central cones and the single resonant harmonic $\mathcal M^{(n_{\rm 1D})}=\widetilde T_{0,p_0}^\xi$ with $p_0=-\xi n_{\rm 1D}$. For the generalized resonance condition Eq.~\eqref{eq:prb_umklapp_resonance}, $p_0$ is replaced by the selected integer harmonic $p$ after the resonant cone pair is chosen. All remaining moir\'e channels enter as a smooth $\mathcal O(u_0^2)$ self-energy, incorporated at leading order through $Z$ and the velocity renormalization recipe in Eqs.~\eqref{eq:S5:SigmaM_def}--\eqref{eq:S5:ren_recipe}.

\noindent\textbf{Caveat near flattening.} When $\eta_l\approx 3u_0^2$, the nominally subleading $\mathcal O(q_x^2/G_{\rm 1D})$ correction can compete with the linear term; the strictly linear description then requires $|q_x|\ll Z|\eta_l-3u_0^2|\,G_{\rm 1D}$.

\section{Near-resonant detuned two-cone theory}
\label{sec:near_resonance}
\label{sec:SI_mismatch}

\begin{figure*}[t]
\centering
\includegraphics[width=\linewidth]{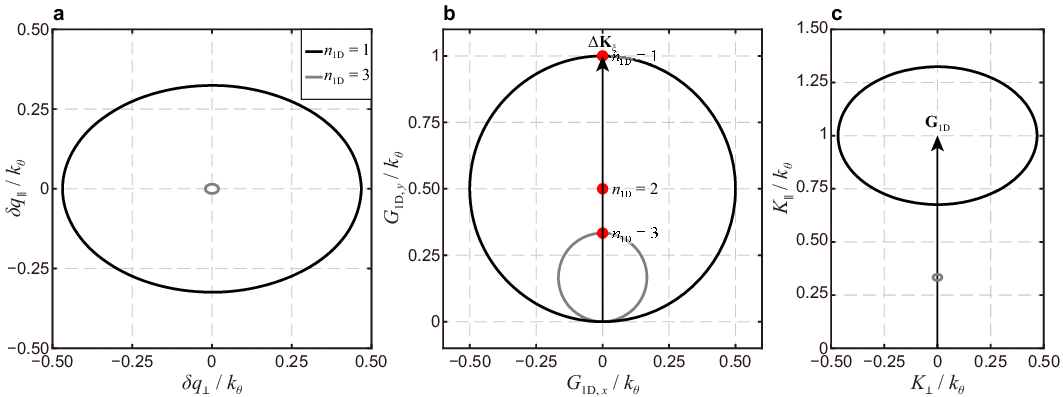}
\caption{
Tolerance in \textbf{a} the momentum-mismatch $\delta\mathbf{q}_\xi$ (Eq.~\eqref{eq:S5:ellipse_master}),
\textbf{b} $\mathbf{G}_\mathrm{1D}$ at fixed-$\Delta\mathbf{K}_\xi$ (Eq.~\eqref{eq:S5:arg_in_G1D}, main-text Fig.~\ref{fig:prb_window}),
\textbf{c} $\Delta\mathbf{K}_\xi$ at fixed-$\mathbf{G}_\mathrm{1D}$ (Eq.~\eqref{eq:S5:filled_ellipse_DK})
for opening a CNP gap in near-resonant configurations of TBG--1D with $\theta=3^\circ$ and $u_1=u_2=1$,
axes are plotted in units of $k_\theta=|\Delta\mathbf{K}_\xi|$.
The black (gray) ellipses in panels \textbf{a} and \textbf{c} bound the gapped regions for $n_{\rm 1D}=1$ ($3$). In panel \textbf{b}, the black (gray) circle denotes the corresponding
$\delta q_\parallel=0$ ridge; the finite-width gapped region occupies only an allowed segment around this ridge.
The red dots in panel \textbf{b} mark $\mathbf G_{\rm 1D}/k_\theta=(0,1)$, $(0,1/2)$, and $(0,1/3)$, corresponding to the $n_{\rm 1D}=1,2,3$ resonant configurations.
A gap opens within the ellipse in \textbf{a} and \textbf{c}, while it opens along the circle (arc) in \textbf{b} due to the nonlinear mapping between $\delta\mathbf{q}_\xi$ and $\mathbf{G}_\mathrm{1D}$.
}
\label{fig:near_resonance_tolerance}
\end{figure*}

Exact Dirac--Dirac momentum resonance requires $\Delta\mathbf K_\xi=\xi n_\mathrm{1D}\mathbf G_{\rm 1D}$ [Eq.~\eqref{eq:prb_resonance}],
which is a measure-zero condition in the two-dimensional design space of $\mathbf G_{\rm 1D}$.
In realistic devices, $\mathbf G_{\rm 1D}$ typically lies in a near-resonant neighborhood of this point but does not coincide with it.
Here we show that the renormalized two-cone theory developed for the resonant case [Eq.~\eqref{eq:S5:two_cone_model}]
naturally extends to this near-resonant regime upon introducing a small momentum-mismatch vector.
The resulting analytic criterion explains why, in a fixed-$\Delta\mathbf K_\xi$ configuration map that scans $\mathbf G_{\rm 1D}$ 
(Fig.~\ref{fig:near_resonance_tolerance}b),
% (Extended Data Fig.~5b),
the gapped region appears as a thick arc whose ridge is not centered at the exact resonant point $\mathbf{G}_\mathrm{1D}=\xi\Delta\mathbf{K}_\xi/n_\mathrm{1D}$.

This appendix treats detuning around the nearest-cone design branch. For a generalized moir\'e-Umklapp branch, the natural local detuning variable would be
\begin{equation}
\delta\mathbf q_{j,\mathbf G_{\rm M},p}^{\xi}
\equiv
\mathbf q_j^\xi+\mathbf G_{\rm M}+p\mathbf G_{\rm 1D},
\label{eq:S_umklapp_general_detuning}
\end{equation}
but the corresponding global design map depends on the selected tuple $(j,\mathbf G_{\rm M},p)$ and is not needed for the quantitative results presented here.

Throughout we focus on the same-chirality regime $\chi_{\rm rel}=+1$ (Appendix~\ref{sec:min_two_cone_model}),
where the $\sigma_z$ component of the intercone coupling provides the leading mass (gap-opening) channel.
All parameters below include the leading remote-moir\'e self-energy through the quasiparticle weight $Z$
and the renormalized velocities $v_{x/y}^{(l)}$ [Eq.~\eqref{eq:S5:ren_recipe}].

\subsection{Momentum mismatch and momentum-mismatched two-cone Hamiltonian}

\paragraph{Momentum mismatch.}
For near-resonant configurations we use the momentum-mismatch vector defined in Eq.~\eqref{eq:S1:mm_def},
\begin{equation}
\delta\mathbf q_\xi \equiv 
\Delta\mathbf K_\xi-\xi n_\mathrm{1D} \mathbf G_{\rm 1D}.
\label{eq:S5:dq_def}
\end{equation}
Let $\hat{\mathbf n}\equiv \mathbf G_{\rm 1D}/G_{\rm 1D}$ be the 1D modulation direction and
$\hat{\mathbf n}_\perp$ its perpendicular unit vector,
\begin{equation}
\hat{\mathbf n}=(n_x,n_y),\qquad
\hat{\mathbf n}_\perp\equiv(-n_y,n_x).
\label{eq:S5:n_def}
\end{equation}
We decompose the mismatch as
\begin{equation}
\begin{aligned}
\delta q_\parallel &\equiv \delta\mathbf q_\xi\cdot\hat{\mathbf n},
\\
\delta q_\perp &\equiv \delta\mathbf q_\xi\cdot\hat{\mathbf n}_\perp,
\\
\delta\mathbf q_\xi &= \delta q_\perp\,\hat{\mathbf n}_\perp + \delta q_\parallel\,\hat{\mathbf n}.
\end{aligned}
\label{eq:S5:dq_components}
\end{equation}

\paragraph{Coordinates adapted to the 1D modulation.}
To avoid mixing different coordinate conventions, in this section we express the two-cone model in the
$(\hat{\mathbf n}_\perp,\hat{\mathbf n})$-adapted components of momentum:
\begin{equation}
q_\perp \equiv \mathbf q\cdot\hat{\mathbf n}_\perp,
\qquad
q_\parallel \equiv \mathbf q\cdot\hat{\mathbf n}.
\label{eq:S5:q_components}
\end{equation}
Equivalently, if $\mathbf q=(q_x,q_y)$ in the fixed lab-frame $(\hat{\mathbf x},\hat{\mathbf y})$,
$q_\perp=-n_y q_x+n_x q_y$, $q_\parallel=n_x q_x+n_y q_y$, 
and the same relations hold for $\delta\mathbf q_\xi$.

\paragraph{Local Pauli matrices.}
We also define Pauli matrices associated with $(\hat{\mathbf n},\hat{\mathbf n}_\perp)$ in each layer,
\begin{equation}
\begin{aligned}
\sigma_\perp^{(l)} &\equiv -n_y \sigma_x^{(l)}+n_x \sigma_y^{(l)},
\\
\sigma_\parallel^{(l)} &\equiv n_x \sigma_x^{(l)}+n_y \sigma_y^{(l)}.
\end{aligned}
\label{eq:S5:sigma_local}
\end{equation}
They satisfy the same Pauli algebra and reduce to the standard form used in Eq.~\eqref{eq:S5:two_cone_model}
at the commensurate resonance of the symmetric-rotation geometry, where $\hat{\mathbf n}=\hat{\mathbf y}$.
(In that case $q_\parallel=q_y$, $\sigma_\parallel^{(l)}=\sigma_y^{(l)}$, and $q_\perp\sigma_\perp^{(l)}=q_x\sigma_x^{(l)}$.)

\paragraph{Momentum-mismatched two-cone Hamiltonian.}
At exact resonance ($\delta\mathbf q_\xi=\mathbf 0$), the low-energy spectrum near charge neutrality is captured by
the renormalized two-cone model $H_\xi^{(\mathrm{2c})}(\mathbf q)$ in Eq.~\eqref{eq:S5:two_cone_model}.
In the near-resonant regime, the same resonant harmonic still provides the dominant intercone hybridization,
but it transfers the residual momentum $\delta\mathbf q_\xi$.
It is therefore convenient to represent the two cones as displaced by $\pm\delta\mathbf q_\xi/2$, giving
\begin{multline}
H_{\xi}^{(\mathrm{2c})}(\mathbf q;\delta\mathbf q_\xi)\\=
\begin{pmatrix}
h_{\xi}^{(1)}\!\left(\mathbf q;- \delta\mathbf q_\xi/2 \right) & \big(Z\mathcal M^{(n_\mathrm{1D})}\big)^\dagger \\
Z\mathcal M^{(n_\mathrm{1D})} & h_{\xi}^{(2)}\!\left(\mathbf q;+ \delta\mathbf q_\xi/2\right)
\end{pmatrix},
\label{eq:S5:H2c_detuned}
\end{multline}
with the anisotropic Dirac blocks
\begin{equation}
\begin{aligned}
&h_{\xi}^{(l)}(\mathbf q;\delta \mathbf{q}_\xi)
\\
&=
-\hbar\big(\xi v_{x}^{(l)}\,(q_\perp+\delta q_\perp)\,\sigma_\perp^{(l)}+v_{y}^{(l)}\,(q_\parallel+\delta q_\parallel)\,\sigma_\parallel^{(l)}\big).
\end{aligned}
\label{eq:S5:h_eff_local}
\end{equation}
Here $(v_\perp^{(l)},v_\parallel^{(l)})$ are given by Eq.~\eqref{eq:S5:ren_recipe} (transverse/longitudinal velocities with respect to the 1D modulation direction),
and $Z\mathcal M^{(n_\mathrm{1D})}$ is decomposed into Pauli channels as in Eq.~\eqref{eq:S5:M_Pauli_decomp}.
Equation~\eqref{eq:S5:H2c_detuned} 
shows that the momentum mismatch $\delta\mathbf q_\xi$ acts as an axial gauge-like vector potential that shifts the two interacting Dirac cones in momentum space.
Equation~\eqref{eq:S5:H2c_detuned} reduces to the commensurate two-cone model when $\delta\mathbf q_\xi=\mathbf 0$.

\begin{figure*}[t]
\centering
\includegraphics[width=\linewidth]{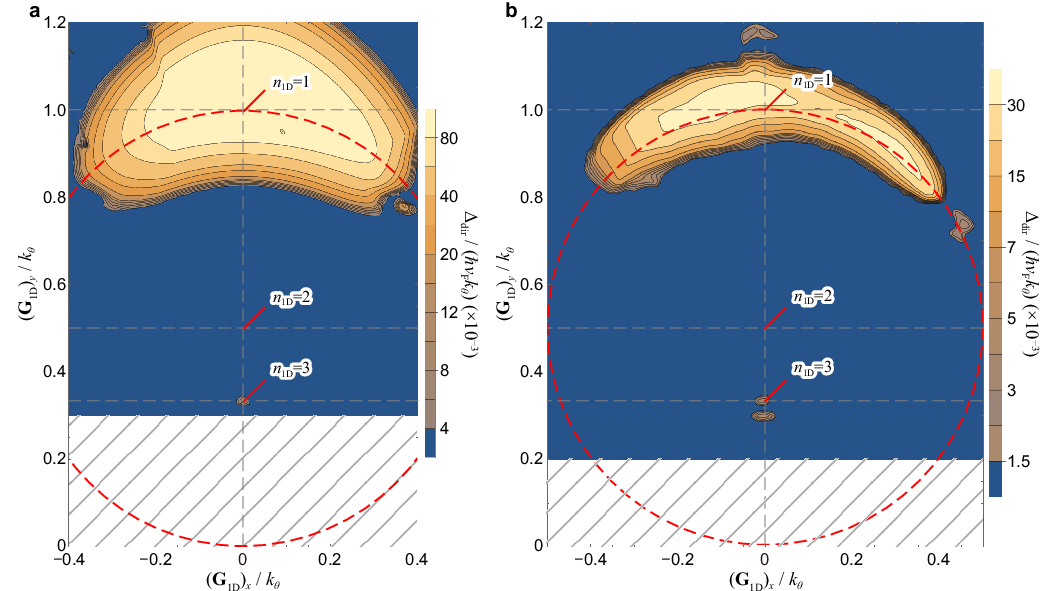}
\caption{
Full-wave continuum maps of the charge-neutrality direct gap $\Delta_\mathrm{dir}$ (in units of $\hbar v_\mathrm{F} k_\theta$, log-scale) versus $\mathbf G_{\rm 1D}$ for a layer-symmetric (in-phase) modulation ($|u_1|=|u_2|=1$ and $\phi_1=\phi_2=0$), shown for \textbf{a} $\theta=3^\circ$ and \textbf{b} $5^\circ$. These maps are analogous to the near-resonant direct-gap map highlighted in main-text Fig.~\ref{fig:prb_window}. The hatched region lies outside the numerical scan. Resonance markers indicate Dirac--Dirac matching for $n_{\rm 1D}=1,2,3$, while the red-dashed ridge circle highlights the near-resonant gapped arc around the $n_{\rm 1D}=1$ resonance. This map corresponds to the default chirality class ($\chi_{\rm rel}=+1$) of the layer-symmetric slice.
}
\label{fig:near_resonance_gap_maps}
\end{figure*}

\subsection{Gapped region in momentum-mismatch space \texorpdfstring{$\delta\mathbf{q}_\xi$}{deltaq}: filled ellipse}

\paragraph{Mismatch energy scale.}
Define the effective velocities
\begin{equation}
v_\perp^{\rm eff}\equiv \sqrt{v_\perp^{(1)}v_\perp^{(2)}},\qquad
v_\parallel^{\rm eff}\equiv \sqrt{v_\parallel^{(1)}v_\parallel^{(2)}},
\label{eq:S5:veff_def}
\end{equation}
and the corresponding mismatch energy scale
\begin{equation}
\varepsilon_{\rm mm}(\delta\mathbf q_\xi)\equiv
\frac{\hbar}{2}\sqrt{\big(v_\perp^{\rm eff}\delta q_\perp\big)^2+
\big(v_\parallel^{\rm eff}\delta q_\parallel\big)^2 }.
\label{eq:S5:emm_def}
\end{equation}

\paragraph{Gap estimate and ellipse inequality.}
When $\chi_\mathrm{rel}=+1$ and $c_z \sigma_z$ is the leading mass channel, diagonalizing Eq.~\eqref{eq:S5:H2c_detuned} gives the estimate
\begin{equation}
\Delta_{\rm dir}^{(\mathrm{2c})}(\delta\mathbf q_\xi)
 \approx
2 \max \bigl( |c_z|-\varepsilon_{\rm mm}(\delta\mathbf q_\xi),0\big),
\label{eq:S5:gap_estimate}
\end{equation}
which is exact when $v_\perp^{(1)}=v_\perp^{(2)}$ and $v_\parallel^{(1)}=v_\parallel^{(2)}$ and $c_x=c_y=0$.
A necessary condition for a finite gap in Eq.~\eqref{eq:S5:gap_estimate} is
\begin{equation}
\frac{\delta q_\perp^2}{Q_\perp^2} + \frac{\delta q_\parallel^2}{Q_\parallel^2} 
< 1,
\quad
Q_\perp\equiv \frac{2|c_z|}{\hbar v_\perp^{\rm eff}},
\quad
Q_\parallel\equiv \frac{2|c_z|}{\hbar v_\parallel^{\rm eff}}.
\label{eq:S5:ellipse_master}
\end{equation}
% \begin{equation}
% \big(v_x^{\rm eff}\delta q_\perp\big)^2+\big(v_y^{\rm eff}\delta q_\parallel\big)^2
% <
% \Bigl(\frac{2|c_z|}{\hbar}\Bigr)^2,
% \end{equation}
i.e., the gapped region in the mismatch coordinates $(\delta q_\perp,\delta q_\parallel)$ is a filled ellipse centered at $\delta\mathbf q_\xi=\mathbf 0$, as 
Fig.~\ref{fig:near_resonance_tolerance}a shows.
% Extended Data Fig.~5a shows.

\subsection{Gapped region in (fixed-\texorpdfstring{$\Delta\mathbf K_\xi$}{DeltaK}) \texorpdfstring{$\mathbf G_{\rm 1D}$}{G1D} space: arc}

The ellipse criterion Eq.~\eqref{eq:S5:ellipse_master} exhibits an arc morphology when scanning $\mathbf G_{\rm 1D}$ at fixed $\Delta\mathbf K_\xi$,
because the mapping from $\mathbf G_{\rm 1D}$ to $(\delta q_\perp,\delta q_\parallel)$ is nonlinear.
For the symmetric layer rotation used throughout (Sec.~\ref{sec:reference_graphene_lattice}),
$\Delta\mathbf K_\xi = k_\theta(0,\xi)^{\mathrm T}$ with $k_\theta\equiv |\Delta\mathbf K_\xi|$.
We introduce the dimensionless scan variable
$\mathbf g\equiv n_\mathrm{1D} \mathbf G_{\rm 1D}/k_\theta=(g_x,g_y)$ and $g\equiv|\mathbf g|=\sqrt{g_x^2+g_y^2}$.
Using Eqs.~\eqref{eq:S5:dq_def}--\eqref{eq:S5:dq_components} one finds
\begin{equation}
\delta q_\perp=\xi k_\theta\Big(\frac{g_x}{g}\Big),
\qquad
\delta q_\parallel=\xi k_\theta\Big(\frac{g_y}{g}-g\Big).
\label{eq:S5:dq_in_g}
\end{equation}
Substituting Eq.~\eqref{eq:S5:dq_in_g} into the ellipse inequality Eq.~\eqref{eq:S5:ellipse_master} gives the compact arc criterion
\begin{equation}
\Big(\frac{g_x/g}{Q_\perp}\Big)^2+\Big(\frac{g_y/g-g}{Q_\parallel}\Big)^2<\frac{1}{k_\theta^2}.
\label{eq:S5:arc_ineq_compact}
\end{equation}
% where $\beta\equiv v_x^{\rm eff}/v_y^{\rm eff}$, $\gamma\equiv 2|c_z|/(\hbar v_y^{\rm eff}k_\theta)$.

\paragraph{Ridge and ``off-centered'' arc.}
For a fixed direction $\hat{\mathbf n}$, changing the magnitude $G_{\rm 1D}$ tunes only the parallel mismatch $\delta q_\parallel$
while $\delta q_\perp=\Delta\mathbf K_\xi\cdot\hat{\mathbf n}_\perp$ is independent of $G_{\rm 1D}$.
Equivalently, at fixed $\hat{\mathbf n}$ the mismatch scale Eq.~\eqref{eq:S5:emm_def} is minimized with respect to $G_{\rm 1D}$ at $\delta q_\parallel=0$.
This defines the ridge (centerline) of the gapped region, along $\hat{\mathbf{n}}$,
\begin{equation}
\frac{g_y}{g}-g=0
\ \Longleftrightarrow \
g_x^2+\Big(g_y-\frac12\Big)^2=\Big(\frac12\Big)^2,
\label{eq:S5:circle_ridge}
\end{equation}
a circle centered at $(0,1/2)$ with radius $1/2$ in the $\mathbf{g}$ coordinates. 
In the $\mathbf{G}_\mathrm{1D}=((\mathbf{G}_{\mathrm{1D}})_x,(\mathbf{G}_{\mathrm{1D}})_y)$ coordinates, this corresponds to
\begin{equation}
    (\mathbf{G}_{\mathrm{1D}})_x^2+\Bigl((\mathbf{G}_{\mathrm{1D}})_y-\frac{k_\theta}{2n_\mathrm{1D}}\Bigr)^2 = \Bigl( \frac{k_\theta}{2n_\mathrm{1D}}\Bigr)^2,
\label{eq:S5:arg_in_G1D}
\end{equation}
a circle centered at $(0,k_\theta/(2n_\mathrm{1D}))$ with radius $k_\theta/(2n_\mathrm{1D})$, 
as Fig.~\ref{fig:near_resonance_tolerance}b shows.
% as Extended Data Fig.~5b shows.
The exact-resonance point $((\mathbf{G}_{\mathrm{1D}})_x,(\mathbf{G}_{\mathrm{1D}})_y)=(0,k_\theta/n_\mathrm{1D})$ lies on this circle but is not its center, explaining why the gapped region in a fixed-$\Delta\mathbf K_\xi$ map is generically off-centered.
Equation~\eqref{eq:S5:arg_in_G1D} for $n_{\rm 1D}=1$ reproduces the ridge of the near-resonant arc in main-text Fig.~\ref{fig:prb_window} and Fig.~\ref{fig:near_resonance_gap_maps}: the circle passes through the $n_{\rm 1D}=1$ resonant point and is centered at the $n_{\rm 1D}=2$ point.

The finite width around the ridge is set by Eq.~\eqref{eq:S5:arc_ineq_compact}; moreover, even on the ridge a gap requires the transverse condition
$|v_x^{\rm eff}\delta q_\perp|<2|c_z|/\hbar$, so only a segment of the ridge circle appears as a bright arc.

% ------------------------------------------------------------
\subsection{Complementary representation: gapped region in (fixed \texorpdfstring{$\mathbf G_{\rm 1D}$}{G1D}) \texorpdfstring{$\Delta\mathbf K_\xi$}{DeltaK} space: filled ellipse}

If, instead, we scan $\Delta\mathbf{K}_\xi$ while holding $\mathbf G_{\rm 1D}$ fixed, the mapping $\Delta\mathbf K_\xi \mapsto (\delta q_\parallel,\delta q_\perp)$ is linear.
Writing $\Delta\mathbf K_\xi = K_\parallel\hat{\mathbf n} + K_\perp\hat{\mathbf n}_\perp$, Eq.~\eqref{eq:S5:dq_components} gives
$\delta q_\parallel=K_\parallel-\xi n_\mathrm{1D} G_{\rm 1D}$ and $\delta q_\perp=K_\perp$.
Then Eq.~\eqref{eq:S5:ellipse_master} becomes the filled ellipse
\begin{equation}
\frac{K_\perp^2}{Q_\perp^2}
+
\frac{(K_\parallel-\xi\,n_{\rm 1D}G_{\rm 1D})^2}{Q_\parallel^2}
<1.
\label{eq:S5:filled_ellipse_DK}
\end{equation}
Thus the filled-ellipse picture is recovered when the scan variables are chosen to be the mismatch components themselves as 
Fig.~\ref{fig:near_resonance_tolerance}c shows.
% Extended Data Fig.~5c shows.

\subsection{Beyond the near-resonant, gapped region}
\label{sec:beyond_the_near_resonant}

Solving Eq.~\eqref{eq:S5:H2c_detuned} at charge neutrality $(E=0)$ for a given momentum mismatch $\delta\mathbf q_\xi$ yields the locations of the band-touching (Dirac) nodes,
\begin{equation}
\mathbf q_\pm=\pm \frac{\delta\mathbf q_\xi}{2}\sqrt{1-\left(\frac{|c_z|}{\varepsilon_\mathrm{mm}(\delta\mathbf q_\xi)}\right)^2},
\end{equation}
provided that real solutions exist. When $|c_z|>\varepsilon_\mathrm{mm}(\delta\mathbf q_\xi)$, no real $\mathbf q_\pm$ exists, indicating that the two nodes annihilate and a gap opens. The critical case $|c_z|=\varepsilon_\mathrm{mm} (\delta\mathbf q_\xi)$ corresponds to node merging at a single momentum point and therefore delineates the boundary of the gapped region in the configuration map. For $|c_z|<\varepsilon_\mathrm{mm}(\delta\mathbf q_\xi)$, two distinct Dirac nodes survive; as $|c_z|/\varepsilon_\mathrm{mm}\to0$, they approach the uncoupled Dirac points at $\mathbf q=\pm \delta\mathbf q_\xi/2$, yielding two separated Dirac cones. This is the off-resonant regime (Sec.~\ref{sec:off_resonance}).

% ------------------------------------------------------------
\subsection{Practical notes}

\paragraph{Fixing $u_l$ vs.\ fixing $v_l$ in a $|\mathbf G_{\rm 1D}|$ scan.}
Because $u_l=v_l/(\hbar v_{\mathrm F}G_{\rm 1D})$ [Eq.~\eqref{eq:prb_dimless}], scanning $|\mathbf G_{\rm 1D}|$ changes $|u_l|$ if $v_l$ is held fixed.
To isolate the geometric momentum-mismatch effect in Eq.~\eqref{eq:S5:arc_ineq_compact},
it is therefore natural to hold $u_l$ fixed (so that $c_z$ and $v_{x/y}^{(l)}$ are independent of $|\mathbf G_{\rm 1D}|$ at this level).
Holding $v_l$ fixed instead promotes $Q_\parallel$ and $Q_\perp$ to slowly varying functions of $G_{\rm 1D}$ through the Bessel factors and $\eta_l$.

\paragraph{Closed-form estimates for arc widths.}
Using Eq.~\eqref{eq:S5:ren_recipe}, 
$v_\perp^{\rm eff}=Zv_{\mathrm F}\sqrt{(\eta_1-3u_0^2)(\eta_2-3u_0^2)}$, $v_\parallel^{\rm eff}=Zv_{\mathrm F}(1-3u_0^2)$, and $u_0\equiv t_0/(\hbar v_{\mathrm F}k_\theta)$.
For odd $n_{\rm 1D}$ in the same-chirality regime, Eq.~\eqref{eq:S:M_parity_rule_theta0} implies $|c_z|\simeq Z t_0|\mathcal J_{+,-\xi n_\mathrm{1D}}|$
with $\mathcal J_{+,p}$ defined in Eq.~\eqref{eq:S4:upm_def}.

A simple geometric readout of Eq.~\eqref{eq:S5:arc_ineq_compact} is obtained from cuts of the $(g_x,g_y)$ map:
for $g_x=0$ and $g_y>0$, one has $|1-g_y|< 2u_0 J_{n_\mathrm{1D}}(|u_+|/2)/(1-3u_0^2)$ (vertical thickness $4u_0 J_{n_\mathrm{1D}}(|u_+|/2)/(1-3u_0^2)$ around the exact-resonance point),
while near $g_y\approx 1$ and $|g_x|\ll 1$ the horizontal extent scales as $|g_x|_{\max}\sim 2u_0 J_{n_\mathrm{1D}}(|u_+|/2)/\sqrt{(\eta_1-3u_0^2)(\eta_2-3u_0^2)}$.

\paragraph{Beyond the two-cone criterion.}
Equation~\eqref{eq:S5:ellipse_master} describes the near-resonant gap mechanism within the renormalized two-cone truncation.
In the full continuum model (Sec.~\ref{sec:full_wave_model}), additional minibands and indirect overlaps can further reduce the global gap,
where we use Eq.~\eqref{eq:S2:Delta_true_def} to diagnose direct band touchings.

\section{Berry curvature and Chern triviality}
\label{sec:chern_appendix}

This appendix defines the Berry-curvature diagnostics used in this work and summarizes the symmetry reason why the charge-neutrality gaps reported here are Chern-trivial.
All Berry-curvature maps and Chern numbers are obtained numerically from the fully periodic single-valley continuum Hamiltonian (Appendix~\ref{sec:full_wave_model}) using a gauge-invariant discretization.

\subsection{\texorpdfstring{$C_{2z}T$}{C2zT} symmetry and Chern-triviality in scalar TBG--1D}

We fix the real-space origin at an AA region of the moir\'e pattern, which is a $C_{2z}$ center of the continuum model.
For a scalar 1D potential of the form
$V_l(\mathbf r)=(|v_l|/2)\cos(\mathbf G_{\rm 1D}\cdot\mathbf r+\phi_l)$,
the potential preserves $C_{2z}$ about the AA origin only when it is even under $\mathbf r\to-\mathbf r$, i.e., when $\phi_l=0$ or $\pi$ (equivalently, when $v_l$ is real and the modulation is cosine-type about the chosen origin).
In this cosine registry, the single-valley Hamiltonian inherits the antiunitary symmetry $C_{2z}T$ within each valley.
Because $C_{2z}T$ leaves the rBZ momentum invariant, it forces the Berry curvature to vanish pointwise in the rBZ and therefore guarantees that the valley Chern number of any isolated set of minibands is zero \cite{po2018origin,Zou2018_PRB,Zhang2019_PRB}.

Starting from this $C_{2z}T$-symmetric limit, we continuously vary the parameters $(\theta,|u_l|,\phi_l)$ explored in this work.
We find that whenever a direct charge-neutrality gap persists across the rBZ, the valley Chern number of the occupied subspace remains $C_\xi=0$.
This is consistent with adiabatic continuity: in the absence of a gap closing, $C_\xi$ cannot change from its $C_{2z}T$-symmetric value.
When the registry phases $\phi_l$ break $C_{2z}T$, the Berry curvature can become finite; however, in the gapped configurations studied here it integrates to zero, so the gap remains Chern-trivial.

% ------------------------------------------------------------
\subsection{Non-Abelian Fukui curvature and valley Chern number}

For a fixed valley $\xi=\pm1$, we diagonalize the periodic continuum Hamiltonian on a uniform mesh $\{\mathbf q\}$ in the rBZ.
Let $\{|u_a(\mathbf q)\rangle\}_{a=1}^{N_{\rm occ}}$ be an orthonormal basis of the occupied subspace at $\mathbf q$.
Following Fukui--Hatsugai--Suzuki \cite{Fukui2005_JPSJ}, we define the overlap (link) matrices along the two mesh directions $i=1,2$,
\begin{equation}
\begin{aligned}
\big[W_i(\mathbf q)\big]_{ab}&\equiv \langle u_a(\mathbf q)\,|\,u_b(\mathbf q+\Delta\mathbf q_i)\rangle,
\\
U_i(\mathbf q)&\equiv \frac{\det W_i(\mathbf q)}{|\det W_i(\mathbf q)|}.
\end{aligned}
\label{eq:S8:link_def}
\end{equation}
The lattice field strength on each plaquette is
\begin{equation}
\begin{aligned}
F_{12}(\mathbf q) 
\equiv 
\mathrm{Arg} \big[
&U_1(\mathbf q)
U_2(\mathbf q+\Delta\mathbf q_1) \\
&U_1(\mathbf q+\Delta\mathbf q_2)^{-1}
U_2(\mathbf q)^{-1}
\big],
\end{aligned}
\label{eq:S8:F12_def}
\end{equation}
where $\mathrm{Arg}$ denotes the principal branch.
The valley Chern number of the occupied subspace is then
\begin{equation}
C_\xi=\frac{1}{2\pi}\sum_{\mathbf q\in{\rm mesh}} F_{12}(\mathbf q),
\label{eq:S8:Chern_fukui}
\end{equation}
which is integer-valued when the occupied subspace is separated from the unoccupied subspace by a nonzero direct gap throughout the rBZ.
% For Berry-curvature maps we plot either the plaquette quantity $F_{12}(\mathbf q)$ itself or the associated density $\Omega_{\rm occ}(\mathbf q)\equiv F_{12}(\mathbf q)/A_{\rm plaq}$, where $A_{\rm plaq}$ is the plaquette area.
For an isolated single band, the Abelian specialization is obtained by setting $N_{\rm occ}=1$ in Eqs.~\eqref{eq:S8:link_def}--\eqref{eq:S8:F12_def}.

\begin{figure*}[t]
\centering
\includegraphics[width=\linewidth]{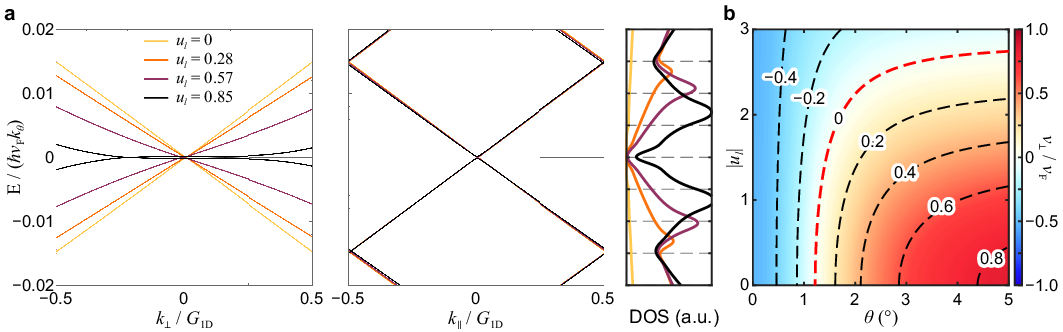}
\caption{
\textbf{Anisotropic band renormalization at off-resonance.}
\textbf{a} Off-resonant band structures ($L_\mathrm{1D}=50~\mathrm{nm}$, $\theta=2^\circ$), along the directions perpendicular (left) and parallel (middle) to $\mathbf G_\mathrm{1D}$, and DOS (right) near the CNP as a function of modulation strength $u_l$, illustrating continuously tunable anisotropy and progressive suppression of the transverse Dirac velocity in the Dirac-point neighborhood without opening a gap.
(b) Map of the renormalized transverse velocity $v_\perp^*/v_\mathrm{F}$ versus twist angle $\theta$ and modulation strength $|u_l|$, highlighting the flattening line (red dashed line) where the low-energy transverse velocity approaches zero (Eq.~\eqref{eq:S3:vstar_incomm}).
Each dashed curve is a contour along which ($v_\perp^*/v_{\rm F}$) equals the value indicated on the curve.
The $|u_l|$ required for band flattening in TBG--1D is smaller than in monolayer graphene under a unidirectional potential ($|u_l|\sim 2.83$) because of the contribution from the moir\'{e} tunneling [Eq.~\eqref{eq:S3:vstar_incomm}].
}
\label{fig:fig_s_i}
\end{figure*}

\section{Off-resonant anisotropic Dirac regime}
\label{sec:off_resonance}

In a generic off-resonant TBG--1D configuration, the Dirac-point mismatch $\Delta\mathbf K_\xi$ cannot be bridged by any integer harmonic of the 1D modulation at low energy, so there is no direct resonant Dirac--Dirac hybridization between the two layers. Consequently, the low-energy spectrum near charge neutrality remains gapless. 
%
% The leading effect of the hybrid moir\'e--1D structure is instead an anisotropic renormalization of the Dirac velocities: in the regime $|E|,\hbar v_\mathrm{F}|\mathbf q|\ll \mathcal E_\theta,\mathcal E_{\rm 1D}$ near the Dirac point, the velocity perpendicular to the modulation can be suppressed beyond the pristine TBG value and tuned to zero by an externally controllable potential. This distinct off-resonant effect shares the same transverse-velocity renormalization that underlies chirality switching in the resonant problem, but it does not rely on a selected resonant mass channel.
The leading effect of the hybrid moir\'e--1D structure is instead an anisotropic renormalization of the Dirac velocities:
Figure~\ref{fig:fig_s_i}a shows that, near the charge-neutrality Dirac point, the dispersion becomes increasingly anisotropic as the modulation strength is increased, 
with the velocity perpendicular to the modulation strongly suppressed while the longitudinal velocity can remain comparatively large.
The suppression of the velocity happens in the Dirac-point neighborhood, within a momentum window on the scale $|\mathbf q|\lesssim \mathcal O(G_{\rm 1D})$ before stronger miniband mixing becomes important.
This off-resonant regime provides a clean route to directional Dirac-cone engineering 
because it is decoupled from the resonant mass-opening channel and does not rely on commensurability with $\Delta\mathbf K_\xi$.
Such a low-energy reshaping of the Dirac cone should manifest as strong transport anisotropy and collimation\cite{park2008electron,li2024transport}, providing an experimentally accessible signature even without opening a mass gap. 
Notably, the same modulation-induced transverse-velocity renormalization that underlies this off-resonant anisotropic Dirac engineering is also what enables chirality switching in the coupled-resonant regime when one dressed cone is tuned through $v_\perp^{(l)}=0$.

\subsection{Coordinate convention and minimal low-energy subspace}
In an off-resonant configuration, the direction of $\mathbf G_{\rm 1D}$ is not locked to any crystallographic axis.
% We mark the direction perpendicular (parallel) to $\mathbf{G}_\mathrm{1D}$ by $\perp$ ($\parallel$), and
% \begin{equation}
% \mathbf G_{\rm 1D}=G_{\rm 1D}\hat{\mathbf y}
% =
% \begin{pmatrix}
% 0\\ G_{\rm 1D}
% \end{pmatrix}.
% \end{equation}
Without loss of generality, we choose coordinates such that the 1D modulation is along the $y$ direction,
\begin{equation}
\mathbf G_{\rm 1D}=G_{\rm 1D}\hat{\mathbf y}
=
\begin{pmatrix}
0\\ G_{\rm 1D}
\end{pmatrix}.
\label{eq:S3:G1D_choice}
\end{equation}

We focus on the low-energy Dirac cone of layer $l=1$ in a fixed valley $\xi$ and define the target (P) subspace as the two-component
sublattice spinor at momentum $\mathbf q$ measured from $\mathbf K_\xi^{(1)}$:
$|P\rangle \equiv |\mathbf q,1\rangle$.
In the off-resonant regime, all states reached from $|P\rangle$ by a single moir\'e tunneling event or a single 1D scattering event are remote, i.e., high-energy, and can be treated perturbatively.

\subsection{Minimal momentum-space Hamiltonian and \texorpdfstring{$P/Q$}{PQ} partition}
We retain the minimal set of remote (Q) states coupled to $|P\rangle$ at first order:
(i) three layer-2 states connected by the first-star moir\'e tunneling $T_\xi(\mathbf r)$,
and (ii) two layer-1 sidebands connected by the 1D potential $V_1(\mathbf r)$,
\begin{equation}
|Q_{\rm M}\rangle \equiv
\begin{pmatrix}
|\mathbf q+\mathbf q_0^\xi,2\rangle\\
|\mathbf q+\mathbf q_1^\xi,2\rangle\\
|\mathbf q+\mathbf q_2^\xi,2\rangle
\end{pmatrix},
\quad
|Q_{\rm 1D}\rangle \equiv
\begin{pmatrix}
|\mathbf q+\mathbf G_{\rm 1D},1\rangle\\
|\mathbf q-\mathbf G_{\rm 1D},1\rangle
\end{pmatrix}.
\label{eq:S3:Q_spaces}
\end{equation}
In the composite basis $(|P\rangle^{\mathrm T},|Q_{\rm M}\rangle^{\mathrm T},|Q_{\rm 1D}\rangle^{\mathrm T})^{\mathrm T}$,
the minimal Hamiltonian reads
\begin{equation}
\begin{aligned}
&H_\xi^{\rm(min)}(\mathbf q)\\&=
\begin{pmatrix}
H_{PP}(\mathbf q) & H_{P Q_{\rm M}} & H_{P Q_{\rm 1D}}\\
H_{Q_{\rm M}P} & H_{Q_{\rm M}Q_{\rm M}}(\mathbf q) & 0\\
H_{Q_{\rm 1D}P} & 0 & H_{Q_{\rm 1D}Q_{\rm 1D}}(\mathbf q)
\end{pmatrix}.
\end{aligned}
\label{eq:S3:H_min_bare}
\end{equation}
Here
\begin{equation}
\begin{aligned}
H_{PP}(\mathbf q)&=h_\xi^{(1)}(\mathbf q),\\
H_{P Q_{\rm M}}&=\bigl((T_0^\xi)^\dagger,(T_1^\xi)^\dagger,(T_2^\xi)^\dagger\bigr)=H_{Q_{\rm M}P}^\dagger,\\
H_{P Q_{\rm 1D}}&=\Bigl(\frac{v_1^*}{4}\sigma_0,\frac{v_1}{4}\sigma_0\Bigr)=H_{Q_{\rm 1D}P}^\dagger,\\
H_{Q_{\rm M}Q_{\rm M}}(\mathbf q)&=\mathrm{diag} \bigl(h_\xi^{(2)}(\mathbf q+\mathbf q_0^\xi),\,h_\xi^{(2)}(\mathbf q+\mathbf q_1^\xi), \\ & \qquad \quad \ \  h_\xi^{(2)}(\mathbf q+\mathbf q_2^\xi)\bigr),\\
H_{Q_{\rm 1D}Q_{\rm 1D}}(\mathbf q)&=\mathrm{diag} \bigl(h_\xi^{(1)}(\mathbf q+\mathbf G_{\rm 1D}),\\ & \qquad \quad \ \ h_\xi^{(1)}(\mathbf q-\mathbf G_{\rm 1D})\bigr).
\end{aligned}
\end{equation}
In this minimal construction, $|Q_{\rm M}\rangle$ and $|Q_{\rm 1D}\rangle$ do not couple directly.

\subsection{L\"owdin downfolding and controlled linearization}
Applying momentum-resolved L\"owdin partitioning to Eq.~\eqref{eq:S3:H_min_bare} yields the projected eigenvalue problem
\begin{equation}
\Bigl[H_{PP}(\mathbf q)+\Sigma_{\rm M}(E,\mathbf q)+\Sigma_{\rm 1D}(E,\mathbf q)\Bigr]\Psi = E\Psi,
\label{eq:S3:lowdin_projected}
\end{equation}
with self-energies
\begin{equation}
\Sigma_{\rm M}(E,\mathbf q)=
H_{P Q_{\rm M}}
\bigl[E-H_{Q_{\rm M}Q_{\rm M}}(\mathbf q)\bigr]^{-1}
H_{Q_{\rm M}P},
\label{eq:S3:SigmaM_def}
\end{equation}
and
\begin{equation}
\Sigma_{\rm 1D}(E,\mathbf q)=
H_{P Q_{\rm 1D}}
\bigl[E-H_{Q_{\rm 1D}Q_{\rm 1D}}(\mathbf q)\bigr]^{-1}
H_{Q_{\rm 1D}P}.
\label{eq:S3:Sigma1D_def}
\end{equation}
%
% \paragraph{Small parameters and linearization.}
Because we are interested in $|E|,\hbar v_\mathrm{F}|\mathbf q|\ll \mathcal E_\theta,\mathcal E_{\rm 1D}$, 
we use the controlled resolvent expansion to first order in $(E,\mathbf q)$:
\begin{equation}
\begin{aligned}
&\bigl[E-H_{QQ}^{(0)}-H_{QQ}^{(1)}(\mathbf q)\bigr]^{-1}  \\
&\approx
(-H_{QQ}^{(0)})^{-1}  \\
&-
(-H_{QQ}^{(0)})^{-1}\bigl(E-H_{QQ}^{(1)}(\mathbf q)\bigr)(-H_{QQ}^{(0)})^{-1}
\end{aligned}
\label{eq:S3:resolvent_linearization}
\end{equation}
for any remote block of the form $H_{QQ}(\mathbf q)=H_{QQ}^{(0)}+H_{QQ}^{(1)}(\mathbf q)$,
with $H_{QQ}^{(0)}\sim \mathcal E_\theta$ or $\mathcal E_{\rm 1D}$ and $H_{QQ}^{(1)}(\mathbf q)=\mathcal O(\hbar v_\mathrm{F}|\mathbf q|)$,
% This approximation drops terms of order $\mathcal O(E^2)$, $\mathcal O(E|\mathbf q|)$, and $\mathcal O(|\mathbf q|^2)$, each suppressed by $\mathcal E_\theta$ and/or $\mathcal E_{\rm 1D}$.
%
% \paragraph{Resulting linearized self-energies.}
The linearized moir\'e and 1D self-energies take the compact form
\begin{equation}
\begin{aligned}
\Sigma_{\rm M}(E,\mathbf q)
&\approx
-3u_0^2\,h_\xi^{(1)}(\mathbf q)
-6u_0^2\,E\,\sigma_0,
\\[1mm]
\Sigma_{\rm 1D}(E,\mathbf q)
&\approx
\frac{|u_1|^2}{8}\Bigl[
-h_\xi^{(1)}(q_x\hat{\mathbf x})
+h_\xi^{(1)}(q_y\hat{\mathbf y}) \\
& \qquad \qquad  -E\,\sigma_0
\Bigr].
\end{aligned}
\label{eq:S3:self_energy_linearized}
\end{equation}
The moir\'e contribution is the standard second-order result of pristine rigid TBG and is isotropic by the $C_3$ symmetry of the first-star set $\{\mathbf q_j^\xi\}_{j=0,1,2}$ \cite{LopesdosSantos2007_PRL, BistritzerMacDonald2011_PNAS}.
By contrast, the 1D contribution distinguishes the $x$ and $y$ directions, reflecting Bragg scattering by $\pm\mathbf G_{\rm 1D}$ \cite{Park2008_NatPhys_1D, BreyFertig2009_PRL}.

\subsection{Renormalized Dirac Hamiltonian and anisotropic velocities}

From Eq.~\eqref{eq:S3:lowdin_projected}, we get the $2\times 2$ effective Hamiltonian near the layer-1 Dirac point
\begin{equation}
h_{\xi,{\rm eff}}^{(1)}(\mathbf q)
=
-\hbar\Bigl(\xi v_x^* q_x\,\sigma_x^{(1)}+v_y^* q_y\,\sigma_y^{(1)}\Bigr),
\label{eq:S3:heff_incomm}
\end{equation}
with the renormalized velocities
\begin{equation}
\begin{aligned}
v_x^* &= \Bigl(\frac{1-3u_0^2-|u_1|^2/8}{1+6u_0^2+|u_1|^2/8}\Bigr)v_{\mathrm F},\\
v_y^* &= \Bigl(\frac{1-3u_0^2+|u_1|^2/8}{1+6u_0^2+|u_1|^2/8}\Bigr)v_{\mathrm F}.
\end{aligned}
\label{eq:S3:vstar_incomm}
\end{equation}
Since $h_{\xi,\mathrm{eff}}^{(1)}(\mathbf{q})$ contains only $\sigma_{x,y}$ terms and therefore no mass term is generated at the retained order, consistent with both pristine TBG and monolayer graphene under a scalar 1D potential.

\subsection{Limiting cases, flattening line, and interpretation}
\paragraph{Pristine-TBG limit.}
For $u_1\to 0$ we recover the familiar isotropic second-order velocity renormalization of rigid TBG,
$\lim_{u_1\rightarrow0} v_x^*= \lim_{u_1\rightarrow0} v_y^*=v_{\rm TBG}^*= v_{\mathrm F}(1-3u_0^2)/(1+6u_0^2)$ \cite{LopesdosSantos2007_PRL, BistritzerMacDonald2011_PNAS}.

\paragraph{Monolayer/1D limit.}
For $u_0\to 0$ the 1D potential yields an anisotropic renormalization:
$\lim_{u_0\rightarrow0}v_x^* = v_{\mathrm F}(1-|u_1|^2/8)/(1+|u_1|^2/8)$ and $\lim_{u_0\rightarrow0} v_y^*= v_{\mathrm F}$,
in agreement with the known behavior of monolayer graphene under a scalar 1D potential \cite{Park2008_NatPhys_1D, BreyFertig2009_PRL}.
%graphene under a unidirectional scalar modulation.

\paragraph{Flattening condition (``magic line'' in $(\theta,|u_1|)$).}
Equation~\eqref{eq:S3:vstar_incomm} shows that the 1D potential provides an independent tuning knob that suppresses the velocity
perpendicular to the modulation.
Complete flattening in the $x$ direction, $v_x^*=0$, occurs at
\begin{equation}
\begin{aligned}
|u_1|&=\sqrt{8(1-3u_0^2)}\\
&\approx
2\sqrt{2}\,
\Bigl[1-3\Bigl(\frac{3t_0 a}{4\pi\hbar v_{\mathrm F}\theta}\Bigr)^2\Bigr]^{1/2},
\quad (\theta\ll1).
\end{aligned}
\label{eq:S3:flattening_u1}
\end{equation}
% Equation~\eqref{eq:S3:vstar_incomm} provides a practical design relation between the twist angle and the required strength of the 1D modulation in the perturbative off-resonant regime.
Compared to the renormalized Dirac velocity of pristine TBG, Eq.~\eqref{eq:S3:vstar_incomm} makes explicit that the 1D modulation provides an additional gate-tunable renormalization of the Dirac velocities.
Increasing $|u_l|$ strongly suppresses the transverse velocity $v_\perp^*$---which can be driven to zero along a ``magic line'' in the $(\theta,|u_l|)$ plane, enabling controllable one-directional Dirac flattening in the low-energy sector.
Figure~\ref{fig:fig_s_i}b visualizes the renormalized transverse velocity and highlights the band flattening line extracted from Eq.~\eqref{eq:S3:vstar_incomm}.

% \paragraph{Longitudinal velocity enhancement.}
% In contrast to $v_x^*$, the velocity parallel to the modulation is enhanced by the 1D potential when moir\'e tunneling is present.
% For $|u_1|\ll1$, expanding Eq.~\eqref{eq:S3:vstar_incomm} gives
% \begin{equation}
% v_y^*
% =
% \Bigl(\frac{1-3u_0^2}{1+6u_0^2}\Bigr)v_{\mathrm F}
% \biggl[
% 1+\frac{9u_0^2}{8(1-3u_0^2)(1+6u_0^2)}|u_1|^2+\mathcal O(|u_1|^4)
% \biggr].
% \label{eq:S3:vy_smallB}
% \end{equation}
% The $|u_1|^2$ correction vanishes in the monolayer limit $u_0\to0$, consistent with $v_y^*=v_{\mathrm F}$ for monolayer graphene under a scalar 1D potential.

\subsection{Neglected terms and validity}

\paragraph{Leading neglected contributions.}
Within the minimal $Q$ space of Eq.~\eqref{eq:S3:Q_spaces} and the linearization
Eq.~\eqref{eq:S3:resolvent_linearization}, we retain only (a) terms up to second order in the couplings
and (b) terms up to linear order in $(E,\mathbf q)$ in the projected Hamiltonian.
Accordingly, we neglect:
(i) higher-order virtual processes in the couplings, such as $\mathcal O(u_0^4)$ and $\mathcal O(|u_1|^4)$,
as well as mixed moir\'e+1D processes that start at $\mathcal O(u_0^2|u_1|^2)$ and require intermediate
states beyond the minimal $Q$ space (e.g., states involving both $\mathbf q_j^\xi$ and $\pm\mathbf G_{\rm 1D}$);
(ii) quadratic-and-higher corrections in $(E,\mathbf q)$ from the resolvent expansion, namely terms of order
$\mathcal O\big((\hbar v_\mathrm{F}|\mathbf q|)^2/\mathcal E_\theta\big)$ and $\mathcal O\big((\hbar v_\mathrm{F}|\mathbf q|)^2/\mathcal E_{\rm 1D}\big)$,
as well as $\mathcal O(E^2)$ and $\mathcal O(E\,\hbar v_\mathrm{F}|\mathbf q|)$ terms (equivalently, higher-order momentum/energy corrections beyond the linear Dirac theory);
and
(iii) the effect of a 1D potential on layer~2, which influences the layer-1 projected Hamiltonian only via mixed
processes involving both moir\'e tunneling and additional layer-2 sidebands
(e.g., intermediate states $|\mathbf q+\mathbf q_j^\xi\pm\mathbf G_{\rm 1D},2\rangle$), so that its leading contribution is of higher order
(schematically $\mathcal O(u_0^2|u_2|^2)$ and beyond) and is neglected in the present minimal second-order treatment.

\paragraph{Validity conditions.}
The results in Eqs.~\eqref{eq:S3:self_energy_linearized}--\eqref{eq:S3:vstar_incomm} are controlled under $|E|,\ \hbar v_\mathrm{F}|\mathbf q| \ll \mathcal E_\theta,\ \mathcal E_{\rm 1D}$,
$|u_0|\ll 1$, $|u_1|\ll 1$
for the second-order expansion in $u_0$ and $u_1$.
Additionally, the configuration must be off-resonant in the sense that no integer harmonic bridges the Dirac mismatch at the low-energy scale.

\section{Feasibility and tolerance estimates}
\label{sec:feasibility_appendix}

This appendix provides order-of-magnitude experimental feasibility estimates for realizing the hybrid moir\'e--1D band engineering studied in this work in an electrostatically patterned device.
We focus on the same-chirality case $\chi_{\rm rel}=+1$, whereas the corresponding case $\chi_{\rm rel}=-1$ can be treated by the same steps and is not discussed further. For $\chi_{\rm rel}=+1$, odd $n_{\rm 1D}$ is the leading gapping class.

We further focus on the experimentally common single-sided electrostatic patterning, where the induced layer potentials are nearly equal and phase-aligned, $|u_1|\approx |u_2|$ and $\phi_1\approx \phi_2$, so that $|u_-| (\approx0)\ll |u_+|(\approx 2|u_1|, 2|u_2|)$ [Eq.~\eqref{eq:S4:upm_def}]. Achieving independent control of the amplitude and phase on the two layers at the relevant length scale, i.e., $|u_1|\neq |u_2|$ and $\phi_1\neq \phi_2$, requires more elaborate device engineering and is currently challenging; 
% (Cite: Alternative platforms---such as patterned dielectrics, block-copolymer self-assembly, moir\'e superlattices employed as remote Coulomb superlattices (Nat. Mater. 23, 189 (2024)), and supramolecular lattices (Nat. Commun. 8, 14767 (2017))---may, in principle, provide an independently tunable potential acting predominantly on the top graphene layer.)
therefore we do not evaluate the feasibility of that general case here and only comment on its qualitative impact on the gap opening in Appendix~\ref{sec:layer_symmetric} for reference.
In fact, $|u_1|\approx |u_2|$ and $\phi_1\approx \phi_2$ are not only favorable for fabrication but also helpful for the gap-opening at resonant configurations; the mass term is controlled by the in-phase combination $u_+$, while the node-shifting term is controlled by the out-of-phase combination $u_-$ (Appendix~\ref{sec:min_two_cone_model}). Therefore single-sided gating that yields $|u_-|\ll |u_+|$ suppresses node-shifting terms and tends to maximize the intended gap within the minimal-wave description.

 \textbf{Period of the unidirectional scalar potential for gap opening at charge neutrality point:}
We assess the experimental feasibility of the period of the 1D potential to open a gap by exact Dirac--Dirac resonance, $\Delta \mathbf{K}_\xi = \xi n_\mathrm{1D} \mathbf{G}_\mathrm{1D}$ with odd $n_\mathrm{1D}$, in Appendix~\ref{sec:S8_2}.

 \textbf{Two practical resonant routes and the key trade-off:}
One may target resonance either with $n_{\rm 1D}=1$ (short $L_{\rm 1D}$) or with odd $n_{\rm 1D}\ge 3$ (longer $L_{\rm 1D}$).
Larger $n_{\rm 1D}$ makes lithographic fabrication more forgiving by enlarging $L_\mathrm{1D}$ (Appendix~\ref{sec:S8_2}) and reduces the attenuation of the potential strength at the graphene plane (Appendix~\ref{sec:S8_3}).
The trade-off is that the resonant mass amplitude is controlled by a higher-order ($n_{\rm 1D}$-th) Bessel coefficient. Thus, higher-order $n_{\rm 1D}$ channels remain parametrically suppressed at weak-to-moderate modulation, so $n_{\rm 1D}=1$ is preferred unless the induced modulation becomes sufficiently large (Appendix~\ref{sec:S8_3}).
This feasibility discussion concerns the nearest-cone design branch. General moir\'e-Umklapp resonances do not remove the same basic trade-off: because $|\mathbf q_j^\xi+\mathbf G_{\rm M}|\ge k_\theta$, a translated branch cannot require a longer $L_{\rm 1D}$ at fixed harmonic order $|p|$ than the nearest-cone branch. Longer-period designs are therefore obtained by increasing the selected integer harmonic order, which suppresses the mass channel at weak-to-moderate modulation through the corresponding Bessel coefficient.

 \textbf{Higher harmonics in realistic waveforms---factors for maintaining the selection rules:}
A realistic patterned gate typically produces a non-sinusoidal waveform containing higher spatial harmonics at $m\mathbf{G}_{\rm 1D}$ ($m\ge 2$) (Appendix~\ref{sec:S8_3}).
The higher ($m\ge2$) harmonics are partially suppressed by dielectric layers, where the suppression is weak for very thin spacers ($1$--$2~{\rm nm}$) or a long $L_{\rm 1D}$ but can become significant for thicker dielectrics ($10~{\rm nm}$) or a short $L_\mathrm{1D}$.
The parity selection rule becomes clear when the even-$m$ harmonics of the potential are suppressed, e.g., by ensuring the potential to have half-translation antisymmetry, which suppresses even harmonics in the waveform.

 \textbf{Feasibility metric, geometric errors and tolerances:}
In globally periodic (rank-2) resonant/near-resonant configurations, our primary interest is a direct gap at or near the CNP $\Delta_\mathrm{dir}$;
we discuss the feasibility in Appendices~\ref{sec:S8_2} and \ref{sec:S8_3}.
Appendix \ref{sec:S8_4} estimates 
tolerance requirements 
of twist angle, 1D modulation period, and alignment for resonant/near-resonant configurations, and compares them with
realistic windows achieved with current fabrication capabilities.
In quasiperiodic (rank-3) settings, where no rBZ exists, our interest lies in the resolvable anisotropic renormalization of Dirac velocities; we discuss the feasibility in Appendix~\ref{sec:S8_5}.

\subsection{Experimental knobs}
\label{sec:experimental_knobs}

\paragraph{Continuum-model parameters and experimental knobs.}
In the symmetric geometry used in this paper, the Dirac-point mismatch is $\Delta\mathbf K_\xi=\xi k_\theta \hat{\mathbf y}$ where its magnitude is
$k_\theta\equiv|\Delta\mathbf K_\xi|=(8\pi/3a)\sin(\theta/2)$ for a given twist angle $\theta$.
The reciprocal lattice vector of the unidirectional potential is 
\begin{equation}
\mathbf G_{\rm 1D}=G_{\rm 1D}(\sin\varphi_{\rm 1D},\cos\varphi_{\rm 1D}),
\label{eq:S8:G1D_param}
\end{equation}
where $\varphi_\mathrm{1D}$ denotes the in-plane orientation of the 1D modulation ($\varphi_{\rm 1D}=0$ corresponds to $\mathbf G_{\rm 1D}\parallel\Delta\mathbf K_\xi$).
% It is useful to introduce the dimensionless couplings $u_0=t_0/(\hbar v_\mathrm{F} k_\theta)$ and $u_l=v_l/(\hbar v_\mathrm{F} G_{\rm 1D})$ ($l=1,2$), where $t_0$ is the moir\'{e} tunneling strength and $v_l$ is the complex amplitude of the layer-dependent 1D scalar potential defined in Eq.~\eqref{eq:prb_Vl}.

We view a device configuration as specified by 
\begin{equation}
(\theta,L_\mathrm{1D},\varphi_{\rm 1D}, d_{\rm diel}),
\label{eq:S8:knobs_tuple}
\end{equation}
where $L_\mathrm{1D}=2\pi/G_{\rm 1D}$ is the period of the 1D modulation, and $d_{\rm diel}$ is the effective separation between graphene and the patterned dielectric.

\subsection{Resonant and near-resonant geometric design and mass channel magnitude}
\label{sec:S8_2}

This section summarizes practical geometric targets for approaching Dirac--Dirac resonance and 
discusses how these targets relate to demonstrated nanofabrication scales.

\paragraph{Exact resonance and target periods.}
Momentum (Dirac) resonance is defined by Eq.~\eqref{eq:prb_resonance}, $\Delta\mathbf K_\xi=\xi n_{\rm 1D}\mathbf G_{\rm 1D}$.
Choosing $\mathbf G_{\rm 1D}\parallel\Delta\mathbf K_\xi$ reduces the resonant condition to a scalar matching,
\begin{equation}
\begin{aligned}
k_\theta=n_{\rm 1D}G_{\rm 1D}\ \Longleftrightarrow\ 
L_{\rm 1D}^{(n_{\rm 1D})}(\theta)&=\frac{2\pi n_{\rm 1D}}{k_\theta} =\frac{3a}{4}\frac{n_{\rm 1D}}{\sin(\theta/2)} \\
&\approx \frac{3a}{2}\frac{n_{\rm 1D}}{\theta},
\end{aligned}
\label{eq:S8:lambda_target}
\end{equation}
where the final expression holds for $\theta\ll 1$ with $\theta$ in radians.
In the same-chirality case $\chi_{\rm rel}=+1$ relevant for our numerics, odd $n_{\rm 1D}$ is the leading gapping class at charge neutrality (Appendix~\ref{sec:exact_resonance}); hence we focus on odd $n_{\rm 1D}$ in the targets below.

\begin{table*}[t]
\centering
\begin{tabular}{c|c|cc|cc|cc}
\hline
$\theta$ &
$\bar{u}_l$ for $v_l=0.26~\mathrm{eV}$ &
$L_{\rm 1D}^{(1)}$ &
$t_0 J_1(\bar{u}_l)$ &
$L_{\rm 1D}^{(3)}$ &
$t_0 J_3(3\bar{u}_l) $ &
$L_{\rm 1D}^{(5)}$ &
$t_0 J_5(5\bar{u}_l)$ \\
\hline
$1^\circ$ & $1.33$ & $21.1$ & $58.1$ & $63.4$ & $47.3$ & $105.7$ & $40.7$\\
$2^\circ$ & $0.665$ & $10.6$ & $34.7$ & $31.7$ & $14.1$ & $52.9$ & $7.24$\\
$3^\circ$ & $0.443$ & $7.05$ & $23.8$ & $21.1$ & $4.81$ & $35.2$ & $1.24$\\
\hline
\end{tabular}
\caption{Target 1D periods $L_\mathrm{1D}^{(n_\mathrm{1D})}$ (nm) for exact resonance [Eq.~\eqref{eq:S8:lambda_target}] (Appendix~\ref{sec:S8_2}), effective modulation strength $\bar{u}_l=|v_l|/(\hbar v_\mathrm{F} k_\theta)$, and the magnitude of the selected mass channel $t_0 J_{n_{\rm 1D}}(|u_l|)=t_0 J_{n_{\rm 1D}}(n_{\rm 1D}\bar{u}_l)$ (meV) in the equal in-phase limit $u_1=u_2\equiv u_l$, $\phi_1=\phi_2$, where the general convention $J_{n_{\rm 1D}}(|u_+|/2)$ reduces to $J_{n_{\rm 1D}}(|u_l|)$.
We evaluated $\bar{u}_l$ and $J_{n_\mathrm{1D}}(n_\mathrm{1D}\bar{u}_l)$ for modulation strength $v_l=0.26~\mathrm{eV}$ with $t_0\approx0.110~\mathrm{eV}$.
We list odd $n_{\rm 1D}=1,3,5$, which are the leading gapping class for $\chi_{\rm rel}=+1$ (Appendix~\ref{sec:exact_resonance}).}
\label{tab:S8:lambda_targets}
\end{table*}

\paragraph{Fabrication scales.}

Experimentally, electrostatically defined superlattice potentials with sub-$100~\mathrm{nm}$ pitch have been realized using several complementary patterning strategies, including arrays of metallic gates down to $100~\mathrm{nm}$ \cite{Kuiri2018_1DGrapheneSL}, block-copolymer templating down to $38~\mathrm{nm}$\cite{Jamalzadeh2025_ACSNano}, and dielectric patterning with sub-$40~\mathrm{nm}$ pitch\cite{Forsythe2018_NatNano, Li2021_NatNano_1DSL}.
Related nanofabrication approaches also demonstrate periodic patterning at even smaller length scales, such as $16~\mathrm{nm}$ by focused ion beam milling\cite{BarconsRuiz2022_NatCommun} and $\sim 5~\mathrm{nm}$ by electron-beam induced deposition\cite{Meyer2008_APL}.
Taken together, these benchmarks suggest that the odd-$n_{\rm 1D}$ target periods in Table~\ref{tab:S8:lambda_targets} are challenging yet broadly compatible with current nanofabrication capabilities; in particular, odd $n_{\rm 1D}\ge 3$ places the required $L_{\rm 1D}$ in the $20$--$100~\mathrm{nm}$ range for $\theta\sim 1^\circ$--$3^\circ$.

\paragraph{Mass channel magnitude and choosing odd $n_{\rm 1D}$ given an achievable $|v_l|$.}
In our framework (Appendices~\ref{sec:dressed_rep}--\ref{sec:near_resonance}), the experimentally controlled modulation is taken to be dominated by the fundamental Fourier component at $\mathbf G_{\rm 1D}$, while the resonant coupling itself arises from the $p_0=-\xi n_{\rm 1D}$ Fourier harmonic of the dressed interlayer tunneling (Appendix~\ref{sec:exact_resonance}).
Here $n_{\rm 1D}$ labels the nearest-cone resonance family; for a generalized moir\'e-Umklapp branch the corresponding design order is the selected integer $p$ in Eq.~\eqref{eq:prb_umklapp_resonance}.
For the experimentally common case of single-sided patterning with nearly equal and in-phase layer potentials, $u_1=u_2\equiv u_l$ and $|u_+|/2=|u_l|$.
Under the commensurate condition $k_\theta=n_{\rm 1D}G_{\rm 1D}$,
$|u_l|=|v_l|/(\hbar v_\mathrm{F} G_{\rm 1D})=n_{\rm 1D}|v_l|/(\hbar v_\mathrm{F} k_\theta)$,
so the odd-$n_{\rm 1D}$ resonant mass weight scales as $J_{n_{\rm 1D}}(|u_l|)=J_{n_{\rm 1D}}(n_{\rm 1D} \bar{u}_l)$ with 
\begin{equation}
    \bar{u}_l\equiv |v_l|/(\hbar v_\mathrm{F} k_\theta).
\end{equation}
Here $\bar u_l$ is a dimensionless measure of the effective modulation strength set by the induced potential amplitude $|v_l|$ at a given twist angle (fixed $k_\theta=|\Delta\mathbf K_\xi|$).
In the weak-modulation regime $\bar{u}_l\ll 1$ (equivalently $|v_l|\ll \hbar v_\mathrm{F} k_\theta$),
\begin{equation}
J_{n_{\rm 1D}}(n_{\rm 1D}\bar{u}_l)\approx \frac{1}{n_{\rm 1D}!}\Big(\frac{n_{\rm 1D}\bar{u}_l}{2}\Big)^{n_{\rm 1D}},
\label{eq:S8:Jn_scaling_fixedv}
\end{equation}
so higher-order odd-$n_{\rm 1D}$ channels are parametrically suppressed and $n_{\rm 1D}=1$ is favored.
As $\bar{u}_l$ approaches order unity, the argument grows proportionally to $n_{\rm 1D}$ and the $n_{\rm 1D}=3,5,\ldots$ harmonics can become appreciable.
For example, patterned-dielectric 1D superlattices report dimensionless strengths up to $u=vL/(\hbar v_\mathrm{F})\sim 7\pi$ for $L\sim 55~\mathrm{nm}$, corresponding to screened modulation amplitudes $v$ on the order of $0.26~\mathrm{eV}$ \cite{Li2021_NatNano_1DSL}.
We emphasize that $|v_l|\approx 0.26~\mathrm{eV}$ is not a structural upper bound of patterned-dielectric platforms, but merely a representative value demonstrated in existing devices and used here as a conservative reference scale.
Taking this value as a reference scale, Table~\ref{tab:S8:lambda_targets} summarizes $\bar{u}_l$ and representative odd-$n_{\rm 1D}$ weights for $\theta=1^\circ,2^\circ,3^\circ$.
In this range, odd $n_{\rm 1D}\ge 3$ channels are already comparable to $n_{\rm 1D}=1$ at $\theta\approx 1^\circ$, remain non-negligible at $\theta\approx 2^\circ$ (with $n_{\rm 1D}=3$ favored over $n_{\rm 1D}=5$), but are strongly suppressed at $\theta\approx 3^\circ$.
Finally, the longer period associated with odd $n_{\rm 1D}\ge 3$ can be advantageous both for fabrication and for reducing finite-distance attenuation of the fundamental component at $G_{\rm 1D}=k_\theta/n_{\rm 1D}$ (Appendix~\ref{sec:S8_3}).

Equation~\eqref{eq:S8:Jn_scaling_fixedv} captures an important design trade-off
when one targets resonance either with $n_{\rm 1D}=1$ (short $L_{\rm 1D}$) or with odd $n_{\rm 1D}\ge 3$ (longer $L_{\rm 1D}$).
While larger $n_{\rm 1D}$ makes lithographic fabrication more forgiving by enlarging $L_\mathrm{1D}$, its mass channels remain parametrically suppressed at weak-to-moderate modulation. Thus, $n_{\rm 1D}=1$ is preferred unless the induced modulation becomes sufficiently large.

\paragraph{Near-resonant window.}
Exact resonance points are measure zero in the $\mathbf G_{\rm 1D}$ configuration map; experimentally, it is sufficient to approach a resonant point so that the momentum mismatch $\delta\mathbf q_\xi^{(n_{\rm 1D})}$ [Eq.~\eqref{eq:S1:mm_def}] lies within the gapped region derived in Appendix~\ref{sec:near_resonance}.
In Appendix~\ref{sec:S8_4} we translate the corresponding $\delta\mathbf q$ window into tolerances in $(\theta,L_{\rm 1D},\varphi_{\rm 1D})$.

\subsection{Electrostatics: finite-distance attenuation of a periodic gate potential and parity leakage from higher (even) harmonics}
\label{sec:S8_3}

\paragraph{Realistic potential waveforms and higher harmonics}
The continuum model in this work assumes that the 1D modulation entering each layer is well approximated by a single cosine,
$V_l(y)=|v_l| \cos(G_{\rm 1D} y+\phi_l)/2$ [Eq.~\eqref{eq:prb_Vl}].
A patterned gate, however, may produce a non-sinusoidal potential profile, i.e.\ a superposition of multiple harmonics at the same reciprocal vector $\mathbf{G}_{\rm 1D}$.
Because the resonant selection rules and minimal models in Appendices~\ref{sec:dressed_rep}--\ref{sec:near_resonance} are derived for a single harmonic, it is useful to parameterize and estimate the harmonic content of a realistic gate-induced potential.

\paragraph{Finite-distance attenuation of the 1D modulation}
The 1D modulation, containing higher spatial harmonics $m$, is generated at a patterned dielectric plane ($z=0$) separated from the graphene plane ($z=d_\mathrm{diel}$) by a finite distance $d_{\rm diel}$ and forms the 1D potential at the graphene plane (for each layer $l$)
\begin{multline}
V_l(y)=\frac{1}{4}\sum_{m=1}^{\infty}\big(v_{l,m}e^{i m G_{\rm 1D} y}+v_{l,m}^\ast e^{-i m G_{\rm 1D} y}\big),
\\ (v_{l,1}\equiv v_l),
\label{eq:S8:Vm_expansion}
\end{multline}
where $v_{l,m}$ are complex Fourier amplitudes of the $m$-th harmonic.
In the absence of electronic screening, the electrostatic transfer from the patterned plane to the graphene plane obeys Laplace attenuation,
\begin{equation}
v_{l,m}\equiv v_{l,m}(z=d_\mathrm{diel})\approx v_{l,m}(z=0)e^{-m G_{\rm 1D} d_{\rm diel}},
\label{eq:S8:Vm_decay_again}
\end{equation}
%=v_{l,m}(z=0)e^{-2\pi m d_\mathrm{diel}/L_\mathrm{1D}}
where $v_{l,m}(0)$ denotes the amplitude of $m$-th harmonic at the patterned plane.
The $z$-direction decay of the $m$-th in-plane harmonic is
\begin{equation}
    m G_\mathrm{1D} d_\mathrm{diel} = 2\pi m d_\mathrm{diel}/L_\mathrm{1D} =m k_\theta d_\mathrm{diel} / n_\mathrm{1D}.
\label{eq:S8:attenuation_factor}
\end{equation}
Consequently, the higher ($m\ge2$) in-plane harmonics decay to the graphene plane faster than the fundamental ($m=1$) harmonic.
Electronic screening further reduces all $|v_{l,m}|$, but the relative harmonic weights are largely set by the waveform at the patterned plane together with the exponential attenuation factor Eq.~\eqref{eq:S8:attenuation_factor}.

Beyond the trade-off discussed in Appendix~\ref{sec:S8_2}, Eqs.~\eqref{eq:S8:Vm_decay_again} and \eqref{eq:S8:attenuation_factor} highlight an additional practical consideration in choosing $n_{\rm 1D}$: for a fixed spacer thickness, shorter periods $L_{\rm 1D}$ (corresponding to smaller $n_{\rm 1D}$ at fixed $k_\theta$) suffer stronger finite-distance attenuation, whereas this effect becomes much less pronounced for very thin spacers. For instance, the factor $e^{-2\pi d_{\rm diel}/L_{\rm 1D}}$ for the fundamental harmonic is $\approx 0.04$ for $L_{\rm 1D}=20~\mathrm{nm}$ and $\approx 0.35$ for $L_{\rm 1D}=60~\mathrm{nm}$ when $d_{\rm diel}=10~\mathrm{nm}$, while it is $\approx 0.73$ and $\approx 0.90$, respectively, when $d_{\rm diel}=1~\mathrm{nm}$.

\paragraph{Parity leakage from even-harmonic contamination.}
With a single harmonic ($m=1$), $\cos\alpha_l(y)$ contains only even multiples of $G_{\rm 1D}$ while $\sin\alpha_l(y)$ contains only odd multiples.
This property is the origin of the clear even/odd $n_\mathrm{1D}$ selection rule for gap opening at leading order in small $\theta$.
If, however, even harmonics are present ($v_{l,2},v_{l,4},\ldots\neq 0$), $\cos\alpha_l$ and $\sin\alpha_l$ generally contain both even and odd Fourier components.
In the resonant two-cone description (Appendix~\ref{sec:exact_resonance}), this appears as small leakage terms that admix the Pauli-matrix subspace associated with the opposite parity of $p$.
A sufficient condition to suppress even harmonics is that the potential flips sign under a half-period translation,
\begin{equation}
V_l(y+L_\mathrm{1D}/2)-\overline{V}_l
=
-\bigl[V_l(y)-\overline{V}_l\bigr],
\label{eq:S8:half_period_condition}
\end{equation}
where $\overline{V}_l$ is the spatial average over one period.

\subsection{Geometric errors and tolerances}
\label{sec:S8_4}
\label{sec:geometric_errors}
Appendix \ref{sec:near_resonance} shows that, near a target resonance, the gapped region in mismatch coordinates $(\delta q_\perp,\delta q_\parallel)$ is approximately the filled ellipse
Eq.~\eqref{eq:S5:ellipse_master}, with semiaxes $Q_\perp$ and $Q_\parallel$ defined there.
Here we translate this condition into practical tolerances on $(\theta,L_\mathrm{1D},\varphi_{\rm 1D})$ and discuss additional requirements from disorder and temperature.

\paragraph{From geometric errors to mismatch components.}
For a fixed integer $n_{\rm 1D}$, the mismatch vector is
$\delta\mathbf q_\xi^{(n_{\rm 1D})}=\Delta\mathbf K_\xi-\xi n_{\rm 1D}\mathbf G_{\rm 1D}$
[Eq.~\eqref{eq:S1:mm_def}].
Linearizing around an ideal design point $(\theta,L_\mathrm{1D},\varphi_{\rm 1D}=0)$ satisfying $k_{\theta}=n_{\rm 1D}G_\mathrm{1D}$ [Eq.~\eqref{eq:S8:lambda_target}], we obtain, to leading order,
\begin{equation}
\begin{aligned}
\delta q_\perp  &\approx k_{\theta}\,\delta\varphi_{\rm 1D},
\\
\delta q_\parallel &\approx \delta k_\theta+k_{\theta}\frac{\delta L_\mathrm{1D}}{L_\mathrm{1D}} + \frac{k_{\theta}}{2}(\delta \varphi_\mathrm{1D})^2,
\end{aligned}
\label{eq:S8:dq_from_dtheta_dlambda_dphi}
\end{equation}
where $\delta k_\theta=(\partial k_\theta/\partial\theta)_{\theta_0}\delta\theta$ and we used $\delta G_{\rm 1D}/G_{\rm 1D}=-\delta L_\mathrm{1D}/L_\mathrm{1D}$.
Equation~\eqref{eq:S8:dq_from_dtheta_dlambda_dphi} shows that a small orientation error generates a linear transverse mismatch $\delta q_\perp$, while its contribution to $\delta q_\parallel$ is only quadratic in $\delta\varphi_{\rm 1D}$.

\paragraph{Tolerance formulas.}
A conservative sufficient condition for remaining in the gapped region is $|\delta q_\parallel|<Q_\parallel$ and $|\delta q_\perp|<Q_\perp$,
with $Q_{\parallel,\perp}$ defined in Eq.~\eqref{eq:S5:ellipse_master}.
Using Eq.~\eqref{eq:S8:dq_from_dtheta_dlambda_dphi}, we obtain the tolerance for each parameter
\begin{equation}
\begin{aligned}
|\delta\theta| &\lesssim \frac{Q_\parallel}{(\partial k_\theta/\partial\theta)_{\theta_0}},
\\
\Big|\frac{\delta L_\mathrm{1D}}{L_\mathrm{1D}}\Big| &\lesssim \frac{Q_\parallel}{k_\theta},
\\
|\delta\varphi_{\rm 1D}| &\lesssim \frac{Q_\perp}{k_\theta},
\end{aligned}
\label{eq:S8:tolerance_box}
\end{equation}
evaluated at the target twist angle.
Here $(\partial k_\theta / \partial \theta)_{\theta_0} =(4\pi/3a)\cos(\theta/2)$,
which is nearly constant in the small-angle regime.
% Using $a=0.246~\mathrm{nm}$, one finds $\partial_\theta k_\theta\simeq 0.297~\mathrm{nm}^{-1}/{^\circ}$ for $\theta\lesssim 3^\circ$.
%
In the two-cone estimate of Appendix~\ref{sec:near_resonance}, the detuning-dependent gap obeys Eq.~\eqref{eq:S5:gap_estimate}.
From Eq.~\eqref{eq:S5:filled_ellipse_DK}, $Q_\perp \approx \Delta_\mathrm{dir}/(\hbar v_\perp^{\rm eff})$ and $Q_\parallel \approx \Delta_\mathrm{dir}/(\hbar v_\parallel^{\rm eff})$, where $\Delta_\mathrm{dir}$ is the estimate of the direct gap size.
% Then, each tolerance is
% \begin{equation}
% |\delta\theta| \lesssim \frac{3a\Delta_\mathrm{true}}{4\pi\hbar v_y^\mathrm{eff}\cos(\theta/2)},
% \qquad
% \label{eq:S8:dtheta_tol_numeric}
% \end{equation}

Below, as an illustrative estimate, we take $v_{\perp,\parallel}^{\mathrm{eff}}\sim v_\mathrm{F}(1-3u_0^2)/(1+6u_0^2)$, i.e., we use the pristine-TBG velocity renormalization.
%as a baseline and defer the additional $v_\perp$ renormalization induced by the 1D potential (Eq.~\eqref{eq:S5:ren_recipe}) to the full numerical analysis.
%
We do not quote tolerance for $\theta=1^\circ$ below, because the velocity renormalization becomes nonperturbatively strong in this range, a simple analytic baseline is not quantitatively reliable for assigning a single number to the tolerance criteria. 
However, the strongly reduced velocity would in fact relax the geometric tolerance.

\paragraph{Twist-angle tolerance}
For $\Delta_\mathrm{dir}=10~\mathrm{meV}$, the twist-angle tolerance $|\delta\theta|$ for a fixed $\mathbf{G}_\mathrm{1D}$ is $\approx 0.10^\circ$ for $\theta=2^\circ$ and $\approx 0.069^\circ$ for $\theta=3^\circ$. 

Large-area mapping at $\theta\approx 2^\circ$ reports variations of order $0.08^\circ$ over hundreds of nm \cite{Benschop2021_PRR}.
Protocols targeting improved homogeneity have reported micrometer-scale regions with twist-angle variations as small as $\sim 0.02^\circ$ \cite{DiezMerida2024_arxiv}.
Thus, although achieving $|\delta\theta|\approx 0.10^\circ$ is experimentally demanding, it appears attainable with state-of-the-art device fabrication and characterization.

\paragraph{1D-period tolerance}
For $\Delta_\mathrm{dir}=10~\mathrm{meV}$, the pattern-period tolerance $|\delta L_\mathrm{1D}/L_\mathrm{1D}|$ for fixed $\theta$ and $\varphi_\mathrm{1D}$ is 
$\approx 4.9\%$ for $\theta=2^\circ$ and $\approx 2.3\%$ for $\theta=3^\circ$. 

Fourier analysis reports less-than-$3\%$ period variations at a $16~\mathrm{nm}$-period lattice in Ref.~\cite{BarconsRuiz2022_NatCommun} and around-$0.1\,\mathrm{nm}$ period variations at a $38.1~\mathrm{nm}$-period lattice in Ref.~\cite{Jamalzadeh2025_ACSNano}.
Thus, such percent-level nonuniformity of $\delta L_\mathrm{1D}/L_\mathrm{1D}$ may be achievable with current advanced fabrication capabilities.

\paragraph{1D-axis alignment tolerance}
For $\Delta_\mathrm{dir}=10~\mathrm{meV}$, the pattern-orientation tolerance $|\delta \varphi_\mathrm{1D}|$ for fixed $\theta$ and $L_\mathrm{1D}$ is 
$\approx 2.8^\circ$ for $\theta=2^\circ$ and $\approx 1.3^\circ$ for $\theta=3^\circ$. 

Direct experimental reports of the angular misalignment $\delta\varphi_{\rm 1D}$ are scarce in gate-defined 1D-superlattice devices, where the superlattice axis is typically set lithographically and used as a reference. A practical estimate for angular misalignment during graphene transfer is $\delta\varphi_{\rm 1D}\sim \delta x/L_{\rm align}$, where $\delta x$ is the overlay/placement error over an alignment length $L_{\rm align}$ (often a few microns), suggesting that sub-degree misalignment is readily achievable in modern lithography.

\subsection{Off-resonant feasibility and one-directional velocity renormalization}
\label{sec:S8_5}

The off-resonant regime derived in Appendix~\ref{sec:off_resonance} does not rely on global commensurability between $\Delta\mathbf K_\xi$ and $\mathbf G_{\rm 1D}$.
Therefore it is not constrained by the strict tolerance requirements of Appendix~\ref{sec:S8_4}.
Instead, feasibility is governed mainly by whether a sufficiently large fundamental 1D coupling $|u_l|$ can be generated while keeping disorder and inhomogeneity small enough to resolve an anisotropic Dirac dispersion.

\paragraph{Design target from the perturbative velocity renormalization.}
For a 1D modulation on one layer, the downfolded Dirac Hamiltonian takes the anisotropic form
Eq.~\eqref{eq:S3:heff_incomm}, with renormalized velocities given by Eq.~\eqref{eq:S3:vstar_incomm}.
The key control knob is the 1D potential strength, which suppresses the velocity perpendicular to the modulation in addition to the inherent velocity renormalization of pristine TBG.
The flattening condition $v_x^*=0$ [Eq.~\eqref{eq:S3:flattening_u1}] occurs at
\begin{equation}
|u_l|=\sqrt{8(1-3u_0^2)}.
\label{eq:S8:u1_flat_condition}
\end{equation}
This condition is most transparent for twist angles above the first magic angle (roughly $\theta\sim 1^\circ$), where $u_0<1/\sqrt{3}$ and the perturbative velocity renormalization provides a reliable baseline. For smaller twist angles, pristine TBG is known to exhibit a sequence of higher-order magic angles at which band flattening re-emerges \cite{BistritzerMacDonald2011_PNAS}, and monolayer graphene under a 1D modulation can likewise show repeated flattening at a series of potential strengths beyond the first threshold \cite{Park2008_NatPhys_1D, BreyFertig2009_PRL, Li2021_NatNano_1DSL}. It is therefore plausible that, when moir\'e coupling and a 1D modulation coexist, additional flattening conditions may also arise for $\theta$ below the first magic angle. In that regime, however, the spectrum is expected to involve stronger multi-wave mixing and becomes less amenable to a simple low-order description, so we do not pursue a detailed characterization here.

Equation~\eqref{eq:S3:flattening_u1} and the zeros in Fig.~\ref{fig:transverse_velocity_flip} show the values of $u_l$ required to achieve a nearly flat dispersion along the direction perpendicular to $\mathbf{G}_\mathrm{1D}$.
In contrast to the resonant case, $L_{\rm 1D}$ is not fixed by momentum matching in the off-resonant regime and can be treated as a design parameter. Using $u_l=v_l/(\hbar v_\mathrm{F} G_{\rm 1D})=v_l L_{\rm 1D}/(2\pi\hbar v_\mathrm{F})$, one can in principle reach larger $u_l$ by increasing $L_{\rm 1D}$, even if the experimentally attainable amplitude $v_l$ is bounded. This comes with an important trade-off: the momentum-space extent of the 1D-induced reconstruction scales with $G_{\rm 1D}$, so choosing a longer period (smaller $G_{\rm 1D}$) reduces the range over which the bands appear flat. 
Overall, the flattening line [Eq.~\eqref{eq:S3:flattening_u1}] is best regarded as a practical design guideline rather than a strict requirement.

\bibliographystyle{apsrev4-2}
\bibliography{refs}

\end{document}